%% file: Instr10y.tex
\documentclass[11pt,manuscript]{aastex63}

\usepackage{graphicx}
\usepackage{amsmath}

\graphicspath{{.}}

\usepackage{color}
\definecolor{bettergreen}{rgb}{0.0,0.5,0.0}
\newcommand{\modified}[1]{#1}

\usepackage{natbib} 
\input{fermiStyle.tex}

\submitjournal{APJ Supplement}
\begin{document}


\correspondingauthor{R.~A.~Cameron}
\email{rac@slac.stanford.edu}
\correspondingauthor{R.~Rando}
\email{riccardo.rando@pd.infn.it}
\correspondingauthor{D.~A.~Smith}
\email{smith@cenbg.in2p3.fr}
\input{authors-merged}


\title{\fermi\ Large Area Telescope Performance After 10 Years Of Operation}

\begin{abstract}
The Large Area Telescope (LAT), the primary instrument for the \emph{Fermi Gamma-ray Space Telescope} (\fermi) mission, is an imaging, wide field-of-view, high-energy gamma-ray telescope, covering the energy range from 30~MeV to more than 300~GeV. We describe the performance of the instrument at the 10-year milestone. LAT performance remains well within the specifications defined during the planning phase,  validating the design choices and supporting the compelling case to extend the duration of the \fermi\ mission. 
 The details provided here will be useful when designing the next generation of high-energy gamma-ray observatories.
\end{abstract}

\keywords{instrumentation: detectors -- instrumentation}




%

\input{intro.tex}

\input{general.tex}

\clearpage

\input{tracker.tex}

\clearpage

\input{calorimeter.tex}

\clearpage

\input{anticoincidence.tex}

\clearpage

\input{other.tex}

\clearpage

\input{conclusion.tex}

\clearpage


\bibliography{Instr10y}

\end{document}

%% file: fermiStyle.tex
\newcommand{\fov}{{\rm Fo\kern-1ptV}}

\newcommand{\secref}[1]{\S~\ref{#1}}

\newcommand{\Tabref}[1]{Table~\ref{tab:#1}}

\newcommand{\tabref}[1]{Table~\ref{tab:#1}}
\newcommand{\Figref}[1]{Figure~\ref{fig:#1}}
\newcommand{\figref}[1]{Figure~\ref{fig:#1}}

\newcommand{\Fermi}{{\textit{Fermi}}}
\newcommand{\fermi}{\Fermi}

\input{thisPaperStyle.tex}

%% file: thisPaperStyle.tex
\newcommand{\acronymlabel}[1]{\label{acronym:#1}}
\newcommand{\acronym}[3][]{\noindent\makebox[70pt]%
  {\ifx\\#1\\#2\else#1\fi\hfill}#3%
  \dotfill\secref{acronym:#2}\\}

\newlength{\enumindent}
\newcounter{fermicnt}

\newlength{\onecolfigwidth}
\newlength{\twothirdscolfigwidth}
\newcommand{\twopanel}[4]{%
  \begin{figure}[#1]
    \centering
    \includegraphics[width=\onecolfigwidth]{#2}\hfill%
    \includegraphics[width=\onecolfigwidth]{#3} 

    #4
  \end{figure}
}

%% file: authors-merged.tex
\author{M.~Ajello}
\affiliation{Department of Physics and Astronomy, Clemson University, Kinard Lab of Physics, Clemson, SC 29634-0978, USA}
\author{W.~B.~Atwood}
\affiliation{Santa Cruz Institute for Particle Physics, Department of Physics and Department of Astronomy and Astrophysics, University of California at Santa Cruz, Santa Cruz, CA 95064, USA}
\author{M.~Axelsson}
\affiliation{Department of Physics, Stockholm University, AlbaNova, SE-106 91 Stockholm, Sweden}
\affiliation{Department of Physics, KTH Royal Institute of Technology, AlbaNova, SE-106 91 Stockholm, Sweden}
\author{R.~Bagagli}
\altaffiliation{former affiliation}
\affiliation{Universit\`a di Pisa and Istituto Nazionale di Fisica Nucleare, Sezione di Pisa I-56127 Pisa, Italy}
\author{M.~Bagni}
\altaffiliation{former affiliation}
\affiliation{Universit\`a di Pisa and Istituto Nazionale di Fisica Nucleare, Sezione di Pisa I-56127 Pisa, Italy}
\author{L.~Baldini}
\affiliation{Universit\`a di Pisa and Istituto Nazionale di Fisica Nucleare, Sezione di Pisa I-56127 Pisa, Italy}
\author{D.~Bastieri}
\affiliation{Istituto Nazionale di Fisica Nucleare, Sezione di Padova, I-35131 Padova, Italy}
\affiliation{Dipartimento di Fisica e Astronomia ``G. Galilei'', Universit\`a di Padova, I-35131 Padova, Italy}
\author{F.~Bellardi}
\altaffiliation{former affiliation}
\affiliation{Universit\`a di Pisa and Istituto Nazionale di Fisica Nucleare, Sezione di Pisa I-56127 Pisa, Italy}
\author{R.~Bellazzini}
\affiliation{Istituto Nazionale di Fisica Nucleare, Sezione di Pisa, I-56127 Pisa, Italy}
\author{E.~Bissaldi}
\affiliation{Dipartimento di Fisica ``M. Merlin" dell'Universit\`a e del Politecnico di Bari, via Amendola 173, I-70126 Bari, Italy}
\affiliation{Istituto Nazionale di Fisica Nucleare, Sezione di Bari, I-70126 Bari, Italy}
% ...

%% file: intro.tex
\section{Introduction}\label{sec:intro}

The \emph{Fermi Gamma-ray Space Telescope} (\fermi) was launched on 2008 June~11 at 16:05 UTC. Power on and commissioning of the \Fermi\ Large Area Telescope (LAT)\acronymlabel{LAT} \citep{REF:2009.LATPaper} began on 2008 June~24.

On 2008 August~4, at 15:43:36 UTC (MJD = 54682.655; DOY 217.655), during orbit 813 after launch, following an early operations checkout period, the LAT began normal full science operations, surveying the entire sky every 2 orbits, or about every 3 hours. 
Ten years after the start of the \Fermi\ science mission, \Fermi\ \modified{had completed 55992 orbits since launch}. The most recent catalog of gamma-ray sources detected by the LAT is the 4FGL catalog \citep{4FGL}, 
using 8 years of data. It was updated to 10 years, called Data Release 2 (DR2)\footnote{\url{https://fermi.gsfc.nasa.gov/ssc/data/access/lat/10yr\_catalog/}}, and in 2021 the LAT collaboration will update the catalog to DR3 with 12 years of data.

Detector technologies were chosen for the LAT that have an extensive history of high-energy physics applications with demonstrated high reliability. These detector technologies significantly improved the LAT's performance compared to previous space-based detectors in the LAT's energy range, and allowed the LAT to meet the mission lifetime goal of 10 years. The LAT is still performing well after 12 years on orbit, and we describe the performance of the instrument at the notable 10-year milestone.

In \secref{sec:general} we describe the overall instrument performance during the 10 years. The on-orbit calibration procedure for the LAT was introduced in \citet{REF:2009.OnOrbitCalib} where calibrations and configurations in the first year of operation were described in detail. Overall performance depends on the raw data from the instrument, but also on event reconstruction. Significant post-launch improvements yielded the ``Pass 8'' analysis \citep{REF:Pass8} and subsequent refinements \citep{improvedPass8}.
In \secref{sec:tracker}, \secref{sec:calorimeter}, \secref{sec:anticoincidence} we detail the evolution of the performance of the three subsystems: tracker (TKR), calorimeter (CAL), and anticoincidence detector (ACD), respectively. Each detector performance description is divided into subsections:
\begin{enumerate}
\item \emph{Calibrations} are quantities used to translate electronics signals into physical quantities (e.g., the energy associated to a value read from an electronic channel). Calibrations can be updated after the data are collected, in which case data would need to be reprocessed.
\item \emph{Configurations} define quantities necessary for optimal data-taking, such as lists of disabled channels. Configurations are set before data acquisition and changes apply only to future data. Configurations of collected data cannot be changed afterwards.
\item \emph{Performance} includes \emph{failures} and \emph{trending}. A failure is irrecoverable damage, such as an unresponsive component having no redundancy. \emph{Trending} of operational parameters and derived quantities monitors how the performance of the LAT as a scientific instrument evolves in time.
\end{enumerate}

\secref{sec:other} summarizes the performance of the trigger and readout systems, the accuracy and stability of the clocks that generate the event timestamps, and management of passages through the South Atlantic Anomaly (SAA) region of intense background radiation.

%% file: general.tex
\section{The LAT instrument after 10 years}\label{sec:general}

After 10 years of on-orbit operation, plus thousands of hours of pre-launch testing, the LAT is performing very well. No major component failures have occurred since launch, with only three notable interruptions to LAT data-taking. In addition, since launch, there have been multiple improvements to LAT performance, both on orbit through updates of the on-board flight software, and on the ground in the data analysis and processing. Twenty updates to the on-board LAT flight software added functionalities, fixed bugs, provided faster task execution, and simplified operation. 
Three major updates of the ground-based analysis software, along with smaller incremental improvements, increased LAT gamma-ray collection efficiency, improved the accuracy of the photon energy and direction measurements, and improved the separation of gamma-ray photons from the significantly larger flux of charged cosmic-ray particles triggering the LAT \citep{improvedPass8}. 

\begin{figure}[htbp]
  \centering
  \includegraphics[width=0.45\textwidth]{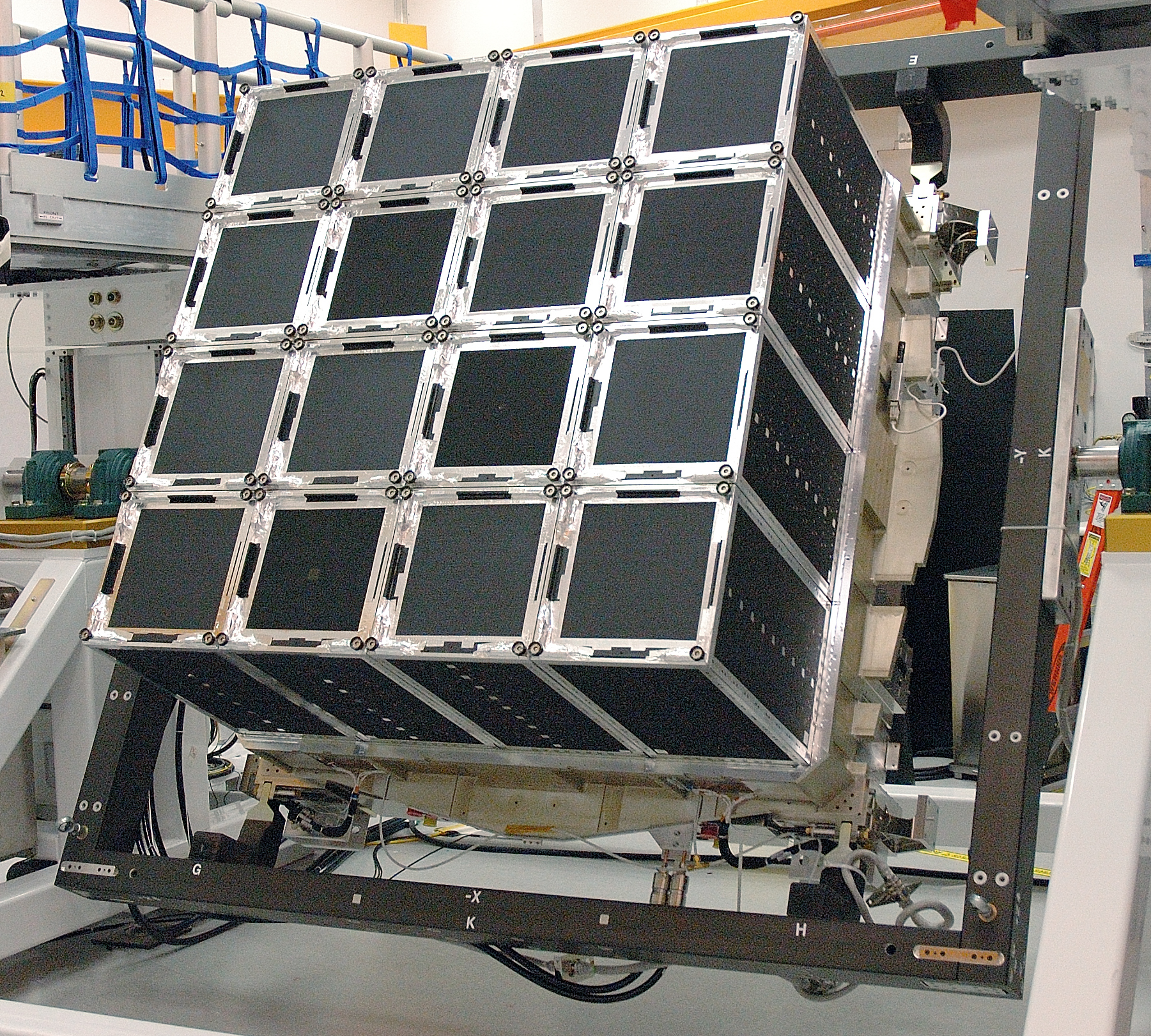}
  \includegraphics[width=0.4\textwidth]{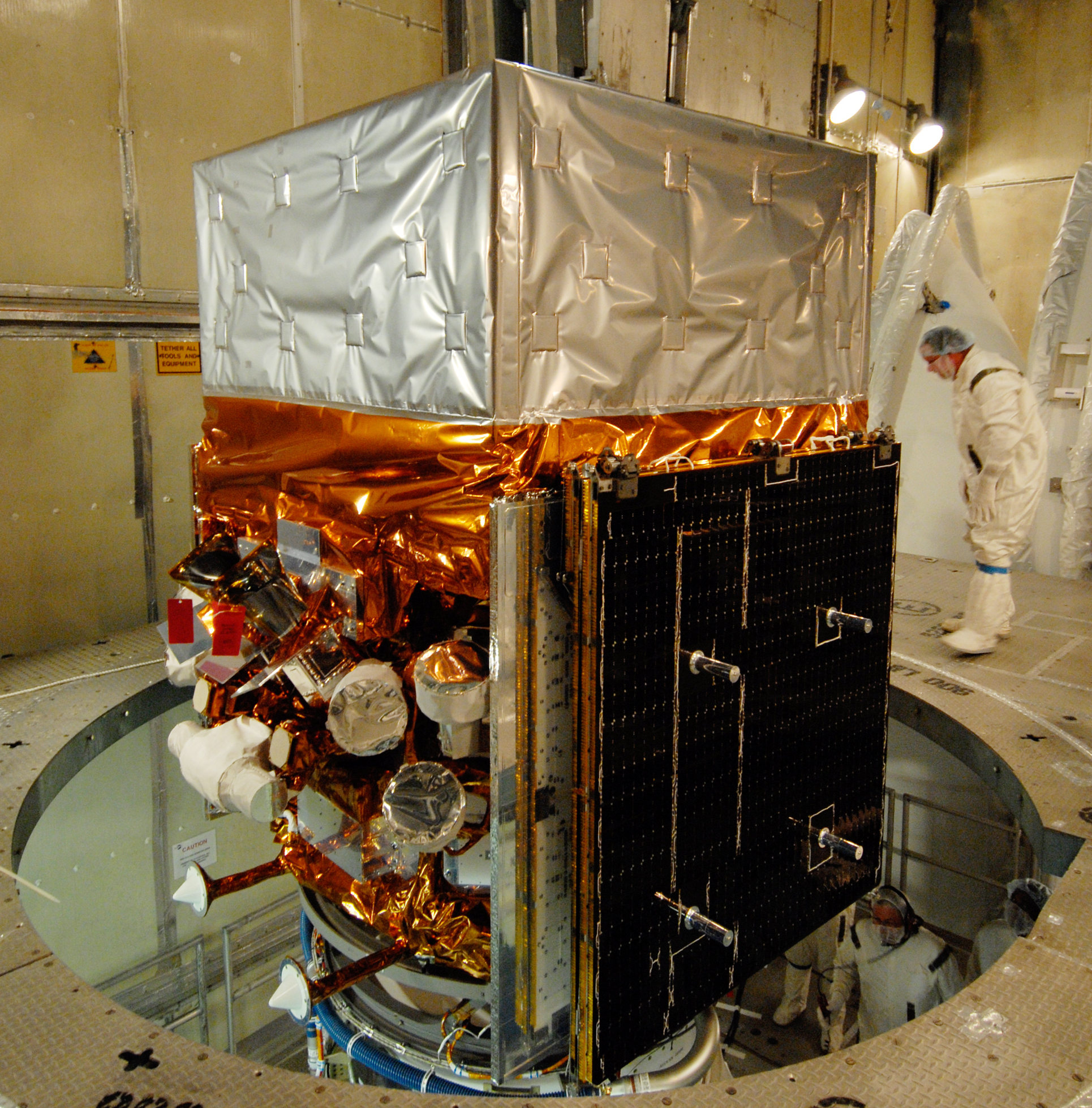}
  \caption{\modified{Left: the LAT during integration, showing the 16 tracker towers. Right: the \fermi\ spacecraft with the LAT on top, integrated on top of the Delta IIH rocket, on the launchpad at Cape Canaveral. Figure 1 of \cite{REF:2009.LATPaper} provides an artist's cutaway drawing of the LAT.}}
  \label{fig:latpicture}
\end{figure}

\input{orbit.tex}
\input{runs.tex}
\input{event-data-dqm.tex}

%% file: orbit.tex
\subsection{In orbit}
\label{subsec:orbital}

\begin{figure}[htbp]
  \centering
  \includegraphics[width=18cm]{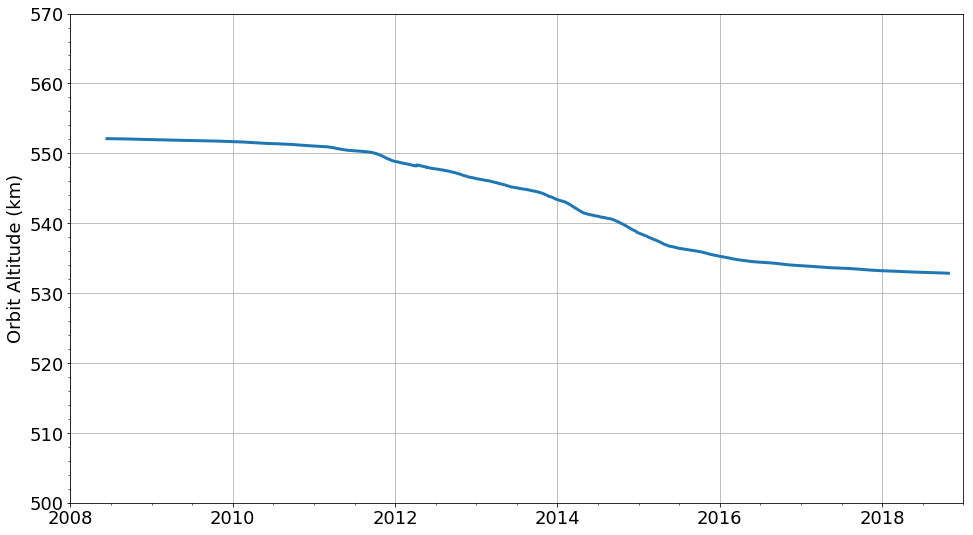}
  \caption{The orbit altitude (km) of \Fermi\ as a function of time, in years.}
  \label{fig:altitude}
\end{figure}

\Fermi\ was launched \modified{and deployed} into low Earth orbit on a Delta IIH rocket from Cape Canaveral in Florida \modified{(\figref{latpicture})}. During launch, available surplus fuel from the rocket was used to reduce the orbital inclination to 25.6$^{\circ}$, from the 28.5$^{\circ}$ latitude of the launch site. The orbit is very close to circular, with an initial eccentricity of 0.0014 that decreased to 0.0012 after 10 years. 
\Fermi's orbit altitude has decreased slightly over the 10 years and is plotted as a function of time in \figref{altitude}. The orbital period changed from the initial value of 95.7 minutes to 95.3 minutes. Altitude loss has not been constant over time. The rate of decrease was greater from 2012 to 2015 during the period of maximum Solar activity, when heating of the Earth's atmosphere is greater and the atmospheric scale height is larger, producing more atmospheric drag on \Fermi. 

\begin{figure}[htbp]
  \centering
  \includegraphics[width=18cm]{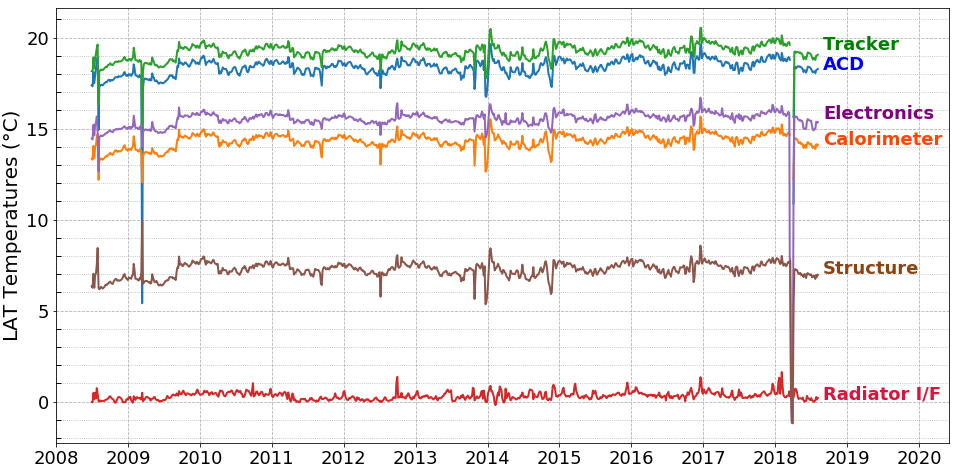}
  \caption{Temperatures ($^\circ$C) of various parts of the LAT as a function of time, in years. Long-term temperatures are quite stable, with small annual variations, and excursions due to fixed orientations and other reasons, and the larger changes and recoveries in 2008, 2009, and 2018 due to LAT power outages.}
  \label{fig:LATtemps}
\end{figure}

LAT temperatures have remained quite stable over the long-term mission. \figref{LATtemps} shows plots of temperatures of various parts of the LAT as a function of time, selected from the large number of sensors. The temperatures vary with position, with the LAT thermal radiators being the coldest, and temperatures rising towards the top of the LAT, through the LAT mechanical support structure, then the LAT calorimeter modules and electronics units, and with the LAT Tracker towers and the ACD being the warmest, and having the largest annual variations. LAT temperatures are maintained within moderate ranges, with the LAT radiators having the most extreme temperatures. Small annual variations in LAT temperatures can be seen, plus short-term variations of various magnitudes due to intervals when \Fermi\ was held in fixed pointings and other reasons. The small step increase in LAT temperatures in mid-2009 comes from the increase in \Fermi's rocking angle: to continuously scan the entire sky, \Fermi\ points at alternating sky hemispheres on each orbit. \Fermi\ rocked $\pm 35^{\circ}$ from the orbital plane in the early mission, and $\pm 50^{\circ}$ thereafter, to improve the temperature of the spacecraft batteries and hence ensure their long-term performance, while maintaining uniform sky coverage.

The large short-term temperature drops and recoveries shown in March 2009 and March 2018, and a smaller short-term change in 2008, are due to LAT power outages during the science mission. Three notable interruptions to LAT on-orbit operations have occurred since launch: on 2008 July 31 (shortly before the start of the LAT science mission), 2009 March 11, and 2018 March 16. Each time, the LAT was automatically powered off by the \Fermi\ spacecraft. The general autonomous safing action for the LAT is to turn off electrical power to the instrument, and to allow its temperature to be kept in a survival temperature range, maintained by thermostat-controlled survival heaters on the LAT structure that are powered by the spacecraft. The LAT can survive for an indefinite period of time in that state. This LAT safe state provides time for the \Fermi\ Flight Operations Team at the \Fermi\ Mission Operations Center (MOC) at NASA's Goddard Space Flight Center (GSFC) and the LAT experts within the LAT Collaboration to diagnose the problem, and devise and implement a fix. 

On 2008 July 31, an intermittent short first occurred in the wiring harness for the spacecraft Thermal Interface Board (TIB) module. The TIB module provides signal conditioning for several LAT temperature sensors on the LAT, to allow the spacecraft to monitor the temperature at various locations. This monitoring allows the spacecraft to perform the autonomous, automatic LAT safing if any monitored temperature is either too low or too high. The TIB harness failure caused several LAT temperature readouts to falsely appear too cold, outside the safe range, and so the spacecraft turned off the LAT power. It was quickly realized that the abrupt very large change in several measured LAT temperatures was not physically possible, and the problem was traced to the spacecraft TIB harness failure. The failure occurred very near the end of the initial on-orbit activation and checkout of the spacecraft and science instruments following launch. Assessment of the exact nature of the failure, followed by restarting the LAT and restoring it to readiness to perform science data collection, took 2 days. The on-board LAT temperature checks affected by the TIB harness failure have since been disabled to prevent further repeats of this problem. The disabling was done knowing that LAT temperatures will change only slowly because of the LAT's large thermal mass, which allows ground operators to see any real temperature changes and take corrective action before the temperatures change too much.

On 2009 March 11, the LAT computer that interfaces to the spacecraft suffered a software error and stopped operating normally. Several minutes later, this led the other two operating event processing computers in the LAT to also stop functioning correctly from a separate software error, as they attempted to process very high event rates during a transit of \Fermi\ through the SAA (see \secref{subsec:saa}), a region of high density of geomagnetically trapped charged particles. During the LAT recovery, the spacecraft went into its safe-mode because of an issue with extended collection of diagnostic data, which then resulted in the LAT being automatically powered off by the spacecraft. Science data collection with the LAT was restarted by ground commanding on March 14. The errors in the LAT computers were determined through captured diagnostic data and fixed in a subsequent LAT flight software update to prevent recurrence.

\begin{figure}[htbp]
  \centering
  \includegraphics[width=16cm]{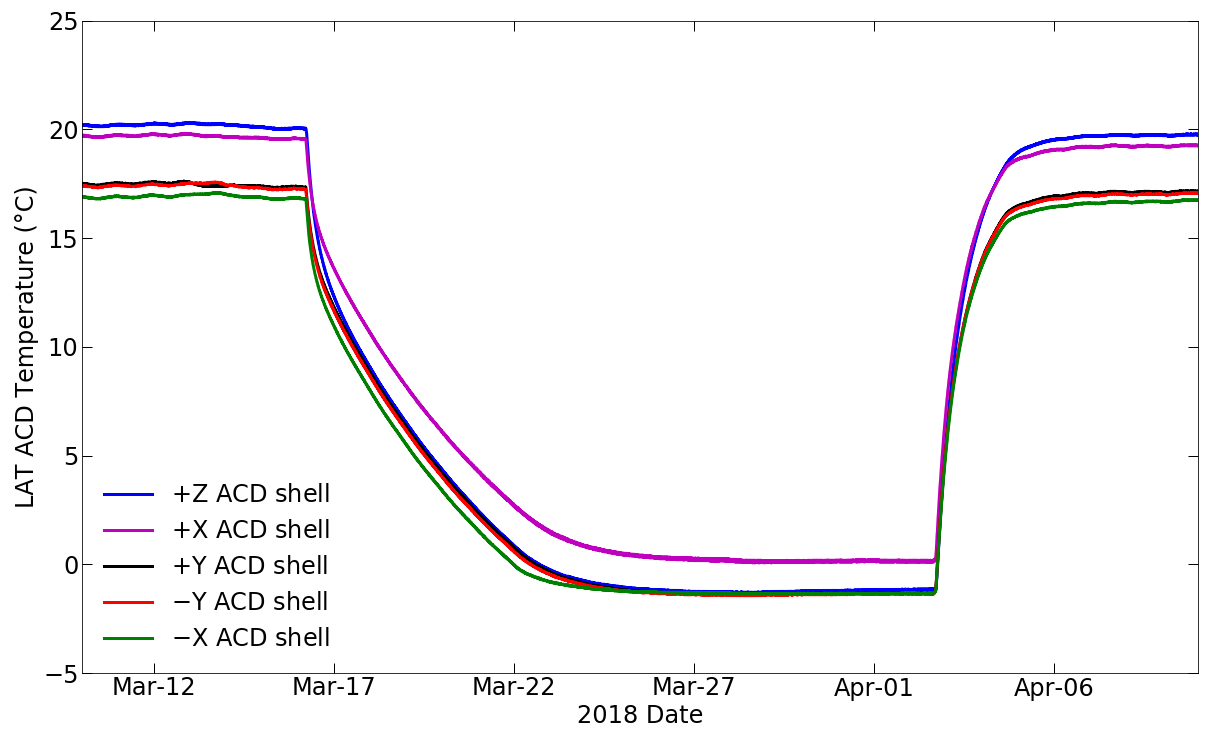}
  \caption{LAT ACD shell temperatures ($^\circ$C) as a function of time before, during, and after the LAT safing event in March 2018. These LAT temperatures change slowly over timescales of days after the LAT was powered down on March 16 and restarted on April 2, because of the large thermal mass of the LAT. The two hotter temperatures are from the Sun-facing +X side (magenta curve) and +Z top (blue curve) of the ACD shell, and the three colder temperatures are from the -Y and +Y sides (red and black curves) and anti-Sun-facing -X side (green curve) of the ACD shell. }
  \label{fig:ACDtemps}
\end{figure}

On 2018 March 16, the \Fermi\ spacecraft went into safe-mode and powered off the LAT, because the $-$Y solar panel rotation drive on \Fermi\ stopped moving. The LAT remained powered off for over 17 days, as the solar panel problem was investigated. The LAT's temperature was kept at survival levels by its spacecraft-powered heaters. Ground commands to power up proceeded without problems on April 2. The first science data were collected the same day, but the first science data flagged as suitable for routine science analysis were collected on April 8, because of the slow thermal response of the LAT calorimeter, which has temperature-dependent performance. \figref{ACDtemps} shows the detailed time profile of LAT temperatures during the March 2018 LAT power outage, demonstrating the slow rate of change of LAT temperatures, because of the slow thermal response of the 1.6 ton LAT calorimeter. The LAT missed a total of 23.3 days of routine science data collection, the longest interruption since launch. 

With one solar panel on \Fermi\ stuck since March 2018, the rocking profile  enabling the LAT all-sky survey was replaced with periods of various alternating rocking angles. This keeps the \Fermi\ power system operating nominally, but with minimal changes to the LAT's long-term sky exposure\footnote{\url{https://fermi.gsfc.nasa.gov/ssc/observations/types/post\_anomaly/}}. Autonomous repointing of the LAT boresight (along the \Fermi\ +Z axis) to temporarily search for high-energy `afterglow' emission from gamma-ray bursts has also been disabled.\looseness=-1

\begin{figure}[htbp]
  \centering
  \includegraphics[width=16cm]{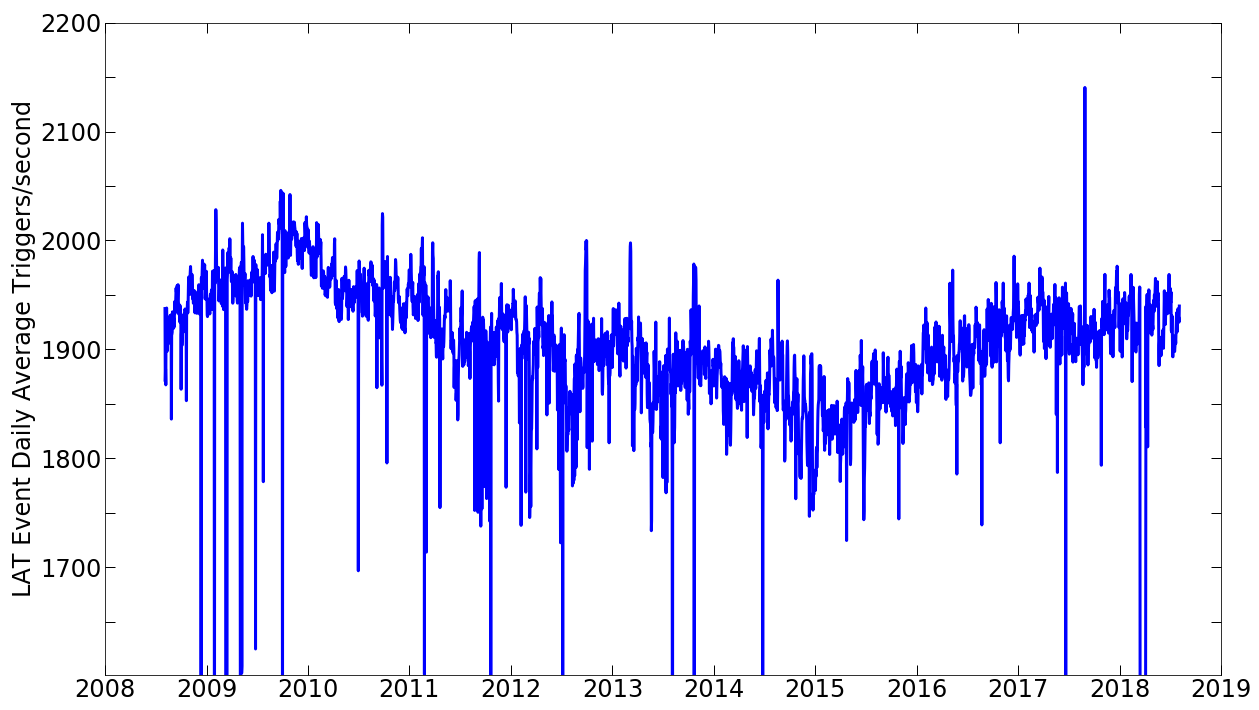}
  \caption{The daily average rate of triggers read out by the LAT versus time, in years. About 20\% of these event triggers pass through software filters in the LAT and are downlinked to the ground for analysis. Short-term variations in the trigger rate are caused by various operational issues, but variations corresponding to the 53-day precession period of the \Fermi\ orbit can be seen, superimposed on the multi-year timescale variation related to the 11-year solar cycle and the resultant variation in the density of trapped charge particles in the \Fermi\ orbit. The spike of high LAT trigger rate in August 2017 was caused by a noisy trigger in CAL module 4, which caused periods of excessive trigger rates before it was disabled by command. }
  \label{fig:trigger-rate}
\end{figure}

\tabref{milestones} gives some numbers describing LAT performance over the first 10 years of the mission. The overall uptime of the LAT for data-taking was 99.8\%, after allowing for time spent transiting the SAA (about 15\% of the time). The 0.2\% of time not used for data-taking or in SAA transits includes the aforementioned 3 major outages, plus down time for upgrades of the LAT flight software, plus regular (twice per year) charge injection calibrations of the LAT detector sub-systems, plus several short unplanned stops in LAT data-taking for various reasons, including unanticipated short transits of corners of the defined SAA region (see \secref{subsec:saa}), configuration verification errors (see \secref{subsec:dataruns}) and infrequent commanding errors. Almost 600 billion triggered readouts of the LAT detectors occurred in the first 10 years, corresponding to a daily average trigger rate of about 1900 triggers per second (\figref{trigger-rate}). Each event readout is then passed through flight software event filters running in the two Event Processing Unit (EPU) computers in the LAT. The EPUs perform simple filtering to discriminate between gamma-ray photons and charged particles, \modified{see \S 3.1.2 in \citet{REF:2012.LATClassification}}. The EPUs discard about 80\% of the events, and losslessly compress the raw data for the remaining 20\% of the events into a data stream with a rate of about 1.5 Mbps, corresponding to an average rate of about 400 triggers per second. These compressed event data are continuously transferred from the LAT to the \Fermi\ spacecraft and stored in a solid-state data recorder. About once per orbit, a radio data link is opened by command between \Fermi\ and the ground, via the TDRSS geostationary data-relay satellite network, to the TDRSS ground stations in White Sands, New Mexico. The data are transmitted at about 40 Mbps via TDRSS, and then transferred to the \Fermi\ MOC at the GSFC (see also \secref{subsec:recon}). The MOC then transfers the raw LAT event data, plus engineering housekeeping data for the LAT and related spacecraft data, to the LAT Instrument Science Operations Center (LISOC) at the SLAC National Accelerator Laboratory (SLAC) at Stanford University. The downlinked data are checked for gaps at both the MOC and the LISOC, and any missing data are re-transmitted from \Fermi\ to the ground. There has been negligible loss of downlinked data during the mission. About 120 billion LAT events have been received at the LISOC and processed there during 10 years. 

After delivery to the LISOC, reconstruction algorithms process each LAT event to search for and reconstruct charged particle tracks and electron-positron pair-production vertices in the tracker, and to find associated energy signals in the calorimeter. Multi-parameter selection filters applied to the reconstructed data distinguish gamma-ray photon events from charged-particle events. Multiple classes of assessed purity of photon selections are used to classify events. The summary data for the reconstructed filtered photons are then delivered back to NASA for immediate public release. LAT photon data are publicly available at NASA's \fermi\ Science Support Center
(FSSC)\footnote{\url{https://fermi.gsfc.nasa.gov/ssc/data}} at the GSFC. About 1.13 billion ``source'' class gamma-ray photons have been detected by the LAT in 10 years and are publicly available through the FSSC. This number of photons corresponds to a rate of about 4 photons per second. However, this photon count includes photons from interactions of Earth limb-grazing cosmic rays in the upper atmosphere: only about 40\% of the detected LAT ``source'' photons have directions less than 110$^{\circ}$ from the local zenith and come from the sky clear of the Earth's limb. About 3.03 billion events with slightly looser classification as photons (``transient'' class events) have also been delivered to the FSSC for public release in the 10 years. These events have slightly more contamination by residual charged particle events, but are suitable for studies of brighter transient gamma-ray events such as gamma-ray bursts, where higher background is not a problem for analysis and source localization, and the sensitivity in these cases is photon limited rather than background limited.

\begin{table}[htb]
  \begin{center}
    \begin{tabular}{lll}
      \hline
      Science Mission start & 2008-08-04 15:43:36 UTC & \\
      Days & 3652 & \\
      Orbits & 55179 & \\
      LAT runtime fraction (including SAA transits) & 99.8\% & \\
      LAT triggered event readouts & $598 \cdot 10^9$ & \\
      Downlinked LAT events & $120 \cdot 10^9$  & \\
      Publicly released LAT photon-like events (transient class) & $3.03 \cdot 10^9$ & \\
      Publicly released Source Class photons & $1.13 \cdot 10^9$ & \\
      \hline
    \end{tabular}
    \caption{LAT Performance Measures after 10 Years of \Fermi's Science Mission.}
    \label{tab:milestones}
  \end{center}
\end{table}

%% file: runs.tex
\subsection{Data collection runs}
\label{subsec:dataruns}
LAT data taking is split into \emph{runs}, each one generally spanning one \Fermi\ orbit. There were 55,259 science runs during the first 10 years of the science mission, spanning 55,179 orbits. Occasionally, longer timespans between short SAA transits are split into 2 separate science runs, so the number of science runs is slightly larger than the number of orbits. Not all of these science runs were used to observe the gamma-ray sky. During 2011 and 2012, 24 observation runs by the LAT were performed to search for Terrestrial Gamma-ray Flashes \modified{\citep{TGFpap}}, in which the LAT boresight was pointed at the orbit nadir, i.e., towards the Earth, accomplished by using a 180$^{\circ}$ rocking angle (angle of the \Fermi\ +Z axis from zenith pointing, perpendicular to the orbit plane). In addition, 55 non-routine pointings of \Fermi\ have been done to observe Targets of Opportunity (ToO) with the LAT\footnote{\Fermi\ Targets of Opportunity are listed at \url{https://fermi.gsfc.nasa.gov/ssc/observations/timeline/too/}}. 
During ToO observations, the \Fermi\ spacecraft inertially points the LAT boresight toward the target when it is not occulted by the Earth, and resumes routine sky survey during target occultations. During 2014, \Fermi's observing strategy was altered to provide more exposure toward the Galactic Center, for increased sensitivity for a search for enhanced gamma-ray emission from the anticipated disruption of the object G2 during its closest approach to the supermassive black hole at Sgr A* \citep{REF:2015.G2atGalCenter}.\looseness=-1

Dividing LAT data-taking into a sequence of runs also has the benefit of enabling a process of regular checking and reloading of the extensive LAT on-board electronic configuration. A large number of data registers in the LAT trigger and dataflow electronics and the detector sub-system electronics define and control data capture and movement in the LAT. This makes the LAT highly configurable, allowing detailed optimization of LAT performance and extensive re-configuration ability in the event of detector or electronics component failures or performance degradation during the mission. 

The LAT operates in a high-flux environment of charged particles (cosmic rays and terrestrial sourced particles) that can cause single event upsets in the electronics. To reliably operate in such an environment, the LAT electronic configuration is regularly automatically verified. Before each new science run, the configuration is loaded, then read back and checked. A check mismatch generates an error message from the LAT, and generally data-taking does not start for that run, although that can be overridden. At the end of each run, the LAT electronic configuration is again read out, and if a mismatch is found, the configuration data are recorded for later analysis on the ground. Additionally, a command to check the electronic configuration is run on the LAT at the end of every transit through the SAA region of high density of charged particles. In the first 10 years of the mission 11 configuration mismatches occurred: 4 in ACD electronics, 3 in Calorimeter electronics, 2 in Tracker electronics, and 2 in dataflow electronics. All these corruptions occurred during transits through the SAA region. The SAA transits also produce most of the corruptions of LAT computer memory (see \secref{subsec:saa}).

In addition to the science runs, charge injection calibrations are periodically scheduled. 
Several predefined charge levels are injected into the detector electronics readouts to calibrate readout response \modified{\citep{TKRreadout}}. Early in the mission they were performed frequently. The LAT's stability having been established, we decreased the rate to twice per year. In all, 80 routine charge injection calibrations across the three ACD, CAL, and TKR subsystems were performed during the first 10 years of the science mission. 

%% file: event-data-dqm.tex
\subsection{Event reconstruction, analysis and storage}
\label{subsec:recon}

\begin{figure}[htbp]
  \centering
  \includegraphics[width=16cm]{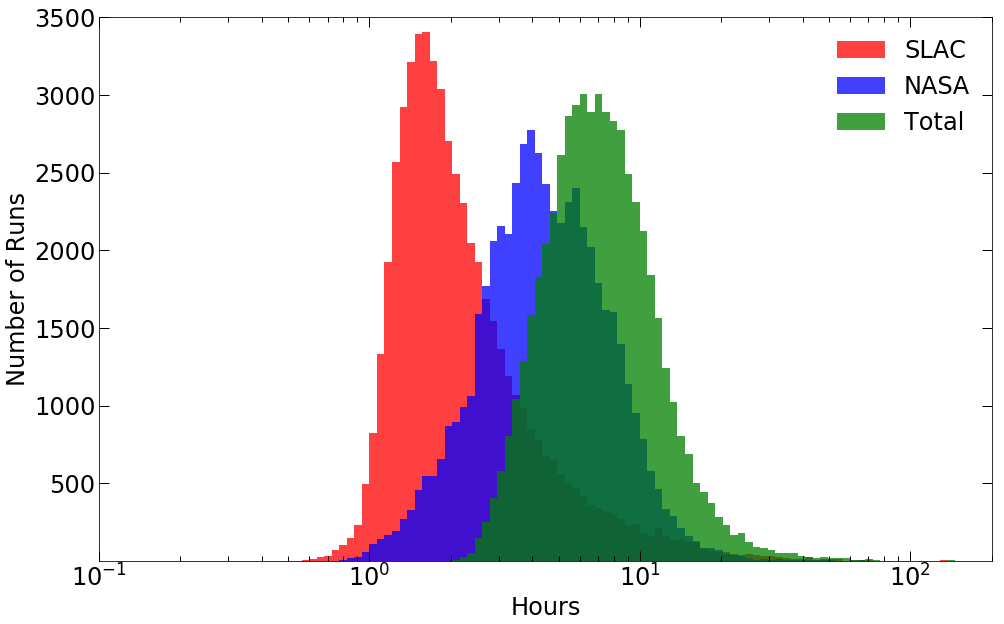}
  \caption{Histograms of data processing times, in hours, for the LAT runs over the 10 mission years, shown with logarithmic time bins. The overall total data processing time histogram is shown in green, which is the sum of data downlink and delivery time through NASA, shown in the blue histogram, plus the subsequent processing time at SLAC, shown in the red histogram.}
  \label{fig:LatencyHistograms}
\end{figure}

The LAT typically generates about 15~GB of compressed raw data every day. The data are downlinked to GSFC (\secref{subsec:orbital}) and delivered to SLAC for processing on the SLAC computer farm where over 3000 CPU cores are available for event processing, including event reconstruction and filtering. The incoming raw data are split into multiple parallel processing jobs, to process the data quickly, so as to release it as soon as possible and to obtain time-critical results for any transient gamma-ray sources. \figref{LatencyHistograms} shows histograms of data processing times for the LAT runs over the 10 mission years. The data processing expands the data volume by a factor of $\sim 50$ with respect to the raw data volume, and 3.7~PB of disk space and tape storage are used for storage and analysis. Legacy data are usually stored on tape, while a copy of the data products from the current version of the reconstruction and analysis pipeline are kept on disk, for fast and easy access. However, in the future, to reduce data storage costs, only the most recent year (or so) of data will be kept on disk, with older data stored on tape only. By far the major part of the data consists of reconstructed events, which are more than 10 times more voluminous than the channel-by-channel digitized electronics data from which the reconstructed event data are derived, and 25 times larger than the tabular photon data in ROOT format\footnote{\url{https://root.cern.ch/}} which are extracted from the reconstructed event data. About 200~MB of high-level photon data and ancillary data in FITS files are sent to the FSSC at GSFC each day.

The 1500 cores available for LAT needs at the IN2P3/CNRS facilities in Lyon, France, are commonly used for Monte Carlo simulations of LAT performance and results, thus assuring the availability of the SLAC computer farm for event reconstruction.


\subsection{Data quality monitoring and trending}
\label{subsec:DQM}
LAT subsystem and ground-support systems performance is continuously monitored through the LAT Data Quality Monitoring (DQM) System, to guarantee the quality of the LAT data delivered for science analysis.
During each step of the ground processing pipeline, histograms of different quantities are generated and stored. The monitored quantities are related to the stability of the detectors (such as pedestals,  gains, and rates of each channel), to the rates of events passing the different trigger conditions and on-board filters, and to high-level reconstruction outputs such as the average photon rate or the event energies. The performance of the timing system is also monitored (GPS lock loss and 20 MHz clock stability, see \secref{subsec:timing}). 

Weekly duty scientists use web-based tools to check all the processed science data from the LAT. Plots of all monitored quantities are automatically generated, and automatic alarms are generated if a monitored quantity deviates from its allowed range. Currently about 12000 parameters are monitored and the DQM system makes about 4100 checks on parameter ranges.

Many of these parameters vary significantly during an orbit as they depend on the geomagnetic cutoff (which influences the rates measurement, driven by the rate of charged cosmic rays) and on the spacecraft attitude (since it influences the arc length of the gamma-ray bright Earth limb in the LAT field of view). For some monitored quantities these dependencies are parameterized as a function of the geomagnetic McIlwain-L and spacecraft rocking angle, leading to normalized quantities that ideally are independent (within ~20\%) of orbit position and spacecraft pointing, and therefore are easier to monitor.

If the DQM system identifies a problem that can potentially influence the data quality, a bad time interval (BTI) is flagged by setting the variable DATA\_QUAL to a negative value for the corresponding time range in the delivered data file. While the alarms are automatically generated by the DQM system, the final decision about a BTI is taken by the duty scientist and members of the LAT collaboration. 

The operational environment of the LAT can influence the quality of the data. The main example is solar flares, which can cause excessive veto signaling from the ACD caused by X-ray absorption in the ACD tiles, reducing the LAT sensitivity to gamma rays. In this case the standard instrument response functions do not describe the detector accurately, and inclusion of data from these time intervals in science analysis can potentially alter the science results. These time intervals are identified by searching for high rates in the ACD and a deficit in the reconstructed photon rate, and are flagged as bad. BTIs due to solar flares are flagged with DATA\_QUAL equal to -1 (see Appendix A of \citet{REF:2012.LATClassification} for a detailed description of the effect of solar flares). The Pass 8 event reconstruction uses the ACD veto signals to mitigate substantially the effect of X-rays in the ACD. Analyses of gamma-ray emission associated with solar flares have been developed to compensate for dead time effects in the LAT during periods of intense X-ray emission in the flares.

\begin{figure}[htbp]
  \centering
  \includegraphics[width=14cm]{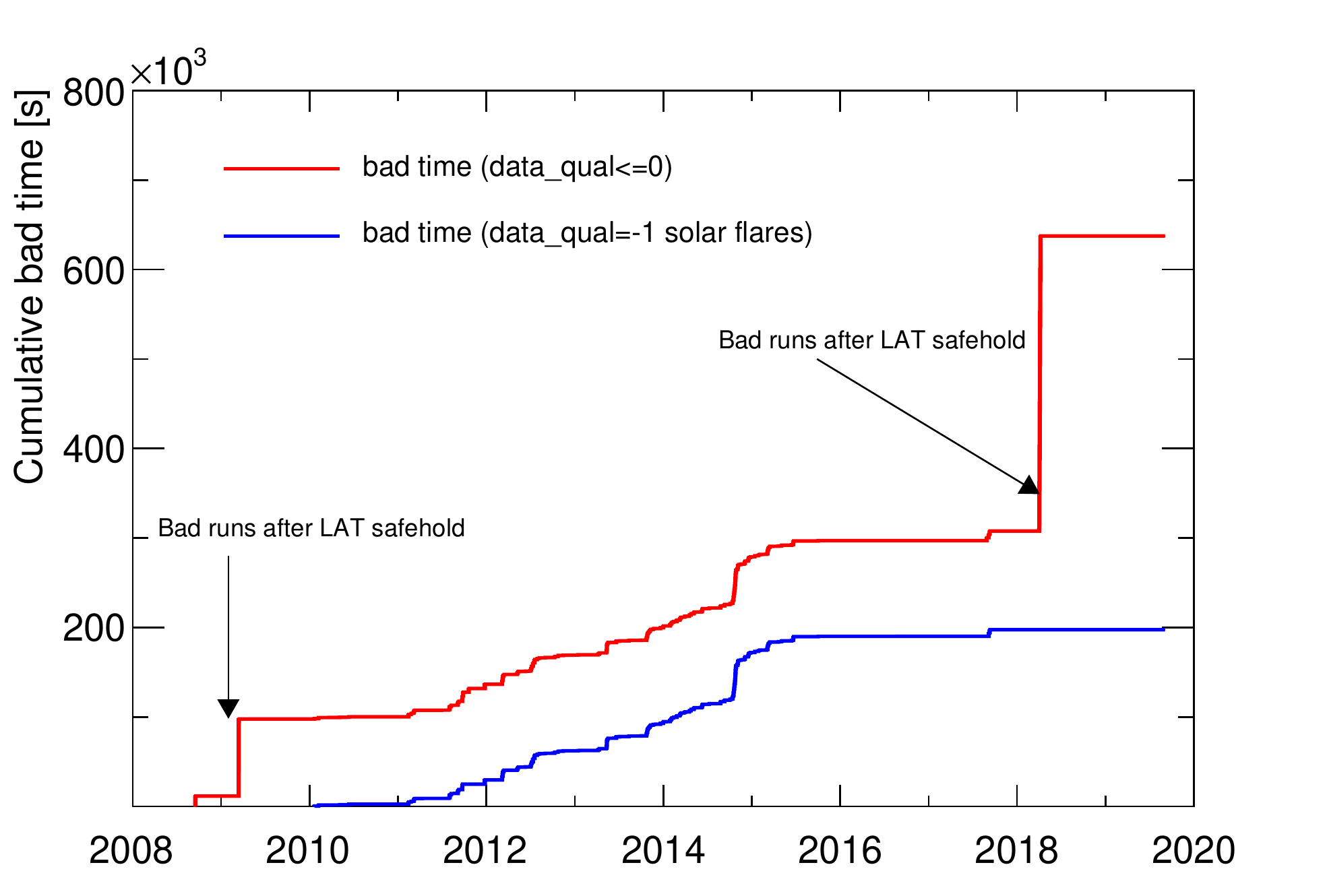}
  \caption{Cumulative number of seconds flagged as bad. The red line shows the total bad time, while the blue line shows bad time associated with solar flares. The approximately steady increase between 2011 and 2015 is due to increased solar flaring during the solar cycle maximum. The two jumps in March 2009 and April 2018 correspond to the LAT temperature stabilizing periods after LAT restarts. The recovery time was much longer in the second case because the LAT was powered off longer, and the temperature decreased to the lower allowed limit.}
  \label{fig:BTIs}
\end{figure}

Figure \ref{fig:BTIs} shows the cumulative sum of BTIs as a function of time, and the bad time due to solar flares. The period of increased solar flaring during the solar cycle maximum between 2011 and 2015 can be clearly seen in the approximately steady rise of bad time during that time. After mid-2015 the combination of the switch to Pass 8 reconstruction and the solar cycle minimum greatly reduced BTIs due to solar flares. The two jumps in March 2009 and April 2018 correspond to the periods after LAT restarts described in \secref{subsec:orbital}, while LAT temperatures were re-stabilizing. Since many instrument parameters depend on temperature, all the runs before the instrument temperatures returned to the nominal range were marked as bad. The fraction of bad data is about 0.2\% of the total LAT dataset. 


Monitoring trends of the \Fermi\ spacecraft and LAT performance and associated ground operations performance is done continuously, to complement and extend the DQM monitoring described above, and to provide a long-term history of key parameters of the LAT and \Fermi\ performance. Review and assessment of spacecraft and instrument performance and operations trends is done at weekly meetings of the spacecraft and instrument operations support teams, and performance and operations trends are also more comprehensively reviewed and assessed at 3-month intervals. Trends of performance-related quantities are assessed within the reporting period, and also over the mission lifetime.

Both direct and derived measurements for the spacecraft, the instruments, and ground processing are monitored for trends. Table \ref{tab:trending} lists some key performance parameters that DQM watches closely. ``Data latency'' is the time interval between raw data collection on the LAT and the resulting processed and filtered gamma-ray photons being delivered to NASA for public release through the FSSC, see Figure \ref{fig:LatencyHistograms}. ``Public data volume'' is the number of gamma-ray photon events released to the public through the FSSC.

\begin{table}[htb]
  \begin{center}
    \begin{tabular}{lll}
      \hline
     Spacecraft & LAT & Ground processing\\ \hline \hline
      Orbit altitude & Power supply voltages & Data processing latency \\
      Onboard data storage use & Power supply currents & Subsystem calibrations\\
      Ground contact periods & Temperatures & Data storage volume \\
      Data downlink latency & Active thermal control & Public data volume\\
      Downlinked data fraction & CPU memory errors  &  Photon classes rates \\
      Spacecraft attitude & Event trigger rate & \\ 
      Spacecraft position & Data acquisition up time & \\
       & Channel and subsystem rates  & \\
       & Channel pedestals (ACD, CAL) &\\
       & Channel calibrations & \\
      \hline
    \end{tabular}
    \caption{Sample of types and categories of parameters trended in Data Quality Monitoring (DQM) and for LAT performance trending.}
    \label{tab:trending}
  \end{center}
\end{table}

%% file: tracker.tex
\section{Tracker}\label{sec:tracker}

The tracker (TKR) is a pair-conversion detector.
The TKR is a $4\times 4$ matrix of towers, each a stack of 18 X-Y silicon (Si) strip detector modules, with interleaved tungsten (W) foils for converting gamma rays to $e^+e^-$ pairs and so initiating electromagnetic showers. The charged particles ionize the silicon as they pass through the layers, providing measurable tracks. The directions of incoming photons or charged particles ("primaries") are evaluated and correlated with the location of the tiles and ribbons of the LAT anti-coincidence detector (see \secref{sec:anticoincidence}) showing co-incidental signals, to enhance charged-particle background rejection. 

The single-sided Si strip detectors, each $8.95$~cm $\times$ $8.95$~cm $\times$ $400$~$\mu$m, are arranged into planes of $4 \times 4$ detectors, with two planes having mutually orthogonal (X-Y) strips in each module. The top 12 modules include tungsten converter foils that are $0.03$ radiation lengths thick. The W foils in the next 4 modules below these are $\sim 6$ times thicker, to maximize conversion efficiency at the cost of additional Coulomb scattering and slightly worse angular resolution. Hits in consecutive TKR layers define the ``primitive'' signal used to build TKR triggers: the TKR generates the majority of LAT trigger requests for gamma-ray physics by triggering when three consecutive X-Y planes are hit (``three-in-a-row''). The last 2 tracking modules closest to the LAT calorimeter have no conversion foils: hits in these layers can participate in but not initiate TKR triggers. 

Each individual Si plane in the TKR contains 1,536 strips, read out by 24 front-end analog ASICs in a daisy-chain configuration. Two digital readout controllers at both ends nominally each deal with half of the front-end ASICs. Strip readout is digital: ``hit'' discriminator thresholds are set to $1/4$ of the charge deposited by a single minimum ionizing particle (a ``MIP"). Recording the time the strip signal remains above threshold provides an estimate of the total charge deposit (i.e., of the total energy lost by ionizing particles traversing a strip). This Time Over Threshold (ToT) measurement is performed on the logical OR of all the strips on a TKR plane and is included in the data stream along with the digital position of the strips above threshold. 
The overall small energy loss in the TKR is especially important for primaries with energy $\lesssim 100$~MeV, which deposit most of their energy before reaching the calorimeter. \citet{REF:2007.TKRPaper} describe the TKR subsystem in more detail.

\input{tkr_calib.tex}

\input{tkr_config.tex}
\input{tkr_failures.tex}

%% file: tkr_calib.tex
\subsection{Tracker calibration}\label{subsec:tkr_calib}
Detailed descriptions of the calibrated quantities and calibration procedure for the TKR are given in \citet{REF:2009.OnOrbitCalib}.  Here we briefly review the calibration constants and discuss how some of the values have evolved during the mission.

The split point in the front-end daisy chain is nominally in the middle of the 24 ASICs, so that two equal groups of 12 ASIC outputs are directed to the two controllers. The split point can be changed, e.g., to isolate a faulty front end. A single readout controller ASIC failed before launch (see \secref{subsec:tkr_failures}), but full readout of the plane was guaranteed by applying a different split point. No failures have occurred on orbit and all other plane readout split points are still set to the middle of the plane. 

\begin{figure}[htbp]
  \centering
    \includegraphics[width=12cm]{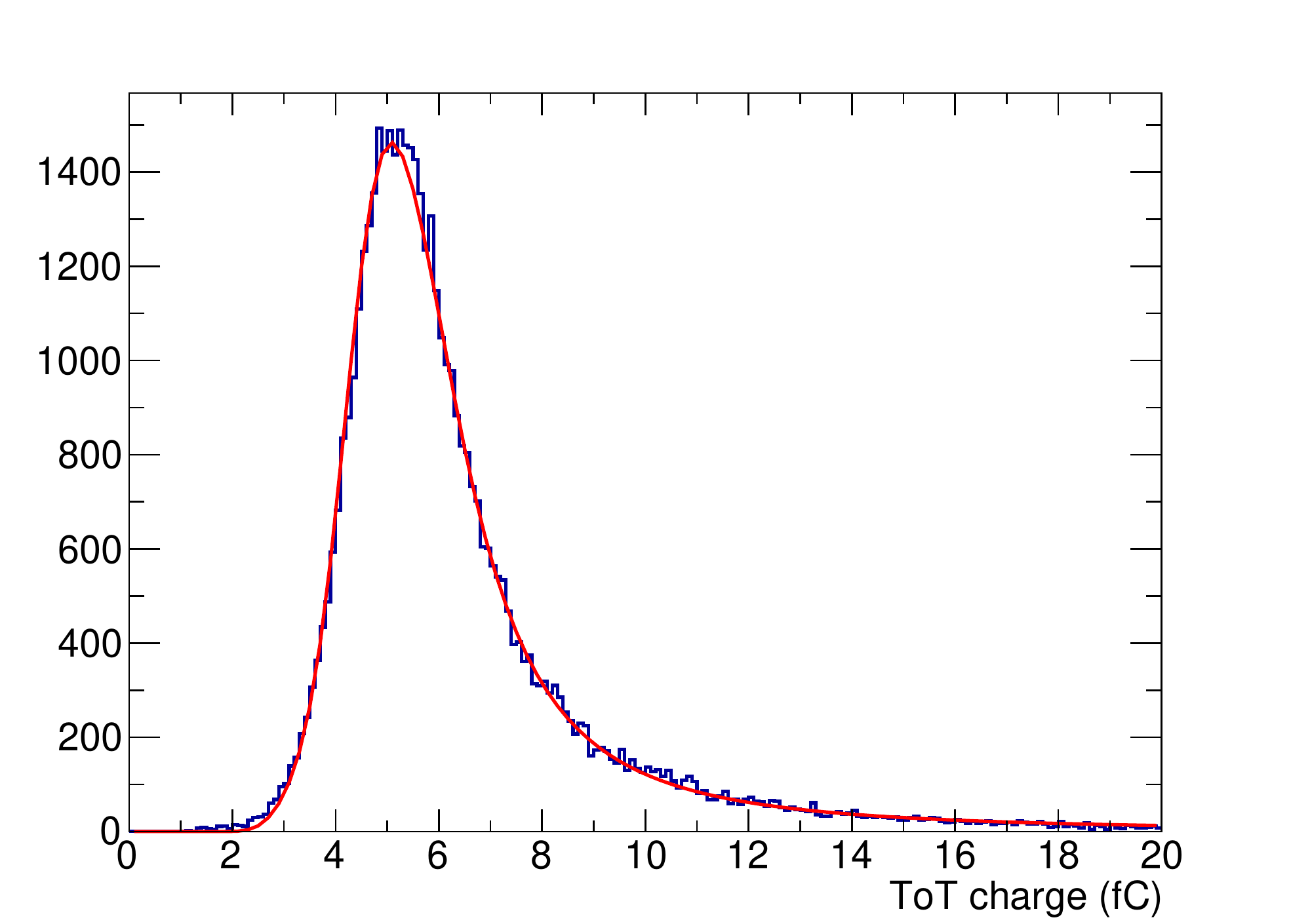}
  \caption{ToT distribution in one run for the whole TKR for a selection of MIP events, and fit to the MIP peak.}
  \label{fig:tkrmipfit}
\end{figure}

The deposited charge in any TKR channel is estimated from the ToT (see \secref{sec:tracker}): time is measured in  20-MHz clock ticks and converted into charge (fC).
Calibrations include one overall energy-scale constant and one scale factor per channel, to allow for channel-by-channel scale adjustments. 
 The overall scale factor affects the energy measurement directly as a multiplicative constant, so good accuracy is necessary. This global energy scale is monitored by observing the energy deposited by minimum-ionizing particles, mostly protons. The ToT values in the MIP dataset are corrected for the incidence angles on the Si planes (a 10\% effect). The distribution is fitted with a Landau curve convolved with two Gaussians, and the location of the MIP peak is determined with a statistical uncertainty around 0.2\%. Systematic effects include charge sharing and a residual effect of the incidence angle (of order $1\%$), evaluated by Monte Carlo simulations and corrected for. A varying amount of non-MIP contamination in different runs causes the largest remaining systematic effect. In \Figref{tkrmipfit} the uncalibrated ToT is shown for one recent run, together with the fitted curve.

 \Figref{tkrchscale} shows the location of the uncalibrated MIP peak as a function of time; each point is the average of several runs, the residual scatter, larger than the statistical uncertainties, is attributed to varying non-MIP contamination in different runs. Rare major changes in the event processing algorithms cause changes in the energy scale that would appear as abrupt offsets in the MIP peak trend: these are quantified and scaled away so they do not appear in the figure.
 A $\sim 10$\% increase in the overall energy scale occurred in the first months (\figref{tkrchscale}, left); this is interpreted as initial radiation damage quickly coming to saturation, \modified{ but a detailed study to understand the root causes of this phenomenon has not been performed}. Since then the additional increase is less than 1\% (\figref{tkrchscale}, right).
 %
 No channel-by-channel variations of more than a few percent have been seen.

\twopanel{htb!}{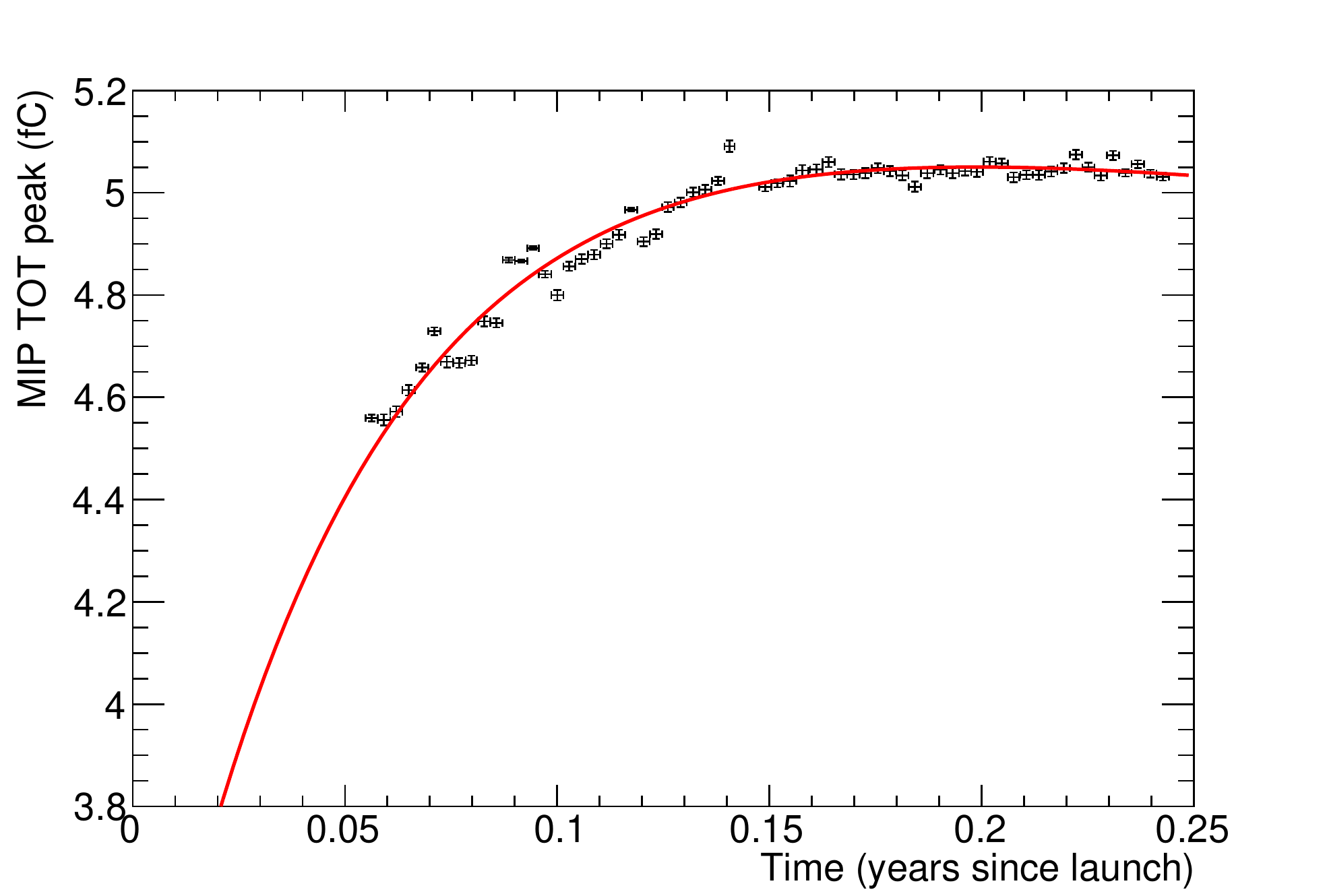}{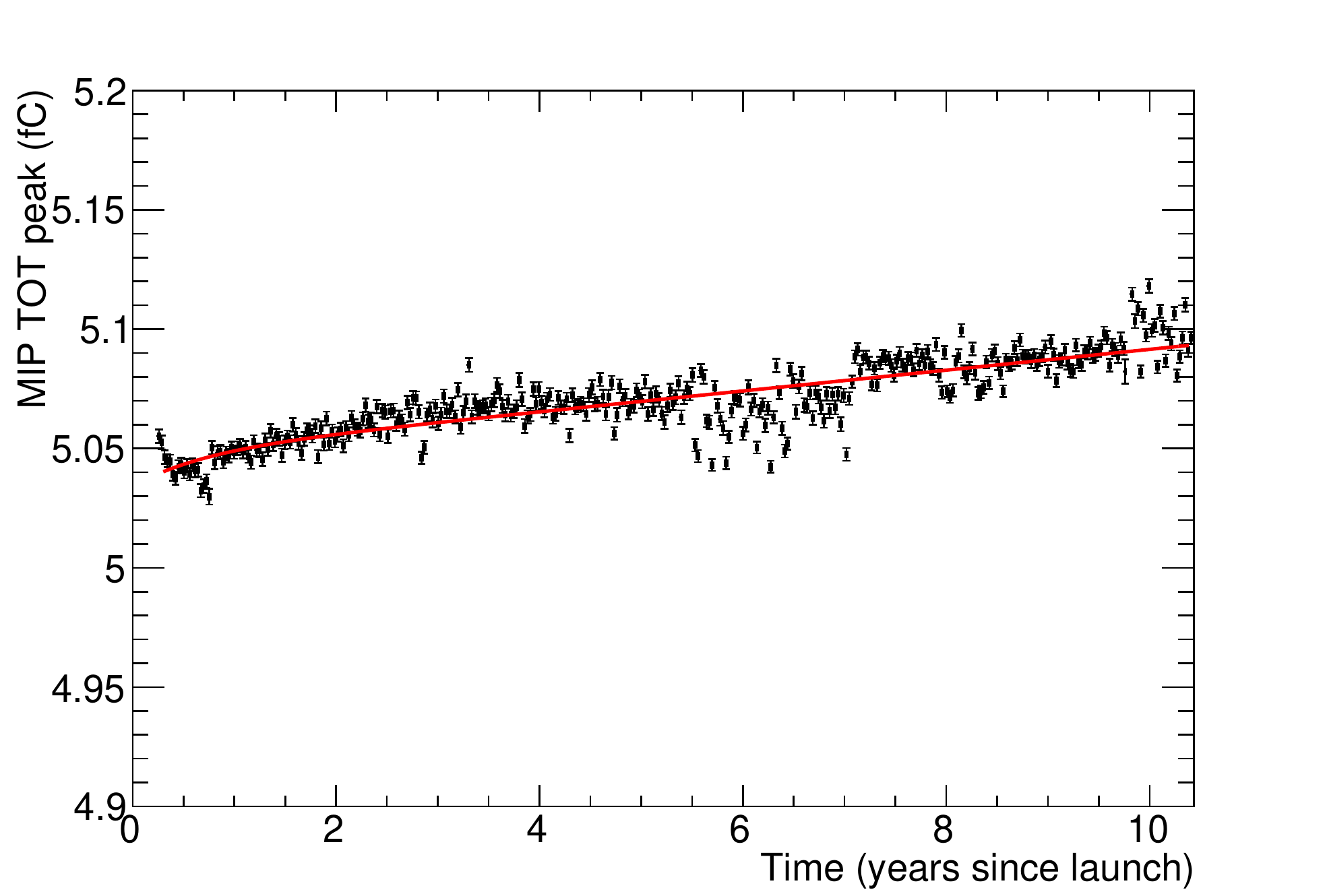}{
  \caption{Trending of the minimum ionizing particle (MIP) peak energy value during the first weeks after launch (left) and for the remainder of the mission (right).}
  \label{fig:tkrchscale}
}

\begin{figure}[htbp]
  \centering
    \includegraphics[width=12cm]{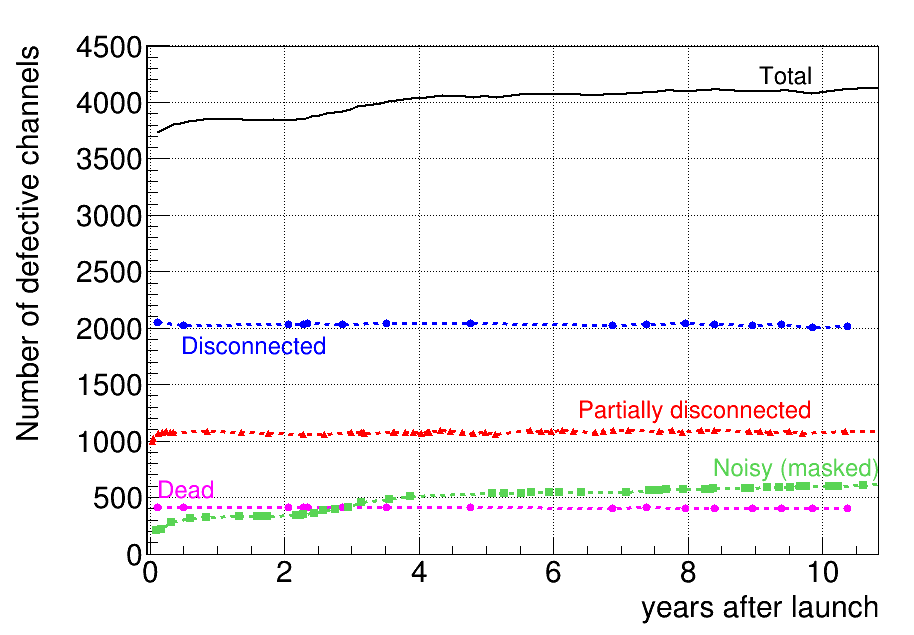}
  \caption{Dead and noisy channels in the tracker as a function of time.}
  \label{fig:tkrbadstrips}
\end{figure}

\begin{figure}[htbp]
  \centering
   \includegraphics[width=12cm]{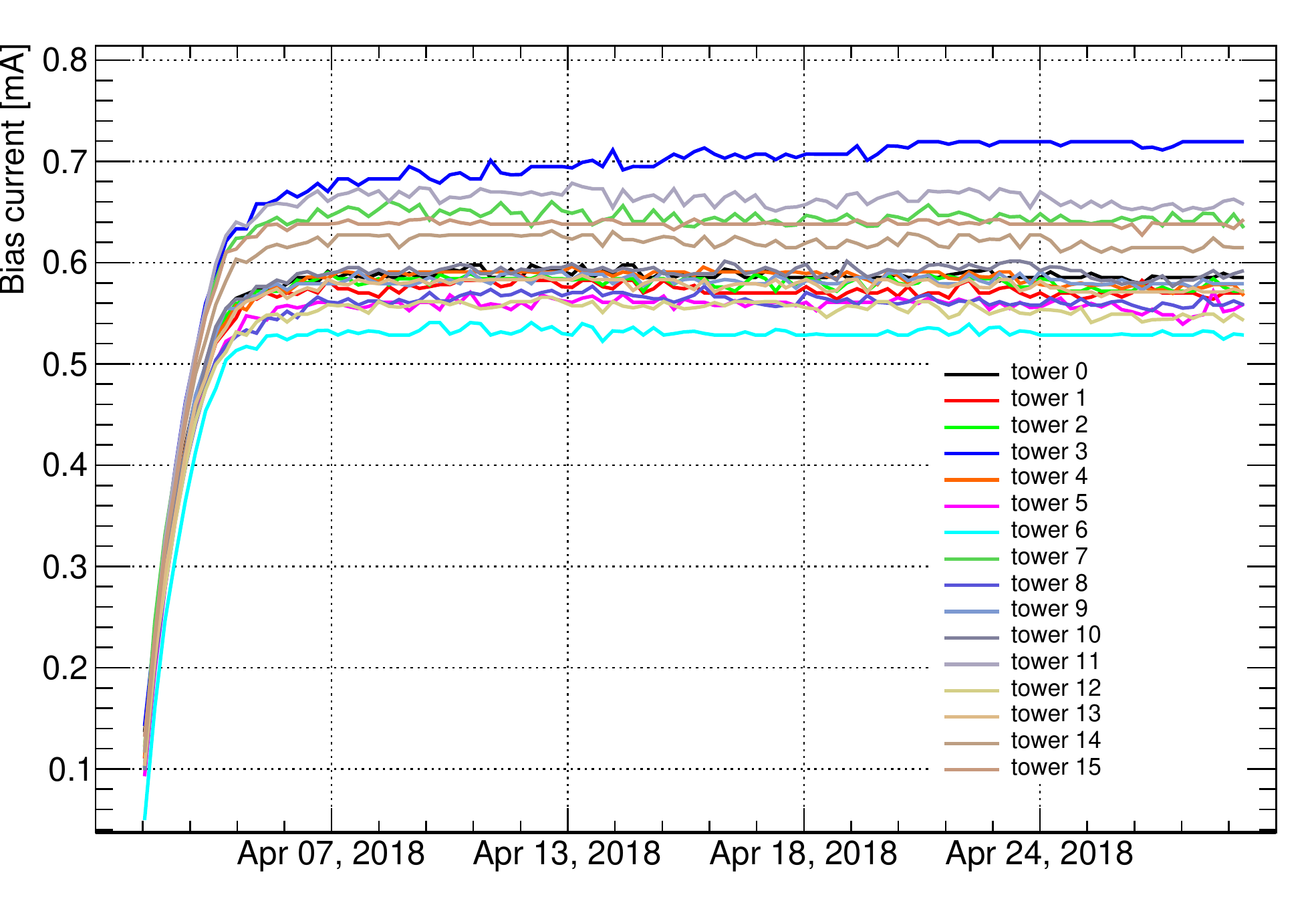}
  \caption{Tracker bias currents for the tracker towers after the \Fermi\ safemode and LAT power off period in March 2018. The bias current for Tower 3 (the top blue curve) shows a slower recovery over a duration of many days compared to the other towers, as the noise in layer 35 of tower 3 slowly recovered to pre-fault levels over that extended time period.} 
  \label{fig:TKRbiascurrent2018}
\end{figure}

We maintain lists of dead and noisy single readout channels in order to flag those that are faulty: see \Figref{tkrbadstrips}.
At launch, 3661 readout strips were declared bad (0.31\% of the total); almost 40\% of these were located in Tower 0, the first tower to be assembled. Most bad strips appeared to be too quiet: 2047 were apparently strips disconnected from the readouts, collecting no signal and having an abnormally low noise level, while 413 channels had a dead preamplifier, showing zero noise and no signal. There were also 998 partially disconnected strips,  where one or more of the wire bonds along the ladder are defective, leading to intermediate noise levels. Finally, 203 strips were electronically disabled for being too noisy (0.02\% of total). 
After 10 years of operation in orbit, the number of bad strips increased to 4087 (0.46\%).
Surprisingly, the number of disconnected and dead channels diminished to 2008 and 405, respectively,
indicating that in some cases the true cause of the malfunction is not clear and in some cases is reversible. The number of partially disconnected strips increased to 1076, while the number of noisy channels increased to 598 (0.07\%), mostly in Towers 0 and 3. 
In Tower 0, most of the noisy strips are in one ladder on silicon layer 14, and were either noisy before launch or became noisy in the first few weeks after launch. In Tower 3, most of the noisy strips are in one ladder on silicon layer 35 at the top of the tower, nearest the ACD shell. In Tower 3, the noisy ladder started becoming noisy in 2010, and more strips in that ladder became noisy over a period of several years. After 10 years, 58\% of the 598 masked noisy strips were on 2 of the 2304 silicon strip ladders in the LAT.
The recovery from the extended LAT power off period in April 2018 provided more evidence of non-standard behavior of the noisy ladder in Tower 3. \Figref{TKRbiascurrent2018} shows the TKR bias currents for the 16 towers after the March 2018 LAT power off period. The bias current for Tower 3 (the top blue curve) shows a slower recovery over a duration of many days compared to the other towers, as the noise in layer 35 of tower 3 slowly recovered to pre-fault levels over that extended time period. 

\begin{figure}[htbp]
  \centering
   \includegraphics[width=\onecolfigwidth]{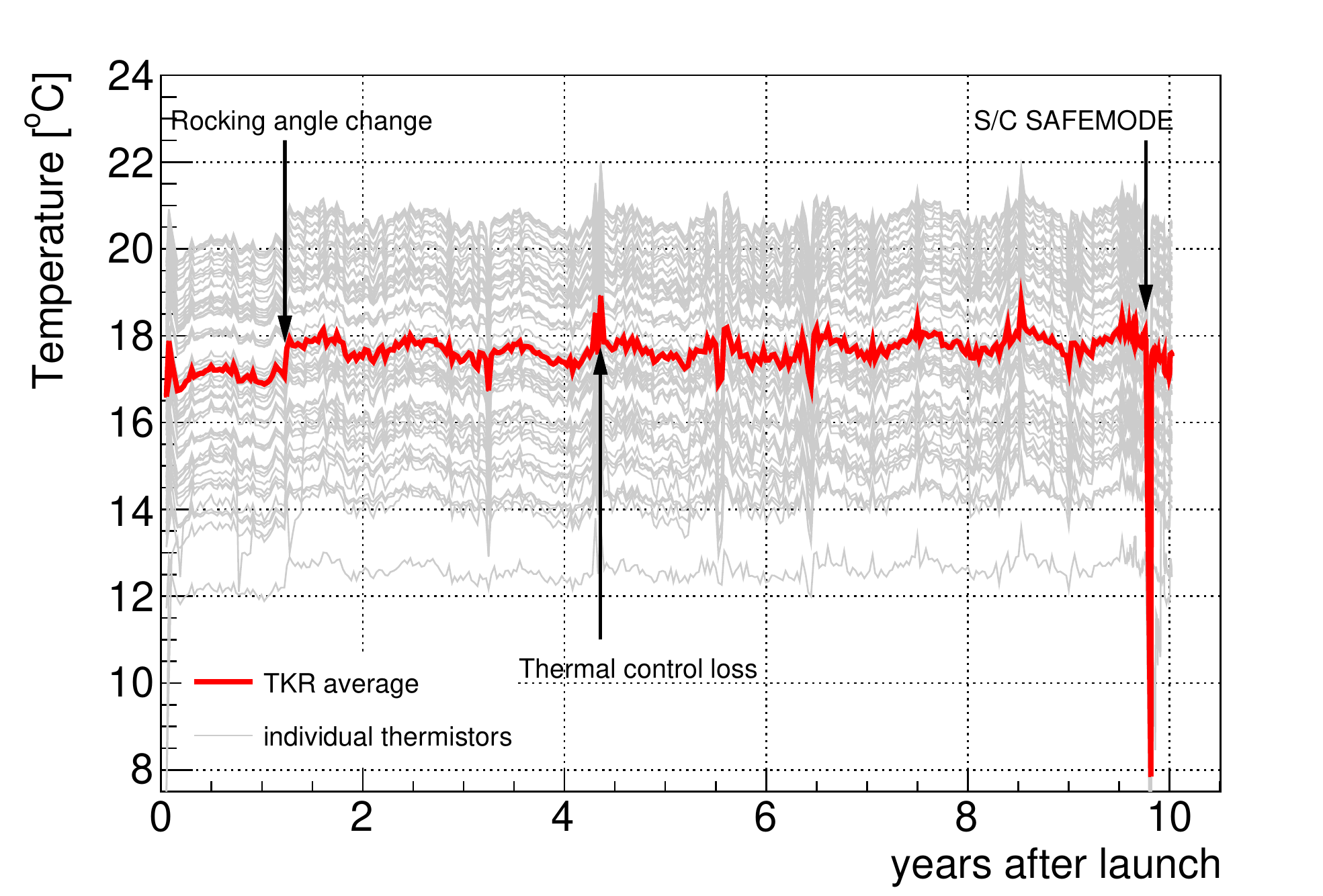}
  
  \caption{Tracker temperature, average (red) and each thermistor (gray) as a function of time (years).}
  \label{fig:tkrtemperature}
\end{figure}

In \figref{tkrnoise}, left, the TKR readout noise is averaged over all channels (ENC is Equivalent Noise Charge). The relative increase is within expectations (a couple of percent). For comparison in \figref{tkrnoise}, right, the long-term bias current increase of each module and the total are shown: the noise increase can be correlated with the overall increase in leakage current, attributed to radiation damage. The rapid variations on top of the slowly increasing trend in the plot can be correlated to temperature variations, as measured by the sensors in the TKR, in \figref{tkrtemperature}.

\twopanel{htb!}{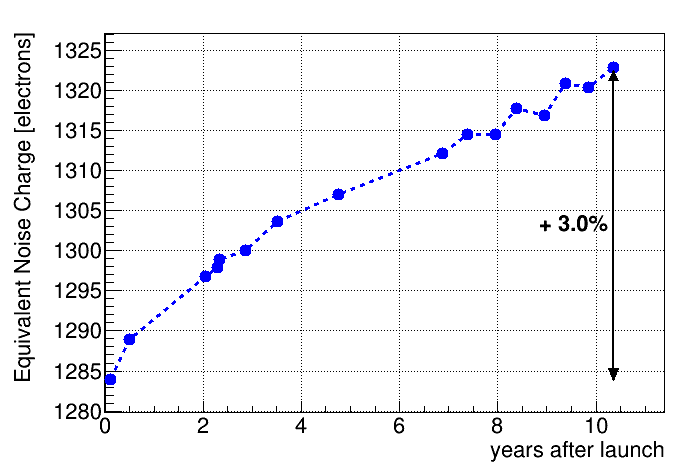}{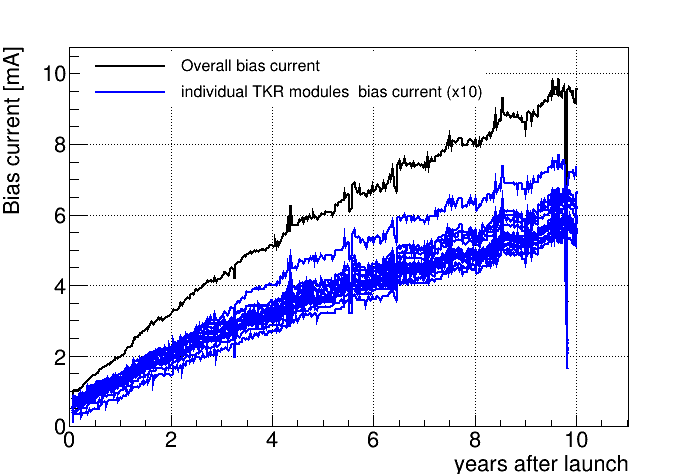}{
  \caption{Left: Tracker equivalent noise charge (ENC) trending (average over all strips; statistical error bars are smaller than markers dimension). Right: Tracker bias current trending.}
  \label{fig:tkrnoise}
}

In \figref{tkrnoisedist}, left, the noise level for each channel in the tracker is shown for two runs, one after the launch and one after almost 10 years of mission. Very low values correspond to the dead or disconnected channels, the uncertainty on the measurement of the noise level for a single channel is $\sim 7$\%. The distribution of the ratio of the values in the later run over the earlier run is shown in \figref{tkrnoisedist}, right: a weak trend towards values greater than 1 (i.e., increasing noise) is apparent. To put the noise figure in context, a MIP releases $\sim 5$ fC, as seen in \figref{tkrmipfit}, i.e., $\sim 31,000$ electron-hole pairs.

\twopanel{htb!}{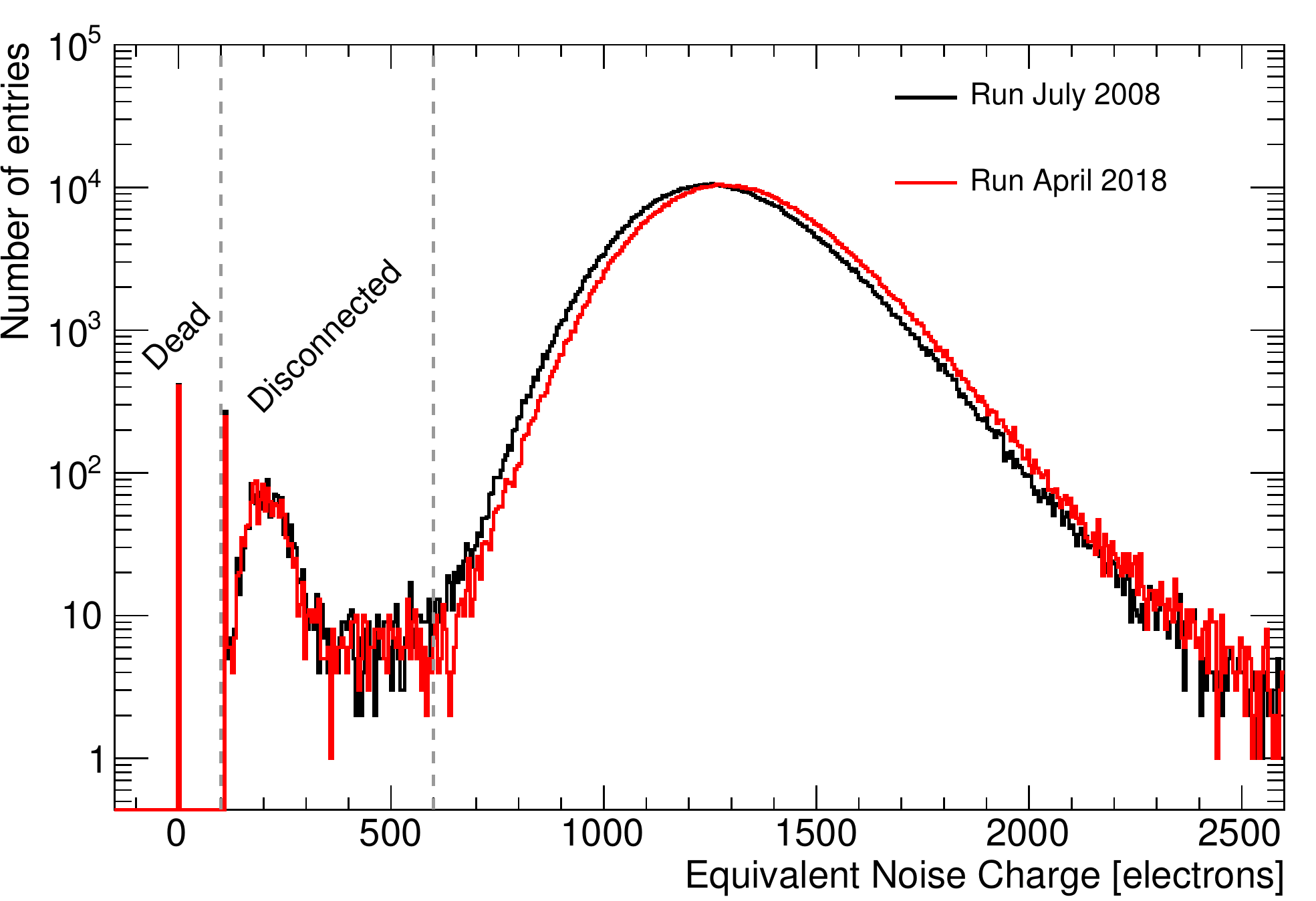}{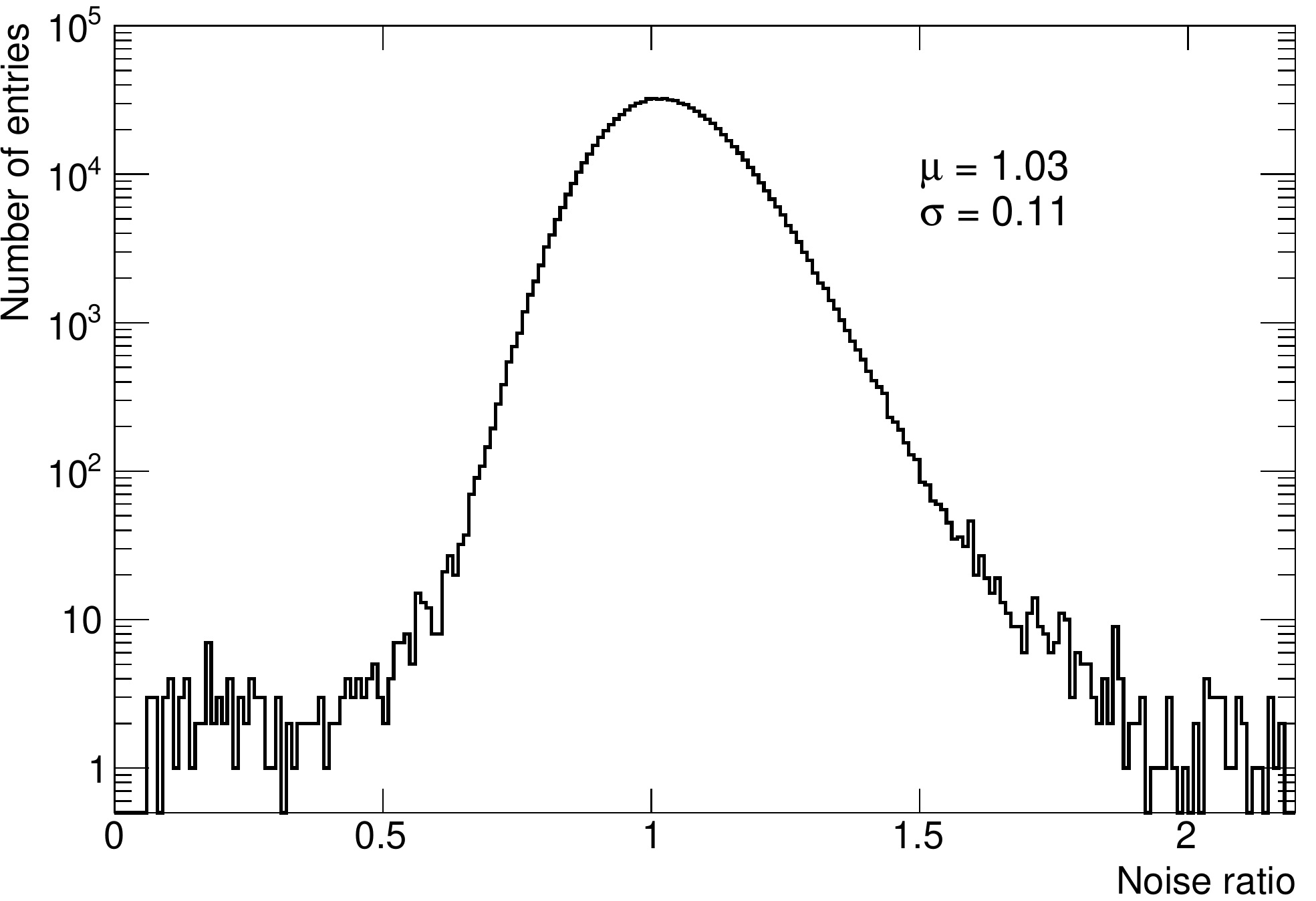}{
  \caption{Left: Tracker equivalent noise charge (ENC) distribution for two runs, one early (black data) and one late (red data) in the mission. Right: The distribution of the ratio of the ENC values for the two runs.}
  \label{fig:tkrnoisedist}
}

The TKR hit efficiency can be estimated by \modified{selecting MIP tracks and searching for missing hits where the tracks cross the active volumes}, see \figref{tkrefficiencies}, left. The efficiency uncertainty is better than $10^{-4}$. An exponential fit shows that the efficiency decreases by $3\cdot 10^{-5}$ per year, with no practical consequences. Similarly, by looking at how many recorded tracks that \modified{should have produced a trigger in an adjacent tower failed to issue one}, we can estimate the TKR trigger efficiency, see \figref{tkrefficiencies}, right. The uncertainty on this estimate is $\sim 10^{-4}$, and an exponential fit gives an increase of trigger efficiency of $2\cdot 10^{-5}$ per year, mostly due to a small increase in the first year. 
Note that the exact values in \figref{tkrefficiencies} are also affected by the details of the reconstruction process: notably with the migration to the Pass 8 reconstruction a systematic shift of $0.1$~\% is scaled away from the data, being irrelevant for performance monitoring.

\twopanel{htb!}{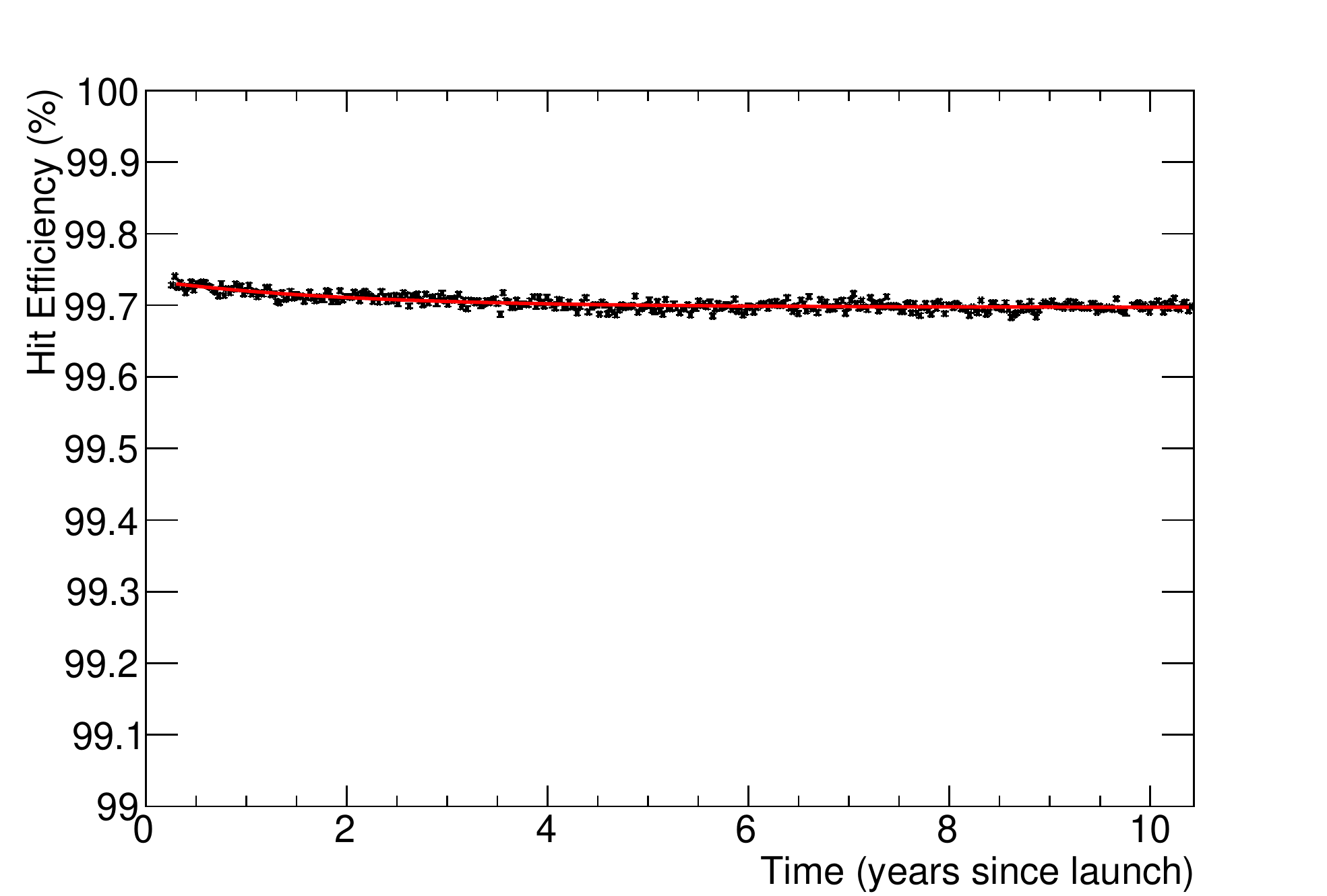}{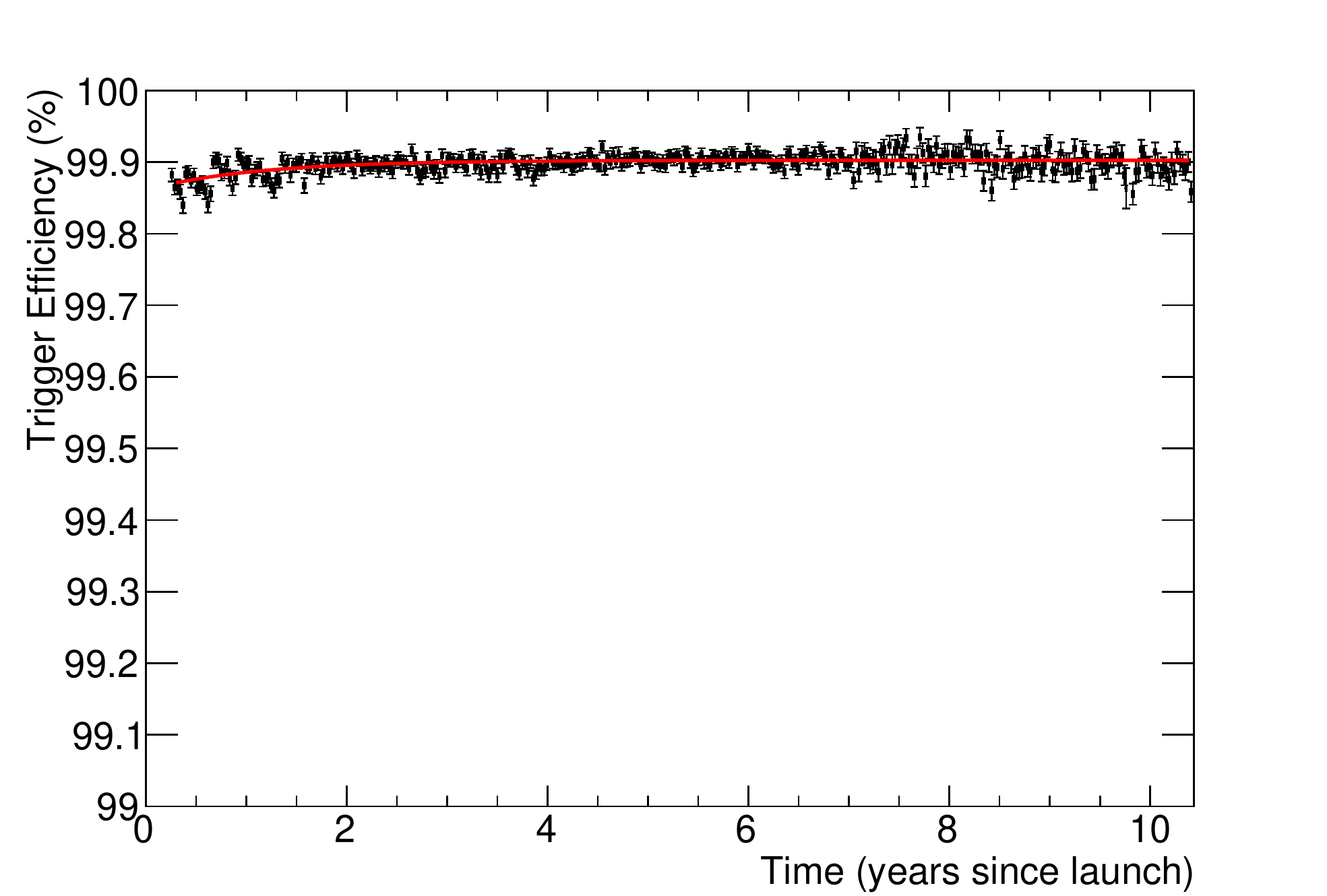}{
  \caption{Left: Tracker hit efficiency. Right: Trigger efficiency. Pass 8 data are scaled for trending purposes. }
  \label{fig:tkrefficiencies}
}

\begin{figure}[htbp]
  \centering
  \includegraphics[width=0.45\textwidth]{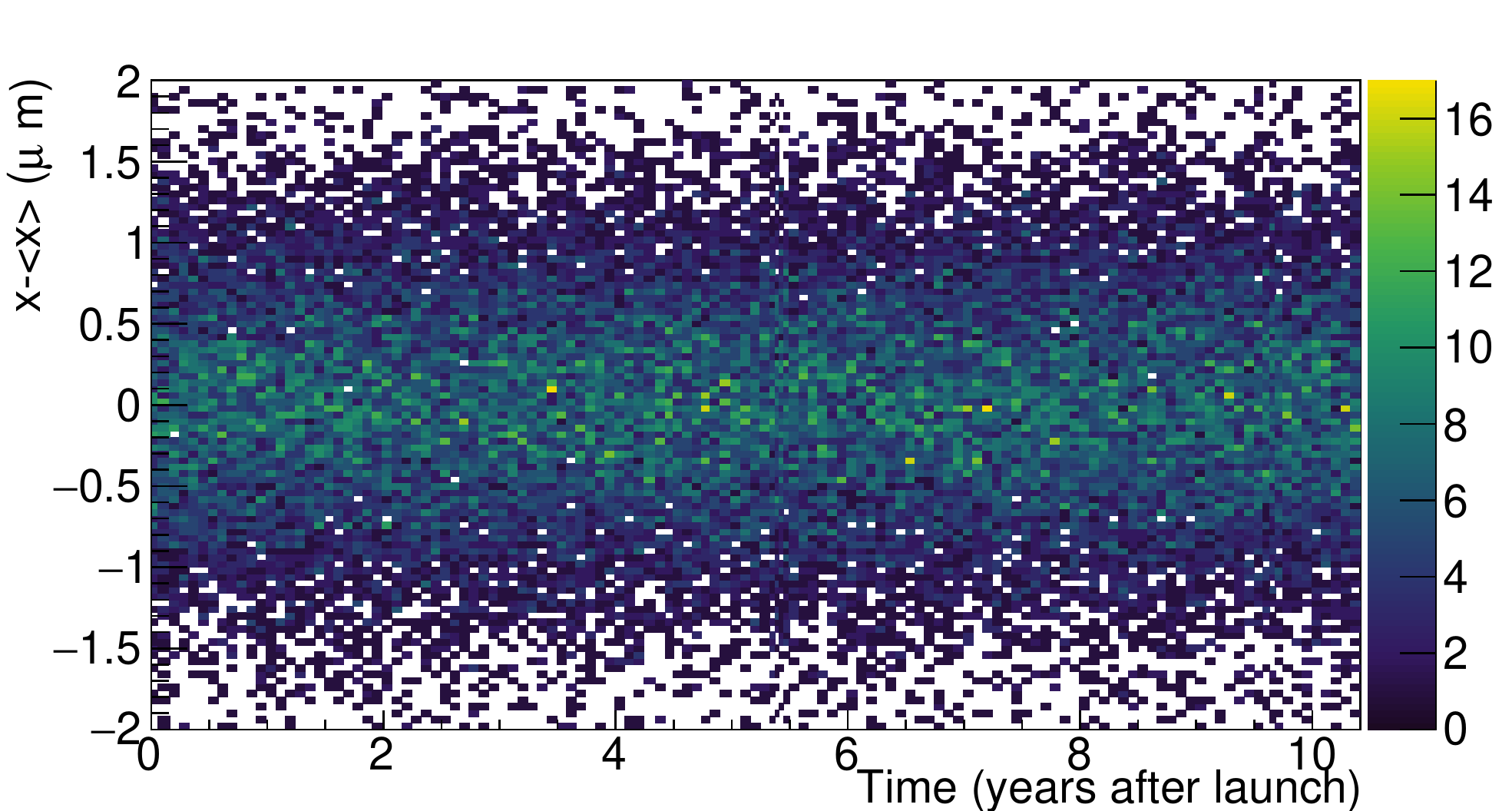}
  \includegraphics[width=0.45\textwidth]{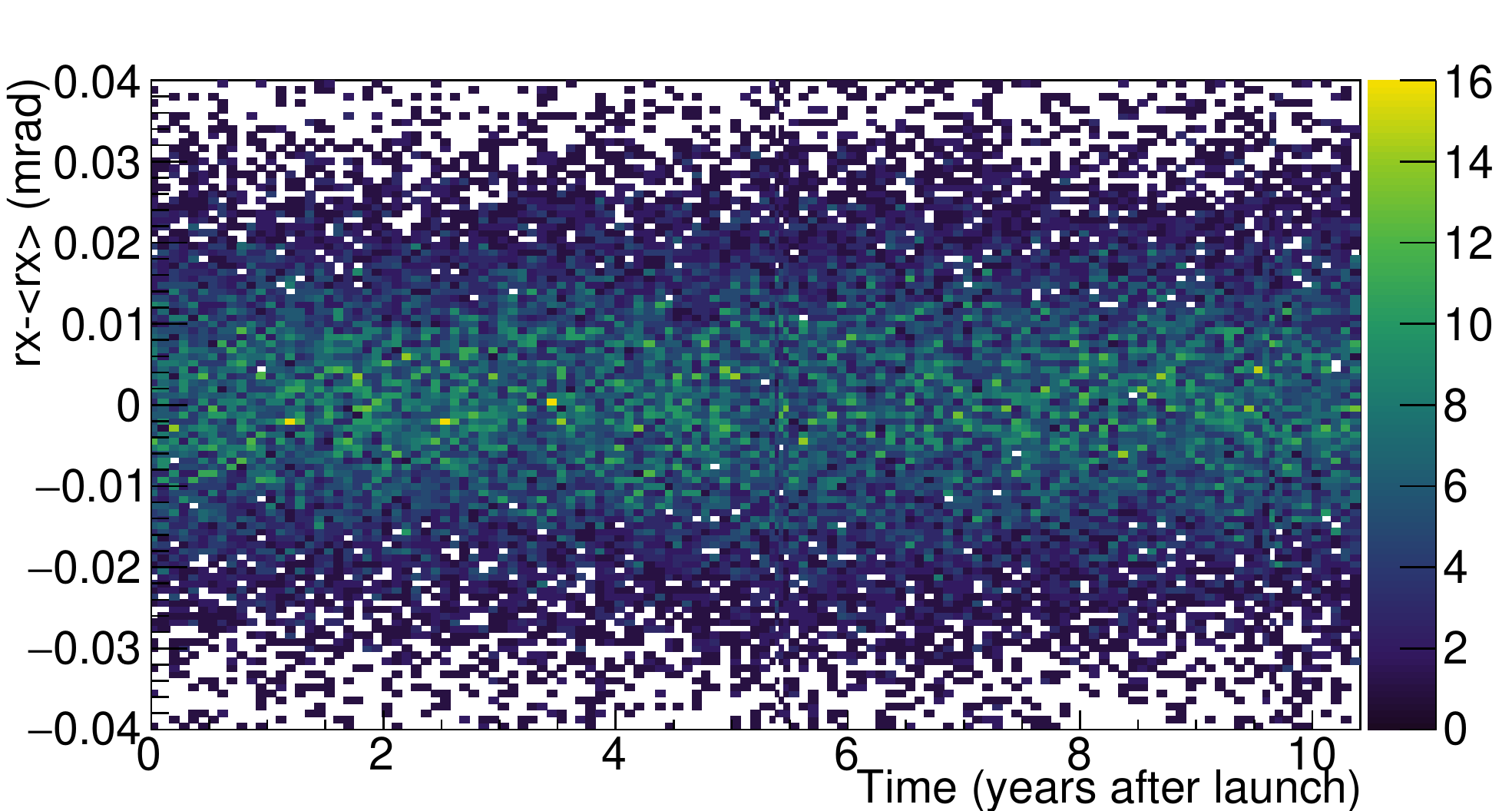}
  \includegraphics[width=0.45\textwidth]{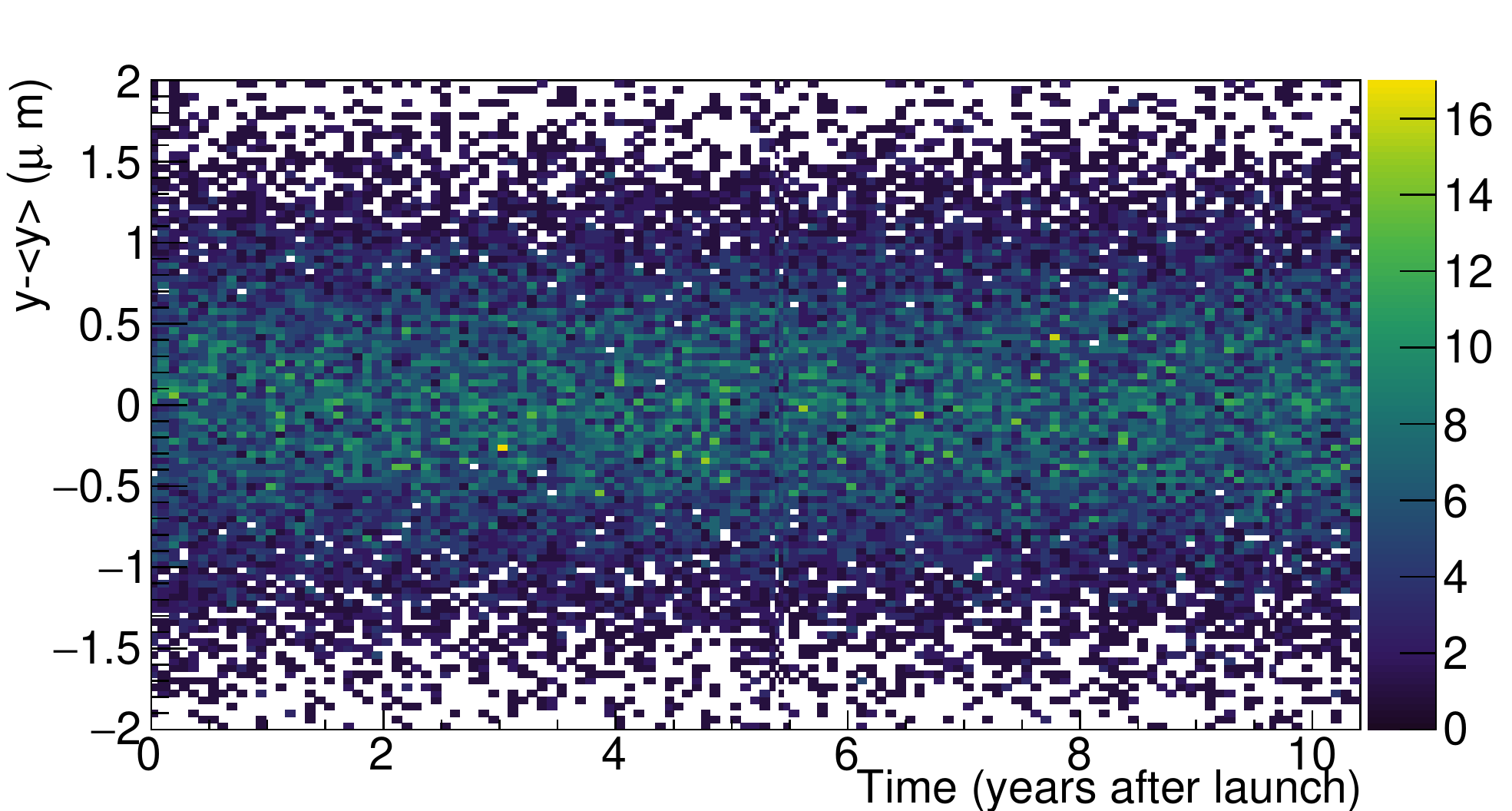}
  \includegraphics[width=0.45\textwidth]{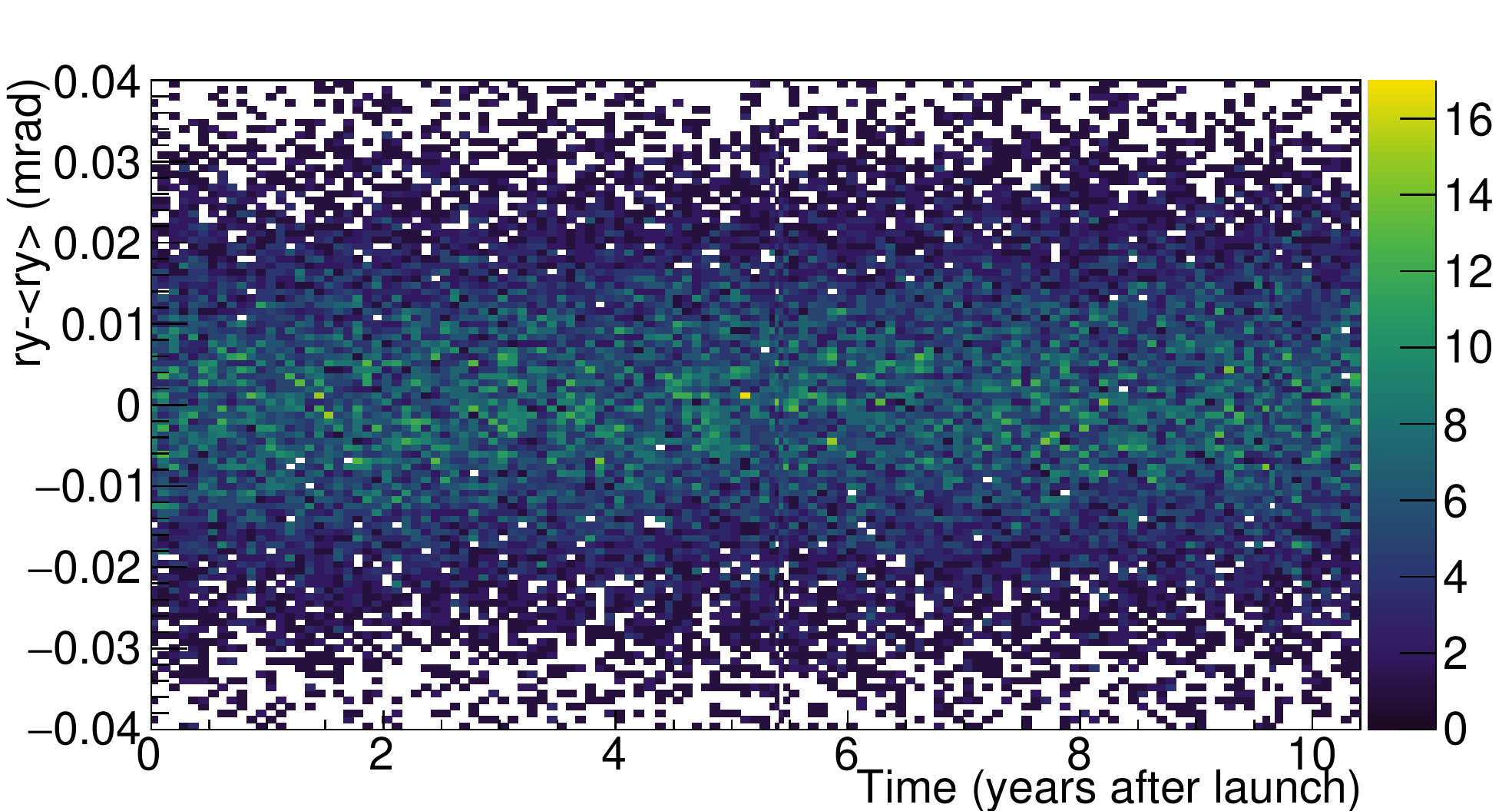}
  \includegraphics[width=0.45\textwidth]{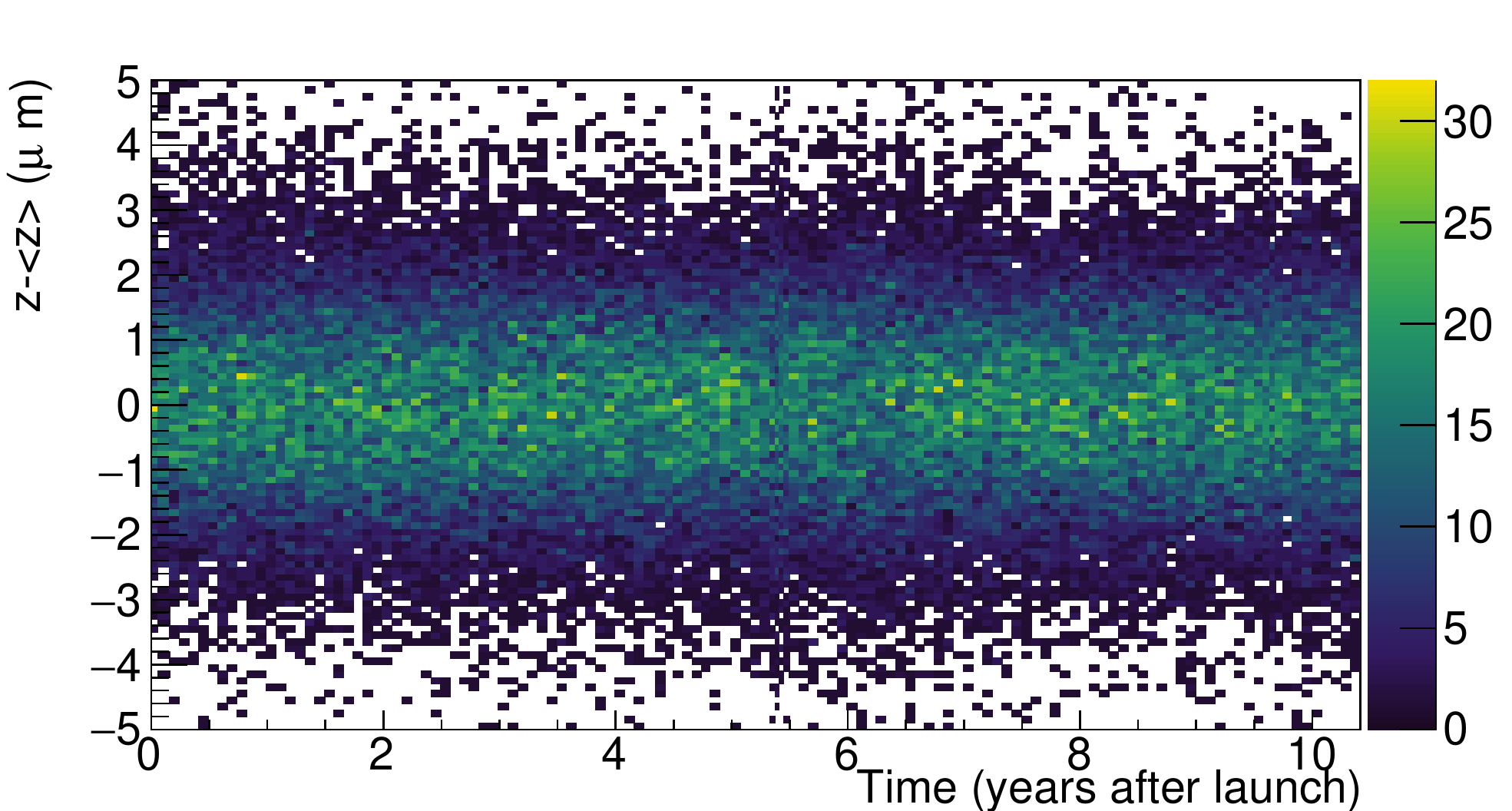}
  \includegraphics[width=0.45\textwidth]{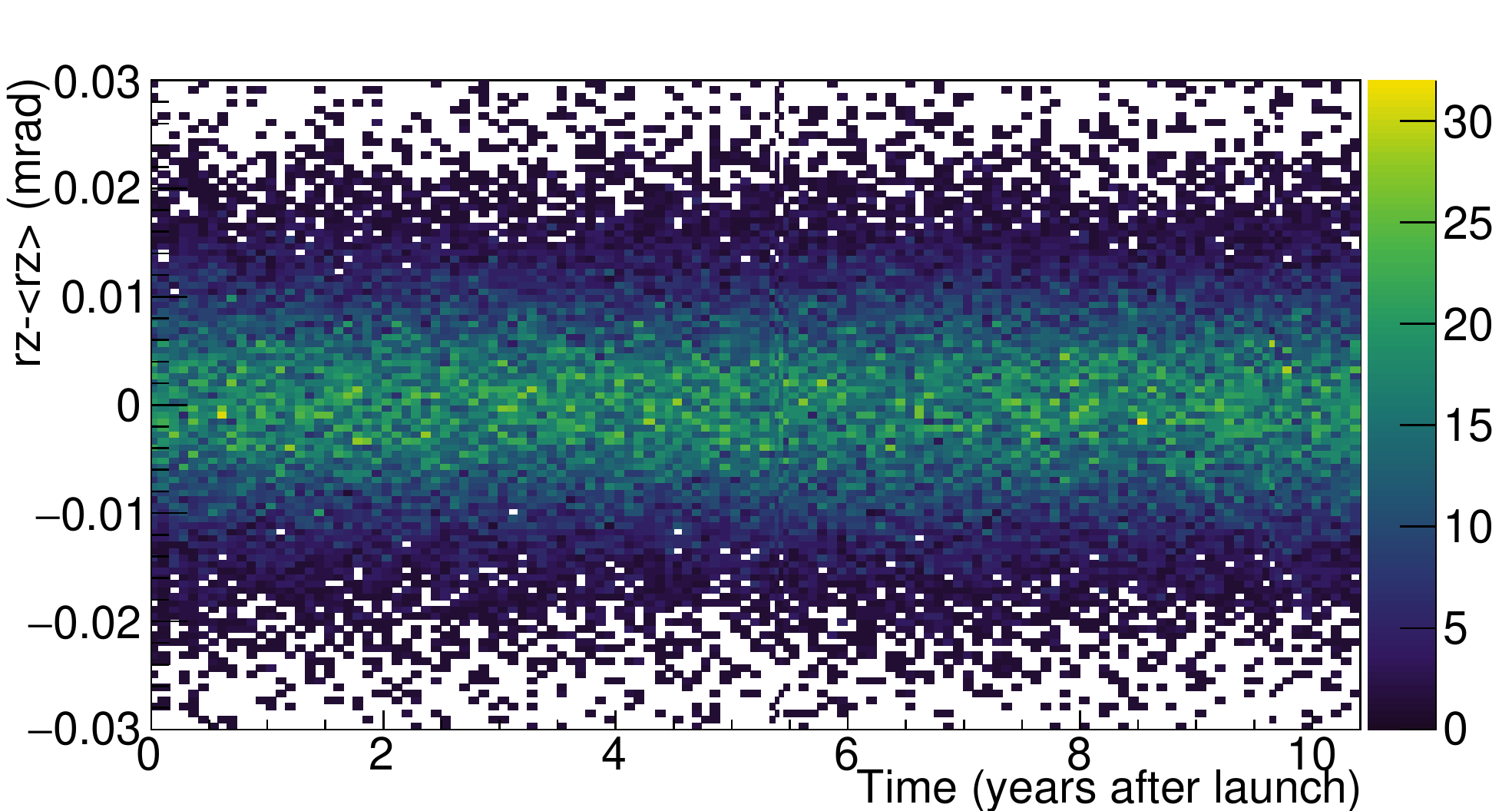}
  \caption{Alignment of the 288 X-Y silicon planes in the LAT reference system, measured 10 times per year: the offsets along the x, y, z directions (left top, middle, bottom) and the rotations around the same axes (right top, middle, bottom).}
  \label{fig:tkralignment}
\end{figure}

The mechanical alignment of the TKR is also regularly calibrated. One rotation matrix aligns the reconstructed direction of each event from the TKR to the spacecraft (and thence the celestial) reference systems. Also, translation and rotation offsets of each Si plane in the TKR mechanical coordinate reference system are accounted for by additional calibration constants.
The relative alignment of the LAT reference system to the spacecraft reference system, derived from the spacecraft star tracker observations, is determined by minimizing the residuals of the measured locations of known gamma-ray sources in the sky. The relative alignment is known with an accuracy better than 5~arcseconds on the three rotation angles, and shows no significant evolution in time.
The relative alignment of each Si plane in a tower with respect to the LAT reference system is given as 3 offset translations and 3 rotation angles. The sensitivity of the measurement is better than $1~\mu$m ($2~\mu$m) for the x, y (z) translations, and than $0.02$~mrad ($0.01$~mrad) for rotations around the x, y (z) axes.  
For each of the 288 X-Y silicon planes in the TKR we compute the time average of the three lateral offsets and of the three rotation angles, and in \figref{tkralignment} we show the deviation from the mean as a function of time of all planes; no evolution can be observed. The overall translation uncertainties are well below the position resolution of a charge cluster given the TKR strip pitch of 228~$\mu$m, and we estimate that spurious rotations would need to be $\sim 0.3$~mrad {\em at the tower level} before affecting the high-energy PSF. 

%% file: tkr_config.tex
\subsection{Tracker configurations}\label{subsec:tkr_config}

The lists of \emph{dead} and \emph{hot} channels, defined in \secref{subsec:tkr_calib}, are also used as configurations. In addition, the {\em trigger thresholds} define the minimum amount of charge to generate a trigger request, with one value for each front-end ASIC. In \figref{tkrthrdist}, left, the distribution of threshold values for each tracker channel is shown for two runs, one after the launch and one after almost 10 years of mission: no difference is evident. The distribution of the ratio, i.e., the values for the later run divided by the values for the earlier run, is shown in \figref{tkrthrdist}, right. The standard deviation is comparable with the uncertainty on the determination of the threshold for a single channel ($\sim 1$\%).



\twopanel{htb!}{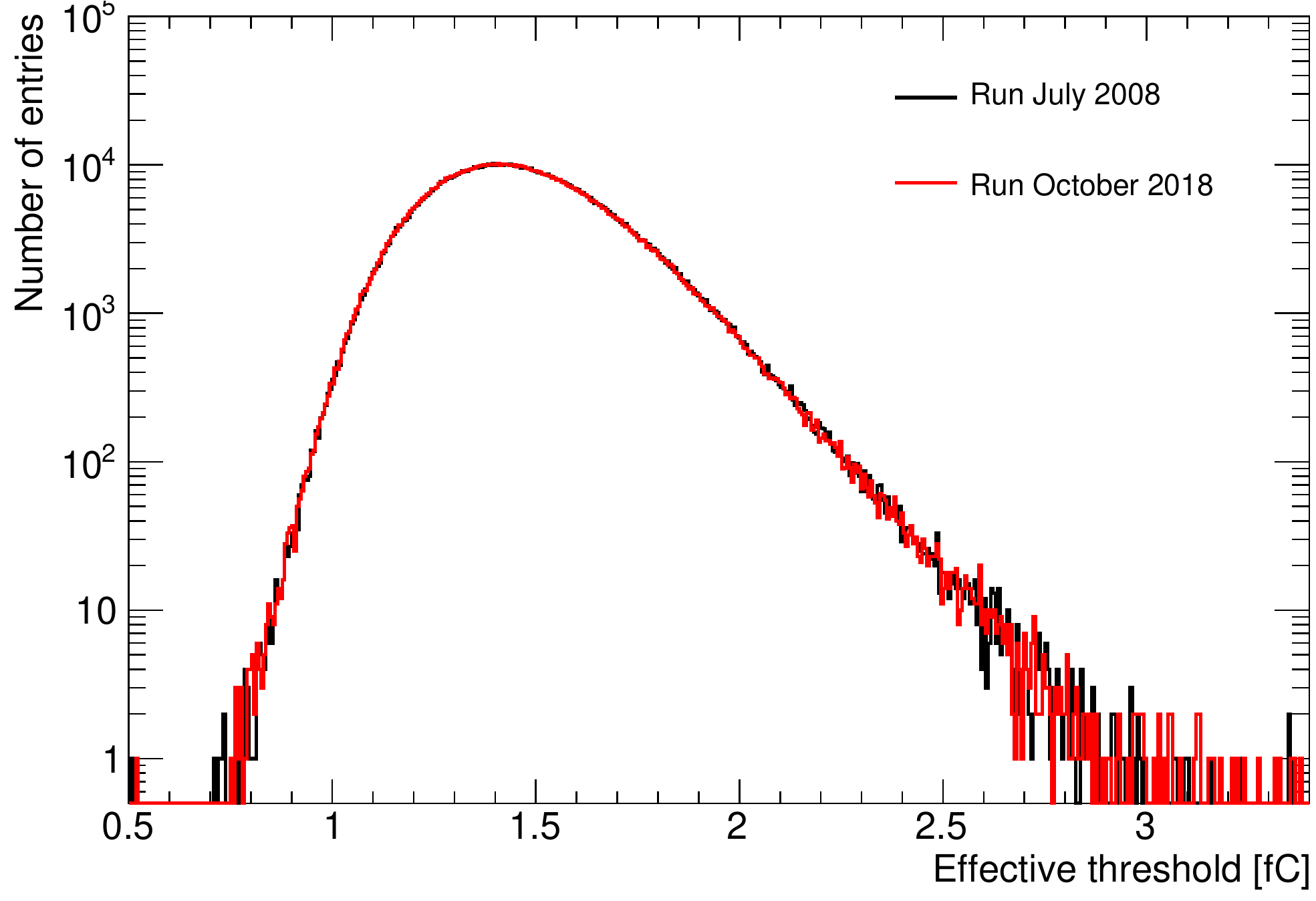}{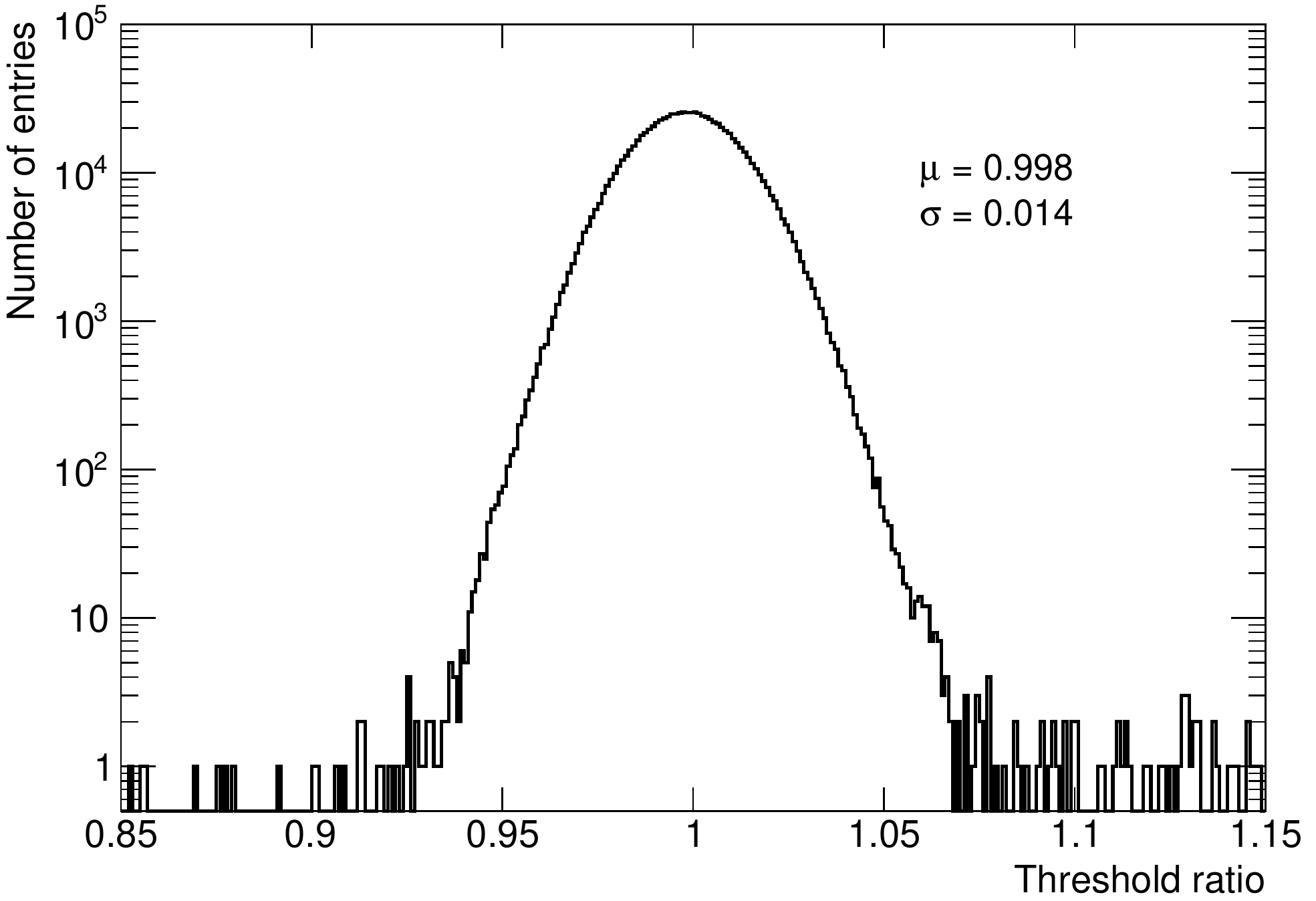}{
  \caption{Left: Tracker threshold distributions for two runs, one early and one late in the mission. Right: The distribution of the ratio of the threshold values for the two runs.}
  \label{fig:tkrthrdist}
}

%% file: tkr_failures.tex
\subsection{Tracker component performance}\label{subsec:tkr_failures}

In 2006, before launch, a single Tracker Readout Controller failed during a LAT power-up during ground testing. Schedule constraints dictated that this failed circuit in silicon Layer 0 of Tower 10 would not be repaired or replaced.
The split point described in \secref{subsec:tkr_calib} was moved such that the entire plane is read by the left-end controller. On orbit, no other tracker component failures have occurred (dead strips are not considered).

%% file: calorimeter.tex
\section{Calorimeter}\label{sec:calorimeter}
The calorimeter (CAL) detector subsystem provides the primary contribution to the energy estimate for all but the lowest energy events, $\lesssim$100 MeV, where depositions in the TKR are important.  The calorimeter is segmented on dimensions comparable to the characteristic longitudinal and transverse length scales of electromagnetic showers, the radiation length and Moliere radius, respectively.  This segmentation allows three-dimensional imaging of shower development and enables good calorimetry ($\Delta E < 20\%$ FWHM) up to $\sim$2 TeV \citep{Bruel2012} in a relatively thin calorimeter (8.3 $X_0$ on axis).  A moments analysis of energy depositions within the CAL locates the energy centroid and provides directional information used in reconstruction of the full event.  Information from the CAL is also used in rejecting cosmic-ray background events through a comparison of the location of the energy centroid and the projected TKR direction and through analyzing the CAL shower topology.

The CAL is a $4 \times 4$ array of modules matched to the array of sixteen TKR towers in the LAT. A Tower Electronics Module (TEM) provides the interface between the each TKR tower and CAL module to the global trigger and data flow electronics system in the LAT. Each CAL module consists of eight layers of 12 CsI(Tl) crystals in a hodoscopic arrangement, with successive layers rotated by 90$^{\circ}$ with respect to each other.  Each of the 1,536 CsI(Tl) crystals is 2.7 cm $\times$ 2.0 cm $\times$ 32.6 cm, read out at each 2.7 cm $\times$ 2.0 cm end-face with a dual PIN photodiode assembly. Each crystal provides three coordinates of the centroid of energy deposition within it. Two come from the location of the crystal within the CAL module. The third, the longitudinal position, comes from measuring the scintillation light asymmetry at the two ends of the crystal.  The longitudinal position has $\sim$mm$-$cm resolution, which is comparable to the transverse dimensions of an individual crystal.  With this configuration, a three-dimensional image of the shower can be constructed from the ensemble of crystal energy measurements and position measurements.

Measuring event energies over the entire five orders of magnitude for which the LAT operates requires that each crystal end be read out in four gain ranges. The two photodiodes at each crystal end have different sizes: the large (small) one is for low (high) energy depositions, called LE (HE). Two readout paths provide readout channels with gains of $\times 1$ (LEX1, HEX1) and $\times 8$ (LEX8, HEX8). Thus the 1,536 crystals have 3,072 end faces read out with 6,144 photodiodes into 12,288 ``spectroscopy'' channels. A single analog front-end ASIC handles signals from the dual photodiode assembly at each crystal end face. 

In addition, each ASIC contains two fast analog trigger discriminators, the Fast Low Energy (FLE) and Fast High Energy (FHE) discriminators nominally set to 100 MeV and 1000 MeV. 
Each CAL module forms two trigger-request primitives, called CAL-LO and CAL-HI respectively, from the logical-OR of the corresponding low-energy and high-energy trigger discriminator outputs of all the front-end ASICs within that module. Each CAL module presents its fast trigger primitives to the central trigger logic of the LAT, the Global-Trigger Electronics Module (GEM), along with trigger primitives coming from the TKR and ACD.  Based on a configurable logic table, the GEM adjudicates whether the presented trigger primitives warrant reading out the LAT detector, and if so generates a Trigger Acknowledge message that is distributed back to each TEM and ACD electronics module.  In response to the Trigger Acknowledge, analog signals from all 3,072 CAL front-end ASICs are digitized; thus the entire CAL is read out for every such trigger.  In the nominal instrument configuration for science data acquisition, only one of the four gain channels at each crystal-end is digitized: the highest gain range that is unsaturated, which preserves the greatest energy resolution in each crystal.  CAL data are sparsified with zero-suppression logic in each front-end ASIC that eliminates energy depositions below a configurable Log-Accept (LAC) threshold, nominally set to 2 MeV.  

See \citet{REF:2010.CALPaper} and \citet{REF:2009.OnOrbitCalib} for more complete descriptions of the CAL and the LAT trigger system.

\input{cal_calib.tex}
\input{cal_config.tex}
\input{cal_performance.tex}

%% file: cal_calib.tex
\subsection{Calorimeter calibration}\label{subsec:cal_calib}
\citet{REF:2009.OnOrbitCalib} detail the calibrated quantities and calibration procedure for the CAL. Here we briefly review the calibration constants and discuss how some have evolved during the mission.

On-orbit calibration of the CAL detector subsystem is performed in 6 Ms intervals (5 times per year).  For each individual CAL crystal, the three quantities that need to be monitored and calibrated over time are the energy scale, the positional light yield dependence, and the pedestal values (i.e., the front-end output for zero input). There are two primary aspects to calibration of the crystal energy scales: monitoring non-linearities in the front-end electronics, and accounting for degradation in the CsI light yield. Twice per year, the electronics response is monitored by injecting pulses of known amplitudes into the CAL input channels and fitting the resulting digital signals. Characterizations of the non-linearities have not changed by more than 0.5\% from values found during on-ground testing. Therefore, we have not updated the electronics calibration constants for any CAL crystals.  

\begin{figure}[htbp]
  \centering
  \includegraphics[width=\onecolfigwidth]{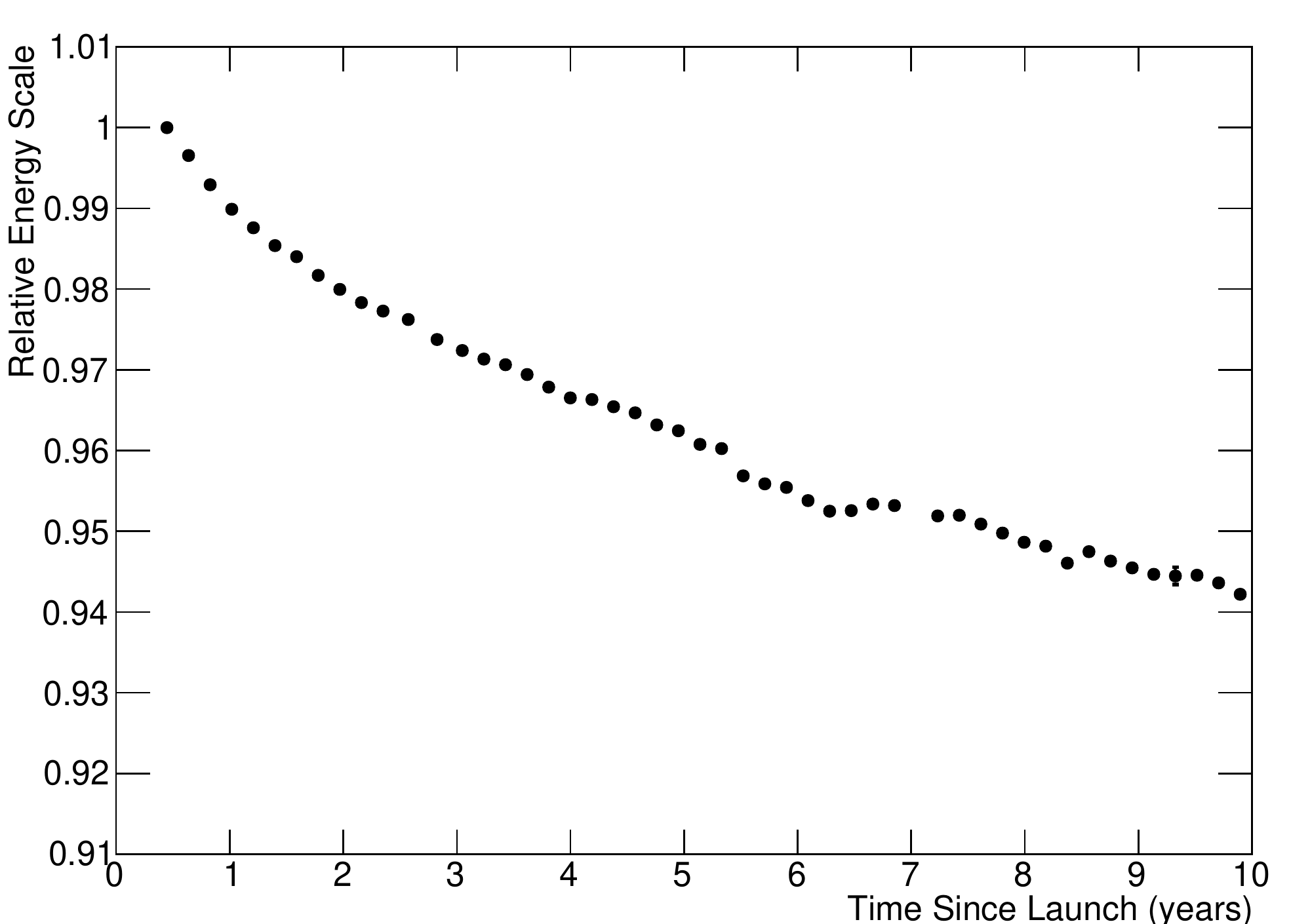}
  \caption{Relative energy scale, averaged over all CAL crystals, as a function of time, referenced to the values just after launch. The calibration at the start of the seventh year was neglected during the Pass 8 software upgrade.}
  \label{fig:calyield}
\end{figure}

Radiation exposure in the \Fermi\ orbit, primarily from the intense particle fluxes within the SAA (see \secref{subsec:saa}) and secondarily from Galactic cosmic rays, degrades the performance of the CsI(Tl) crystals over time.  Radiation exposure induces visible and infrared absorption bands in the CsI(Tl) that reduce the scintillation light output and increase the attenuation of scintillation light as it propagates along the length of the crystal \citep{REF:1992.Woody, REF:1997.Kazui, REF:1999.Chowdhury}.  Both of these changes are observed in the CAL. The light yield for a given energy deposition is calibrated on orbit using cosmic-ray events identified by reconstruction algorithms. Since launch, we have observed a decrease in the average energy scale of $\lesssim$1\% per year, with the degradation rate decreasing with time, see \Figref{calyield}. On-orbit calibration minimizes the effect on reconstruction of event energy and shower parameters.

\begin{figure}[htbp]
  \centering
    \includegraphics[width=\onecolfigwidth]{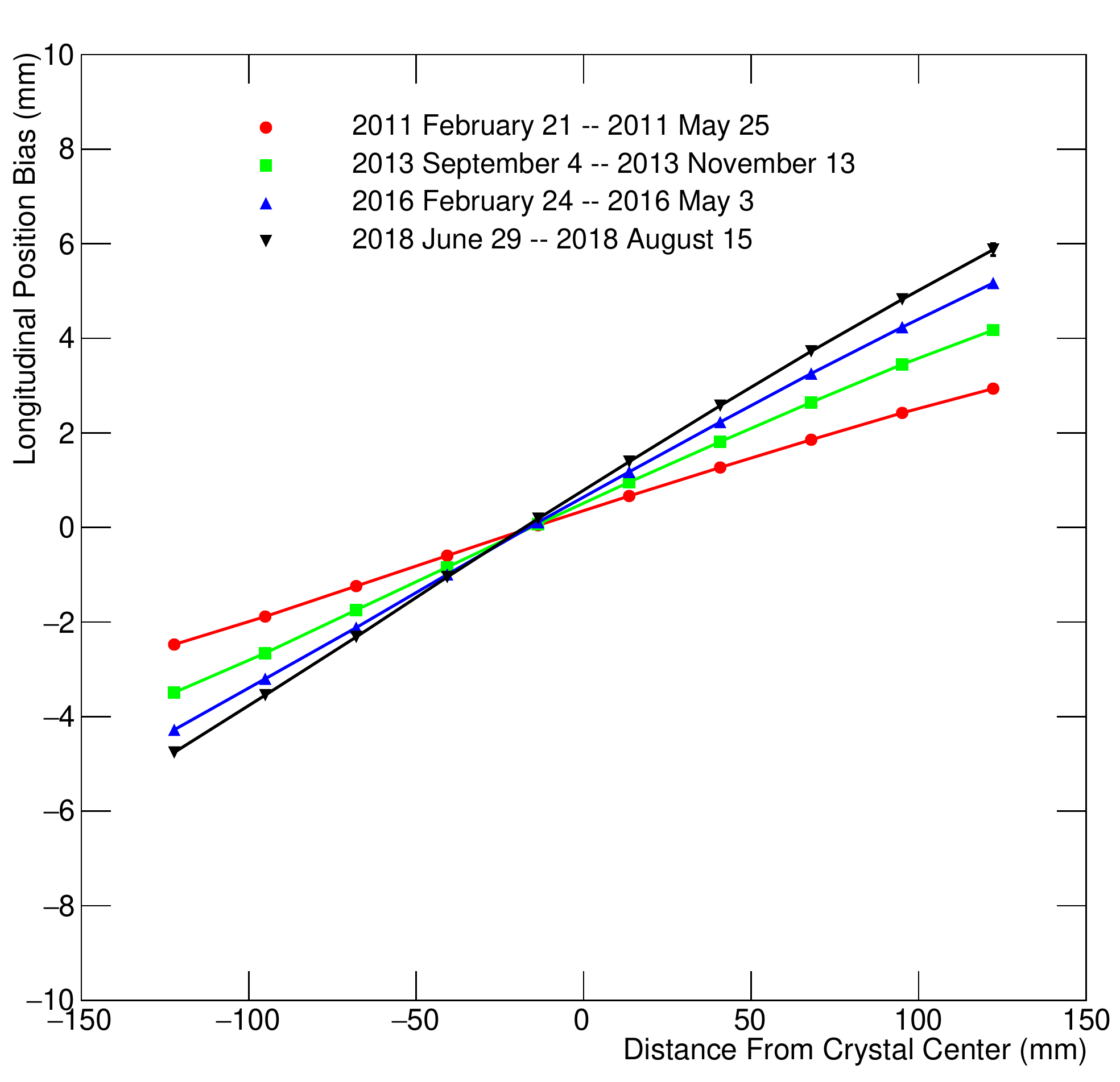}
  \caption{Average longitudinal position `bias' (see text) as a function of distance from crystal center during four different 6 Ms time periods corresponding to approximately 2.5 years (red circles); 5 years (green squares); 7.5 years (blue upward-pointing triangles); and 10 years (black downward-pointing triangles) since launch.}
  \label{fig:calasymevo}
\end{figure}

The light asymmetry, used to infer where along the crystal an interaction occurred, is the logarithm of the ratio of signals from diodes at opposite crystal ends. The conversion from asymmetry to longitudinal position uses parameters that were determined for each crystal before launch and are regularly updated at each 6-Ms calibration epoch. With two different diodes at each end there are four possible readout combinations; accordingly four asymmetry maps are necessary. Each crystal is separated into twelve bins evenly spaced longitudinally, and the asymmetry is calibrated using cosmic rays. The maps of the light asymmetry in each crystal are generated from a fit to the inner ten bins, while the bins closest to the photodiodes are ignored because of the known change in light collection visible in Figure 7 of \citet{REF:2009.LATPaper}.
As the attenuation of scintillation light within each crystal increases (i.e., as the attenuation length decreases) over time from radiation exposure, the light asymmetry increases.  If this change in asymmetry were not accounted for, the reconstructed event position would be biased increasingly with time. 
\Figref{calasymevo} shows an example of how the `bias' between true event position and measured event position would have evolved through the mission had the increasing attenuation not been accounted for, i.e. had the crystal light asymmetry maps not been updated with time. The data show the longitudinal position bias, averaged over all crystals, when reading the larger diodes at both ends: the increasing light attenuation would have introduced a bias of up to $\sim$1 mm per year in reconstructed event positions. However, the regular calibration updates to the light asymmetry maps account for the changing attenuation and remove this bias in the event reconstruction on the ground. Results are similar for the other three possible diode combinations.

The pedestals are offset voltages which define the zero point for the energy scale of each readout channel for each crystal.  The pedestal values will drift if the satellite pointing changes in such a way that temperatures change in the CAL.  The pedestals are constantly monitored using periodic triggers issued by the LAT at 2 Hz, with new average pedestal values calculated every 6 Ms.

\begin{figure}[htbp]
  \centering
  \includegraphics[width=2\onecolfigwidth]{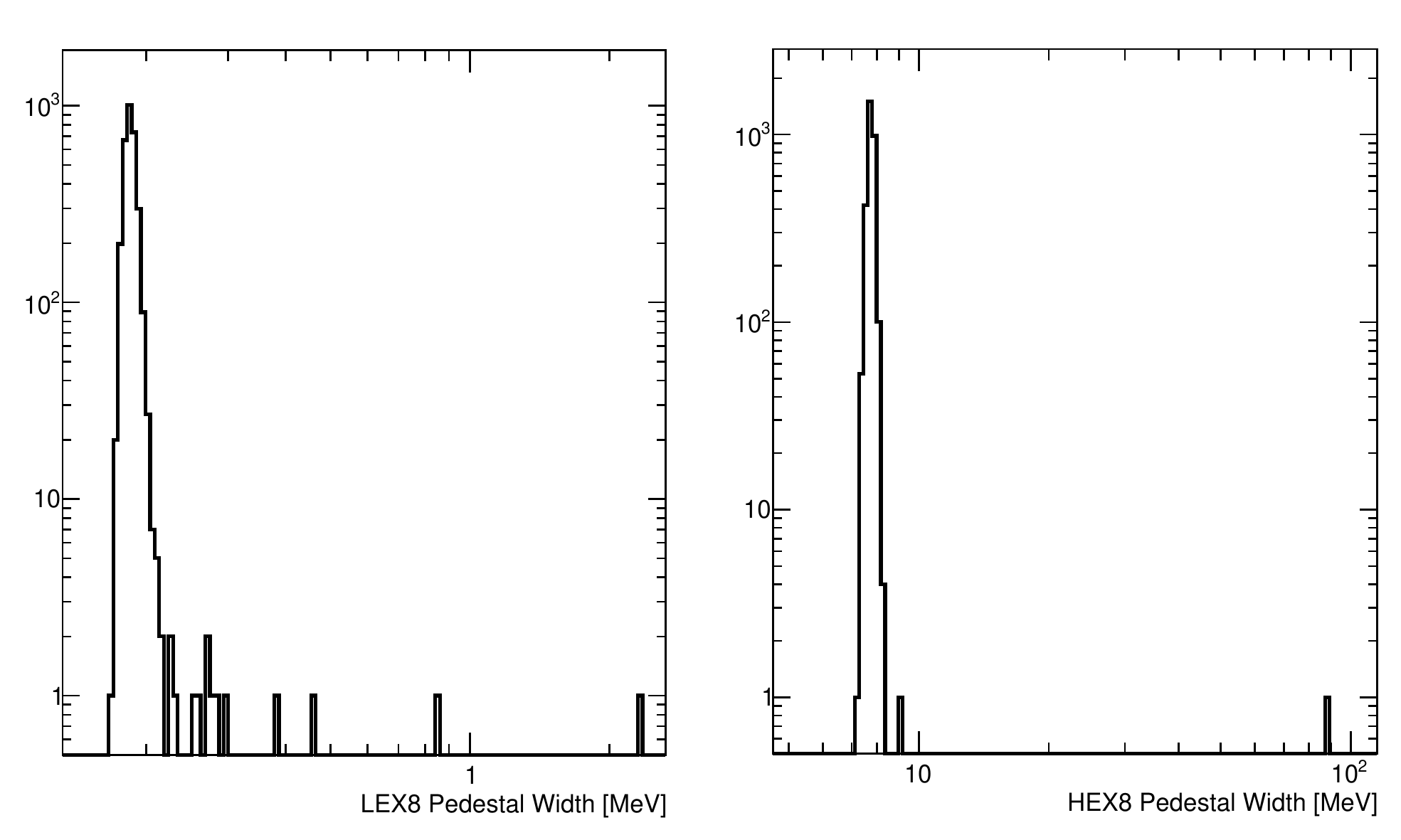}
  \caption{Distribution of electronic pedestal widths for the CAL low-energy photodiodes (left) and high-energy photodiodes (right) converted to MeV.  The distributions combine readout channels from both faces of each crystal and are for the 6 Ms calibration interval at the end of year 10 of the mission. The outlier channels shown in each plot are discussed in \secref{subsec:cal_perform}.}
  \label{fig:calpedwidth}
\end{figure}


%% file: cal_config.tex
\subsection{Calorimeter configurations}

More than 24,000 registers within or directly related to the CAL must be set in order to configure the CAL for science data acquisitions.
\secref{subsec:dataruns} described the re-initialization process that occurs roughly every orbit. About half of these configuration registers are intended to be set at particular energies (e.g., trigger threshold registers), and therefore the register settings can be expected to need to be varied as the instrument response evolves with time on orbit.  Other registers would not be expected to change except in response to a component failure (e.g., a trigger enable mask for an individual readout ASIC would be changed if the trigger threshold discriminator were to become noisy).  Still other registers, e.g., those defining the crystal readout order, should remain static throughout the \Fermi\ mission because the definition of the nominal science data acquisition has not changed.

Registers defining the CAL threshold discriminators are listed in \Tabref{cal_config}, followed by descriptions of the parameters. Each discriminator was calibrated -- i.e., the correspondence between register setting and threshold value in energy units was measured -- before launch and verified during on-orbit commissioning \citep{REF:2009.OnOrbitCalib}. Prior to the start of nominal science operations in August 2008, each register was set to an intended value in energy units. While the  threshold settings have been constant for all nominal science data acquisitions since then, the register settings for each channel have changed as the crystal scintillation light yield has declined with cumulative radiation exposure on orbit (see the discussion in \secref{subsec:cal_calib}).

\begin{table}[htb]
  \begin{center}
    \begin{tabular}{lll}
      \hline
      Name & Purpose & Intent  \\
      \hline\hline
      LAC & Zero-suppression threshold & 2 MeV\\
      FLE & Low-energy trigger threshold & 100 MeV \\
      FHE & High-energy trigger threshold & 1000 MeV\\
      ULD & Range-selection upper-level & -5\%  \\
    \end{tabular}
    \caption{Calorimeter settings applied to each of the 3072 crystal end faces.}
    \label{tab:cal_config}
  \end{center}
\end{table}

The Log-Accept, or LAC, zero-suppression threshold sparsifies CAL data by qualifying for inclusion into the CAL data stream only those crystals for which at least one end-face registers a pulse in excess of the nominal setting of 2 MeV, shown as the threshold ``intent'' in \Tabref{cal_config}. The LAC settings for all CAL channels have been adjusted 3 times: twice in the early science mission to apply flight calibrations as the temperature of the CAL slowly settled, and once in 2011, to account for the declining scintillation light yield from the CAL crystals and maintain the 2 MeV intent for the LAC settings, keeping the actual LAC threshold near the intended 2 MeV value despite the reduced light output.

The digital-to-analog converter that controls the LAC threshold setting has a step size of $0.44$ MeV, which sets a natural cadence for adjusting the setting.  The nominal 2 MeV LAC threshold is well above the typical pedestal noise (pedestal width) of $\sim$0.3 MeV.
 Figure \ref{fig:calpedwidth} shows the pedestal widths for all 3072 LEX8 and HEX8 channels. If a front end becomes noisy, the threshold will sometimes be exceeded, and the noisy channel will appear at a rate in the CAL data stream that increases with the noise, i.e., with the pedestal width.  Crystal occupancy (the LAC rate) is reported continuously by Data Quality Monitoring ( \secref{subsec:DQM}) as are the pedestal widths.  LAT event reconstruction \citep{REF:Pass8,improvedPass8} is robust against a modest number of CAL noise hits, so there is no immediate need to respond to noisy channels when they occur.  Over the first ten years in orbit, the typical pedestal width has remained constant, although several channels show increased noise, detailed in \secref{subsec:cal_perform}.  To accommodate the increased noise in those few cases, we have raised the LAC threshold to its maximum possible value on 3 channels, effectively disabling them.  Because a given crystal is qualified for inclusion in the CAL data stream by the LAC discriminator on either end-face, and because the energy deposited in the crystal is calculated from the geometric mean of the energy calculated at each end-face, the modest number of noisy channels has no effect on the science performance of the CAL.

The FLE trigger serves only a secondary role in LAT triggering. The CAL-LO trigger-request primitive formed from the logical OR of the FLE discriminators is not allowed to open a trigger window, i.e., it cannot by itself initiate a LAT event readout; thus, for example, a side-entering gamma-ray that misses the TKR and deposits $>100$ MeV in one or more crystals (but not more than 1000 MeV in any crystal) will \textit{not} cause a LAT event to be generated.  The CAL-LO primitive is instead used by the GEM solely to help classify an event -- for example, if asserted, as one of the trigger conditions required to qualify an event as a heavy cosmic ray to be used for calibration of the four CAL gain ranges.  A minor error or drift in an FLE threshold has no effect on the science performance of the CAL.

In contrast, FHE \textit{does} independently initiate a LAT event readout, regardless of the state of TKR or ACD trigger primitives.  Importantly, this independent trigger condition dramatically reduces the self-vetoing of high-energy gamma rays by secondary shower particles splashing back into the ACD and preserves the effective area of LAT at high energies.  An error or drift in an FHE threshold, or a disabling of an FHE discriminator, can therefore affect LAT performance. However, the high redundancy provided by the segmentation of the CAL into 1,536 crystals with 3,072 FHE discriminators means that the electromagnetic shower of a high-energy gamma ray typically has many opportunities to register $>1000$ MeV in single discriminator.

The range-selection Upper Level Discriminator (ULD) settings have not been changed since launch.  This has no effect on the CAL performance.  As the light yield has declined with time, the equivalent energy at which a given CAL front end presents a particular gain range for digitization has increased, but the gain scales remain calibrated, and the event energy is properly reconstructed.

The values of the above parameters were set before launch. They are checked every 6 Ms. The change in light yield is small enough that gamma-ray detection, filtering, and subsequent transmission to the ground are unaffected. Thus, no settings changes have been made since launch. The light yield changes are updated in the ground calibrations, so that the final physical quantities (energy, longitudinal position, and so on) are stable.

%% file: cal_performance.tex
\subsection{Calorimeter component performance}\label{subsec:cal_perform}
During 10 years of on-orbit operations, only a single hardware component has failed in CAL, a single preamplifier among the 12,288 in the CAL front-end electronics.  Signals from the high-energy photodiode from the corresponding crystal end face stopped being received in July 2010, with no increase in pedestal noise visible in periodic trigger events.  However, by August 2017, the noise had increased enough to solicit an FHE trigger request with some frequency. This channel (Tower 4, Layer 2, Column 4, at the +X end) appears above 800 MeV in the right-hand panel of \Figref{calpedwidth}. At that time, we disabled automatic range selection and FHE triggers from the affected front-end electronics chip.  No root cause for the failure has been identified.  Reconstruction of any event that showers in the calorimeter and either misses this crystal or deposits less than 1 GeV in this crystal is unaffected by this failure.  The overwhelming majority of LAT events is thus utterly unaffected by this failure.  For a high-energy event that showers in the calorimeter and deposits more than 1 GeV of energy in this crystal, the reconstructed event energy and direction will be incorrect.  The magnitude of the errors on reconstructed total event energy and direction has not been estimated but it is expected to be quite small.

 Of the 3072 LEX8 pedestal widths in the left-hand panel of \Figref{calpedwidth}, 12 are outside of the body of the distribution and 5 are $>2\times$ the average value, after 10 years on orbit. Before launch, there were 4 out-of-family LE channels. Besides the failed channel discussed above, only one HE channel is out-of-family. These noisy channels have no discernible effect on the quality of the science data.

%% file: anticoincidence.tex
\section{Anti-Coincidence Detector}\label{sec:anticoincidence}

The Anti-Coincidence Detector (ACD) is the first and foremost means of rejection of the predominant background of charged cosmic rays detected by the LAT. Cosmic ray detections outnumber the gamma-ray signal by 3 to 5 orders of magnitude across the \emph{Fermi}-LAT's energy range.
As charged particles pass through individual ACD scintillators, they deposit energy via ionization which is converted to optical light and transmitted via wavelength shifting (WLS) fibers to ACD detector elements. 
Signals in the ACD are used in both the on-board event veto and ground-based event reconstruction for discriminating between the charged cosmic-ray background and the desired gamma-ray signal.

The ACD is composed of 89  plastic scintillator tiles and 8 scintillating fiber ribbons, each component being read by 2 photomultiplier (PMT) tubes, for a total 194 channels. The dynamic range of each PMT is enhanced by a double readout with different gains for a low range and high range, for a total of 388 ranged readouts. The ACD covers five sides of the \emph{Fermi}-LAT extending far enough to cover the entire TKR subsystem.  
The segmentation of the ACD is used to minimize signal variation across the large area of the LAT and self-veto due to `backsplash' electrons and positrons induced by gamma rays propagating upwards through the TKR and ACD after interacting with the CAL subsystem. 
The ACD particle detection efficiency, estimated through a combination of ground measurements and simulations, exceeds 0.9997.

The ACD can issue two trigger primitives: one signals the passage of an ionizing particle, with a low discriminator value typically used as a veto in defining the gamma trigger; the second indicates a heavily ionizing particle event, useful for calibrating the instrument. 
For more details see \citet{REF:2007.ACDPaper}.

\input{acd_calib.tex}
\input{acd_config.tex}

\input{acd_failures.tex}

%% file: acd_calib.tex
\subsection{ACD calibration}\label{subsec:acd_calib}

The calibration constants for the ACD system are actively monitored and evaluated biweekly.  
The ACD has proven to be very stable over 10 years of operation and there has not been a need, as of yet, to update the monitored calibration constants on board; the calibration constants used for ground processing have been updated once in September 2012.
For the ACD detector elements, the scintillator tiles and ribbons, four calibrations are actively monitored and are required to determine energy deposited: pedestals, coherent noise, Low Range Gain, and High Range Gain. 
Detailed descriptions of each of these calibrations can be found in \citet{REF:2009.OnOrbitCalib}.

\begin{figure}[htb!]
\centering
\includegraphics[width=1.5\onecolfigwidth]{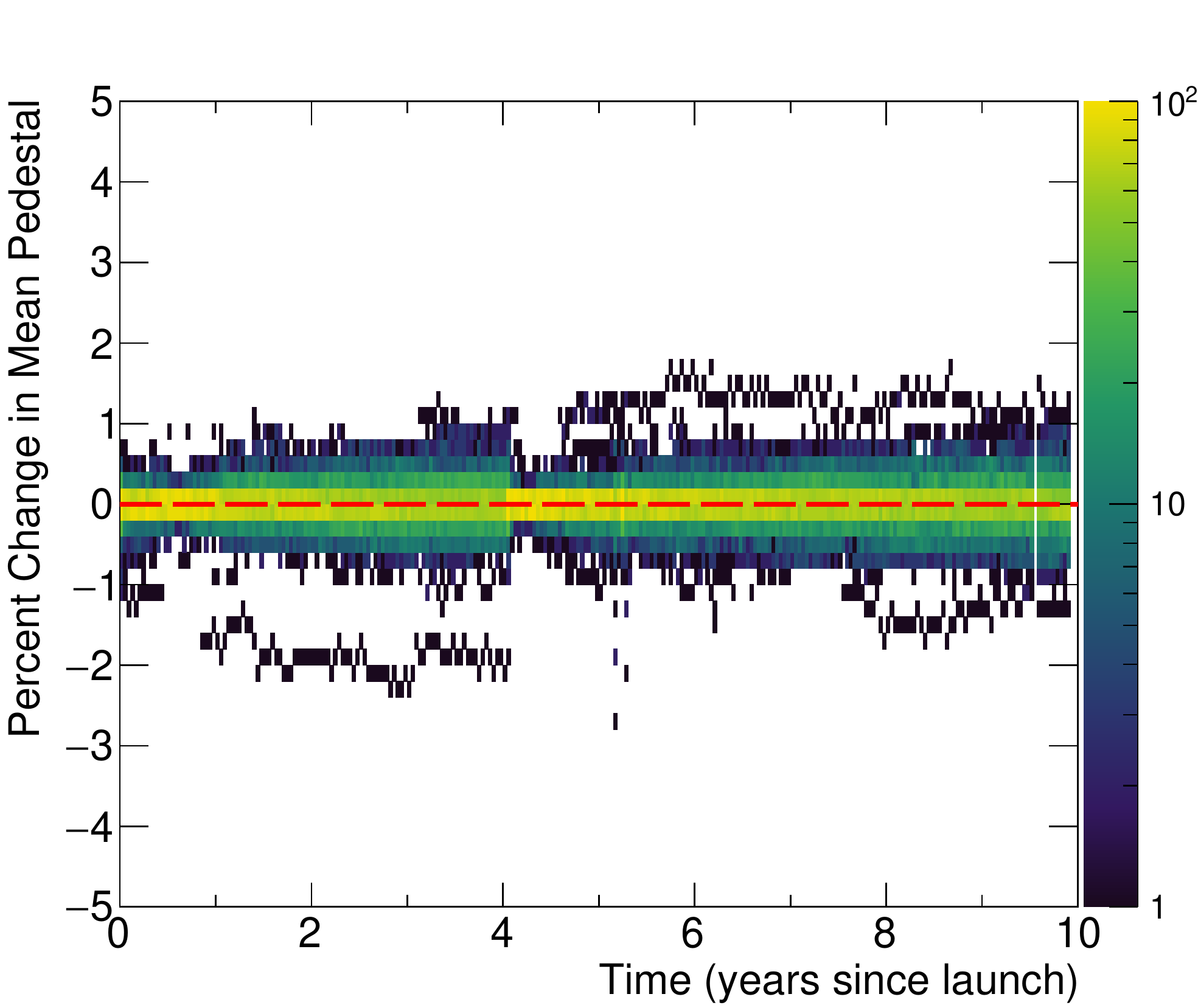}
  \caption{Trending of the ACD low-range pedestal. The x-axis is in units of years since launch. The y-axis is in units of percent change from the reference calibration. The z-axis (color bar) is the number of PMT channels. The red dashed line shows zero change, to guide the eye.  The jump in pedestal values at the four-year mark is typical of ACD calibration updates.}
  \label{fig:acd_ped}
\end{figure}

The electronic readout pedestals define the zero of the energy scale for each channel. The ACD low-range pedestal is measured using a 2 Hz periodic trigger for which a single orbit provides over 10,000 samples. The ACD high-range pedestal requires specialized charge injection and therefore cannot be monitored as actively. Figure \ref{fig:acd_ped} shows the low-range pedestal peak trending of all 194 channels over 10 years. Some channels show a few-percent drift of either low-range or high-range pedestals but nevertheless the on-orbit calibrations are within an overall 5\% allowable drift specified for the ACD. The effect of the updated calibrations for ground data processing can barely be seen 4 years after launch.

Figure \ref{fig:acd_single_ped} compares the reference pedestal calibration constants and the pedestal calibration constants after ten years. Clearly the ACD low-range pedestals are and have remained stable over the 10 years of operation. The ACD low-range and high-range pedestals are highly correlated with the ACD temperature and during the safe-mode event in March 2018, several channels took days to return to pre-safe-mode values. 
\twopanel{htb!}{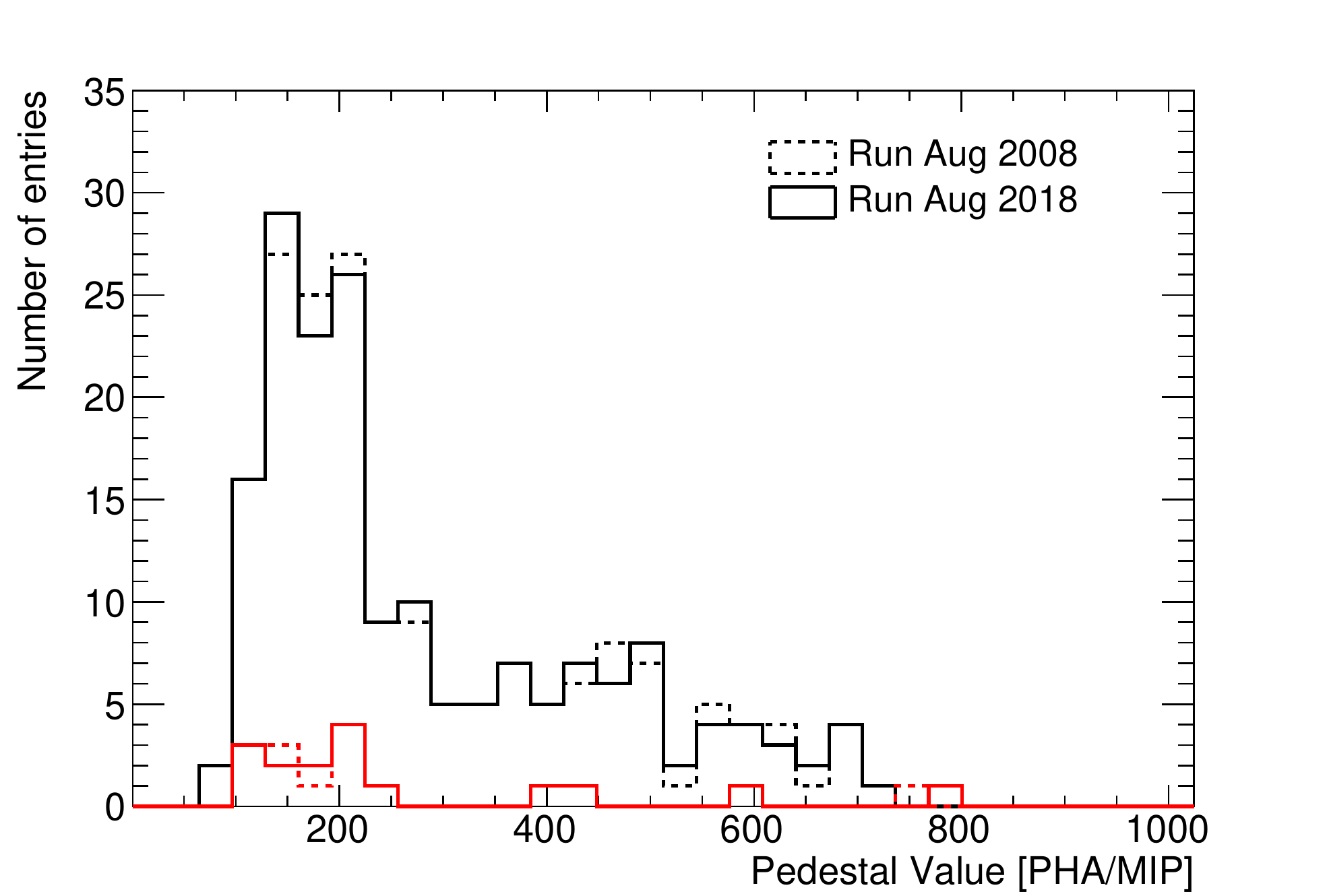}{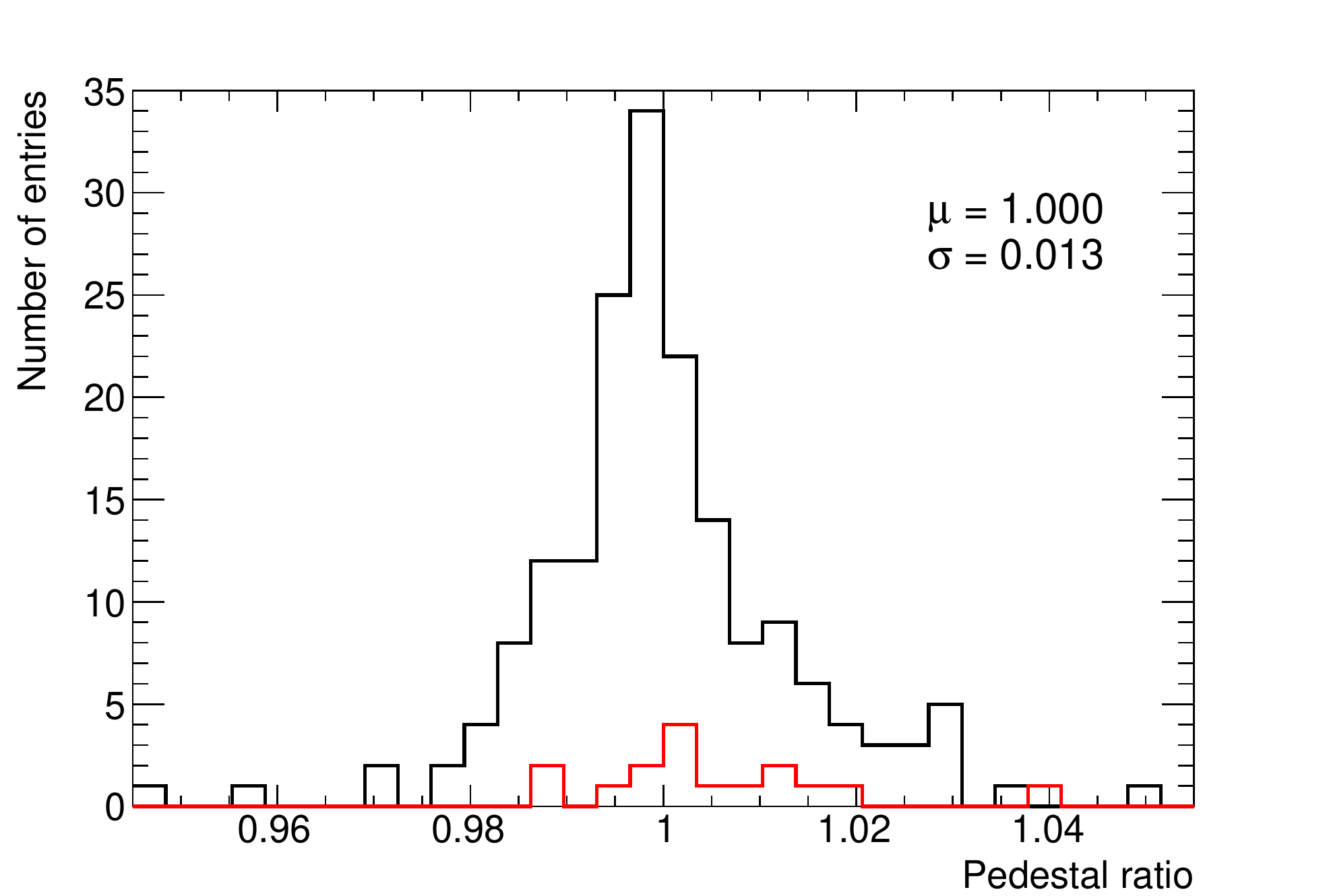}{
  \caption{Comparison of the ACD channel pedestal values for a reference run in the early mission with a run 10 years into the mission (left) and the pedestal value ratios for the two runs (right). Channels associated with ACD tiles are shown in black while channels associated with ACD ribbons are shown in red.}
  \label{fig:acd_single_ped}
}

Coherent noise is modeled and corrected for in an offline analysis, to better estimate the energy deposition in the ACD elements. This is used specifically for the low range of the ACD readout and represents a small correction of up to 0.5\% increase or decrease in the energy deposited in a single ACD detector element. The three variables used to measure the coherent noise, amplitude, decay and phase, show little change over ten years. Amplitude and phase show almost no change over ten years, under 1\% drift. Decay lifetime shows slightly more drift, about 2\% over ten years.

The gain represents the conversion from measured signal, in Pulse Height Amplitude (PHA) units, to units of minimum ionizing particles (MIPs) and equivalently deposited energy for each channel.  
The ACD low-range gain is measured using cosmic-ray protons with a confidently reconstructed track pointing towards the ACD tile or ribbon which recorded the signal.  
The ACD high-range gain is measured using cosmic-ray carbon events identified within the first three layers of the deposited energy in the CAL and having a confidently reconstructed track.
The signal, in either range, is then pedestal subtracted and path-length corrected and fit with a Gaussian distribution to represent the signal and a line to represent a background of poorly reconstructed events.  
The peak of the Gaussian distribution is used as the on-orbit gain for that respective channel and range.  
The gain measurement is only accurate within 5\% and is not meant to be a precise measurement of all factors in the energy deposited in an ACD scintillator element.

Figure \ref{fig:acd_mip} shows the low-range gain peak and width trending of all 194 channels over 10 years. The effect on the gain spread due to the updated calibrations for ground data processing is particularly evident.
Figure \ref{fig:acd_single_gain} compares the reference low-gain calibration constants from the early science mission, for all the ACD channels, and the equivalent low-gain calibration constants from late in the 10-year
mission. Both Figure \ref{fig:acd_mip} (left) and Figure \ref{fig:acd_single_gain} (right) clearly show an overall average reduction in channel gains, with an overall slow change of about $\sim$3\% over the 10 years of the mission. Nevertheless that average calibration change is still within the required overall 10\% change limit, which represents less than 0.1\% in the particle detection efficiency for single ACD detector elements. 

\twopanel{htb!}{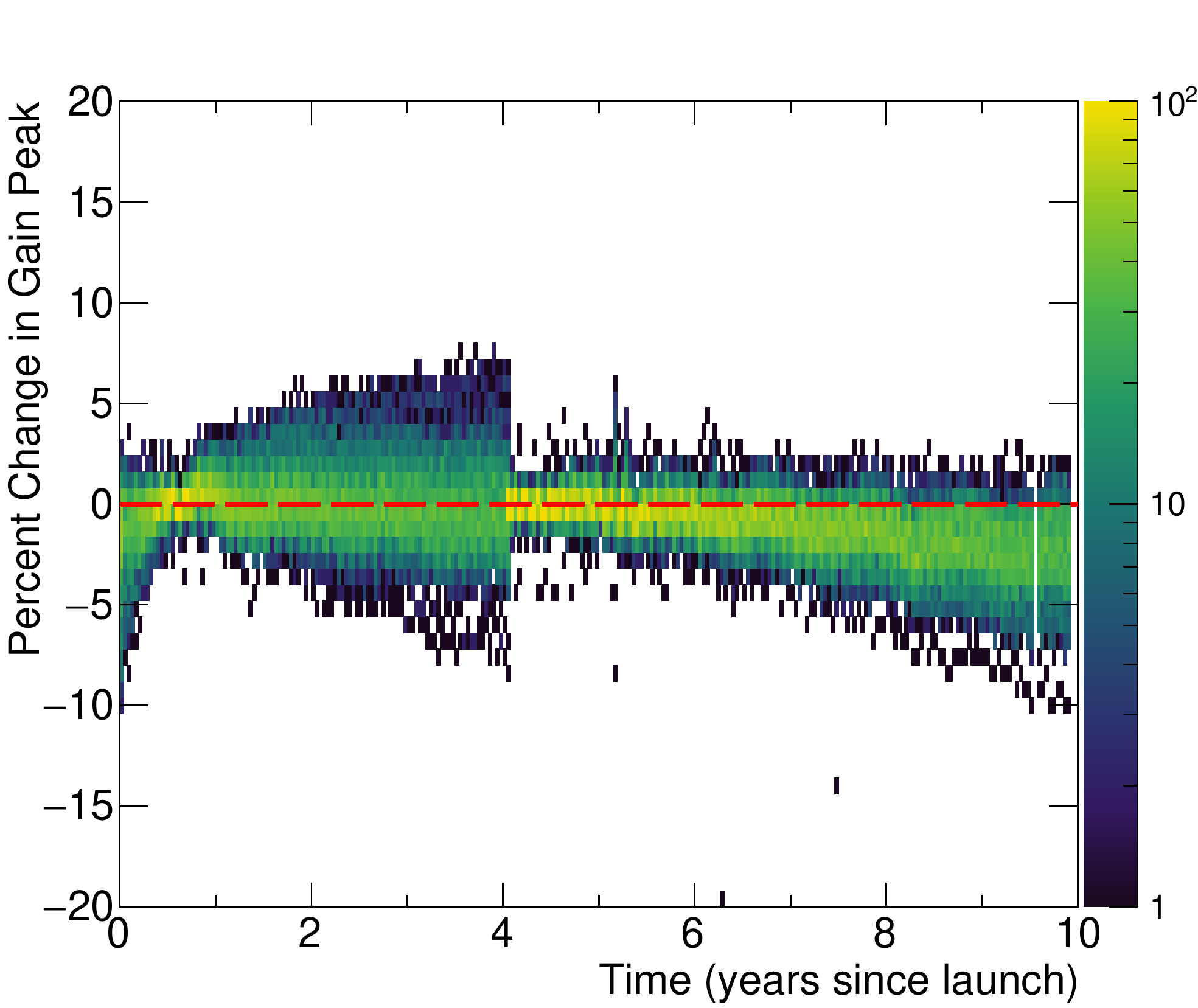}{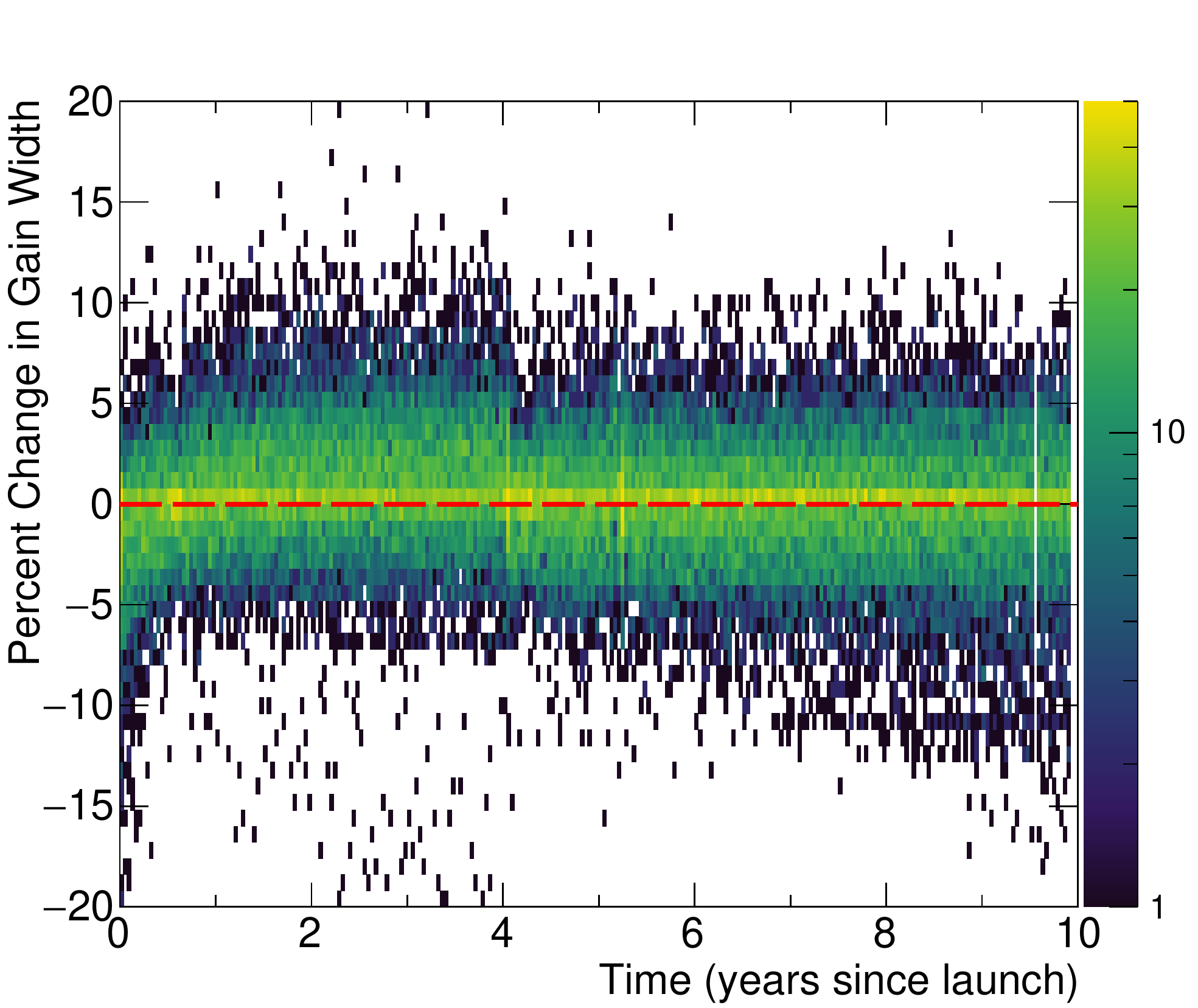}{
  \caption{Trending of low-range ACD gain peak (left) and the low-range ACD gain width (right). The x-axes are in units of years since launch. The y-axes are in units of percent change from the respective reference calibration. The red dashed lines show zero change to guide the eye. The jump in gains at the four-year mark is typical of ACD calibration updates.}
  \label{fig:acd_mip}
}

\twopanel{htb!}{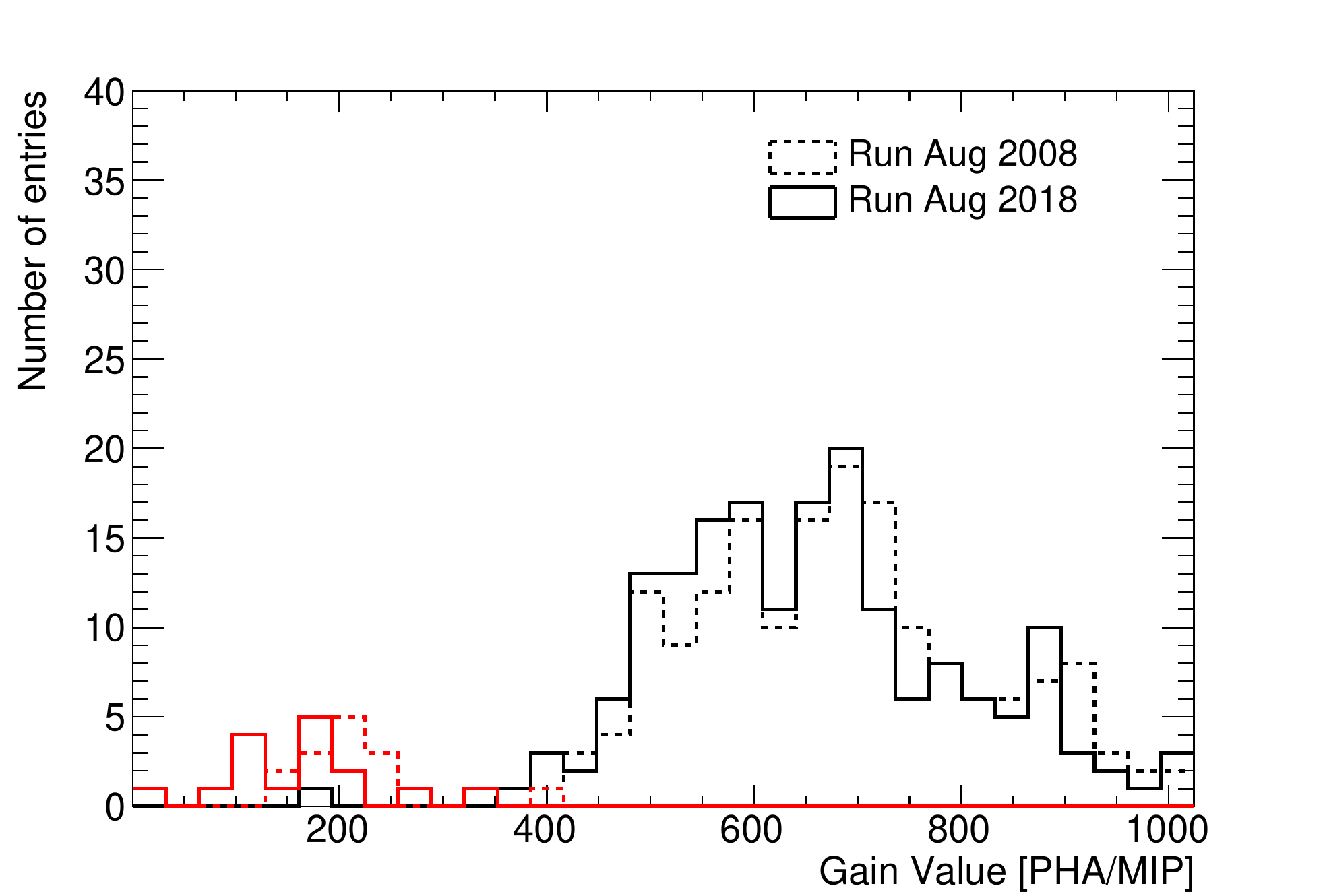}{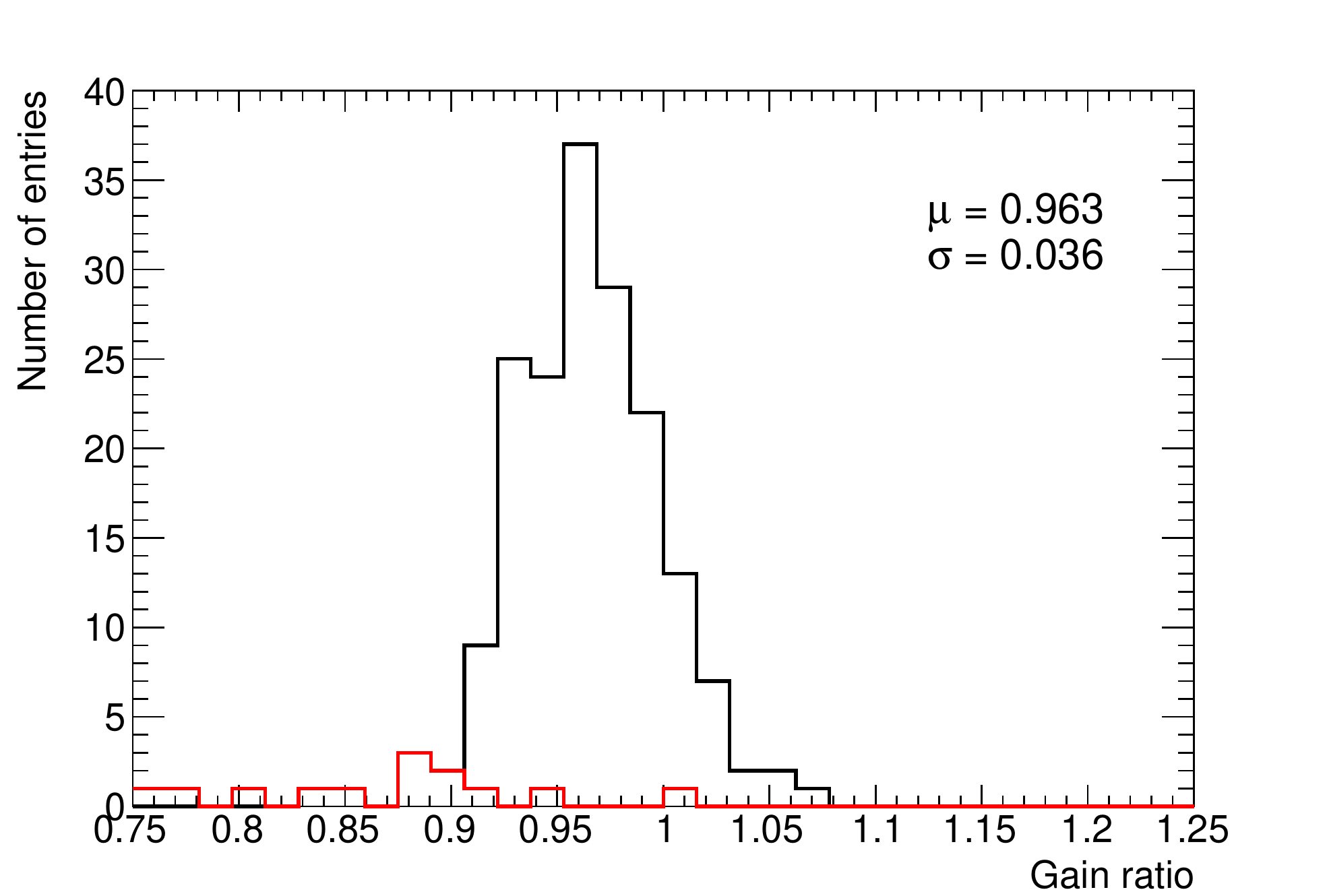}{
  \caption{Comparison of the ACD channel gains for a reference run in the early mission with a run 10 years into the mission (left) and the gain ratios for the two runs (right). Channels associated with ACD tiles are shown in black while channels associated with ACD ribbons are shown in red.}
  \label{fig:acd_single_gain}
}

%% file: acd_config.tex
\subsection{ACD configurations}

The ACD pedestals described in \secref{subsec:acd_calib} are part of the ACD configuration.
The two remaining parts of the ACD configuration are the veto and high-energy discriminator, which define the minimum energy to issue a low-energy and high-energy trigger primitive. 
These calibrations are set on-board and measured on the ground using events specifically collected and downlinked for diagnostics, described in \S 3.1.2 of \citet{REF:2012.LATClassification}.
For the low range, we measure the 50\% efficiency in PHA units between events with and without a veto trigger.  

For the high range, this configuration is complicated by the fact that the high-energy discriminator is OR'ed between 18 ACD channels.
As a result, a given channel can have its associated high-energy discriminator bit set even if that channel did not trigger the high-energy discriminator.  
The process is similar to that of the veto discriminator: we measure the 95\% efficiency in PHA units between events with and without a high-energy discriminator trigger. 
The 95\% efficiency is used to compensate for the multiplexed signals.  
These values are set to 0.4 MIPs for the low range and 25 MIPs for the high range.  




%% file: acd_failures.tex
\subsection{ACD component performance}

Prior to launch two ACD channels were identified during testing as having larger than average noise and one appeared to be dead.  
One noisy channel is associated with a side bottom tile and the other is associated with a ribbon. These two channels are operational but have been running with reduced performance since launch. The dead one is associated with a ribbon; most likely this is a loose optical contact between the ribbon and the PMT, occurred during one of the several thermal tests during integration. Since the ribbon can be read from the other end this does not affect the overall performance of the ACD.

After 10 years of operation there have been no component failures within the ACD in either the detector or the electronic readout chain.  
The ACD is being monitored for degradation of both detector elements and the electronics readout chain through continuous observations of its low-range pedestals and gains via the active data monitoring system and through bi-weekly recalculation of calibration constants.

%% file: other.tex
\section{Trigger, Timestamp Accuracy, South Atlantic Anomaly}\label{sec:other}

In addition to the detectors, the LAT's overall performance depends on some other subsystems. These include the trigger electronics and computers, the accuracy and stability of the system clocks, and the response and resilience of all the electronics to the hostile charged-particle environment around \Fermi, especially during transits through the SAA region. Below we discuss these elements.

\input{trigger.tex}
\input{timing.tex}

\input{saa.tex}

%% file: trigger.tex
\subsection{Trigger and readout}\label{subsec:trigger}

The trigger system relies on several configurations, listed in \tabref{trig_config}, to ensure optimal behavior of the LAT readout. All are time delays: since the system clock runs at 20 MHz, the granularity is  50 ns (1 tick). Optimal delays were set on ground, and updated on orbit in the initial checkout phase of the mission, but not updated since then \modified{\citep{REF:2009.OnOrbitCalib}}.

\begin{table}[htb]
  \begin{center}
    \begin{tabular}{lll}
      \hline
      Configuration & Dimensions & Updates \\
      \hline\hline
      Time coincidence window & 1 & None \\
      Fast trigger delays & 1 per subsystem & None \\
      Latch delays  & 1 per subsystem & None \\
    \end{tabular}
    \caption{Configurations of the trigger system; \modified{values are discussed in the text.}}
    \label{tab:trig_config}
  \end{center}
\end{table}

The {\em Time coincidence window} specifies the amount of time the system waits while collecting trigger requests after the first one received starts the procedure. The  value \modified{was set to 700 ns during the initial calibration phase after launch and not changed since}. 

Trigger signals from TKR, CAL and ACD to the GEM follow different time paths. One {\em Fast Trigger Delay} per subsystem synchronizes them. \modified {On orbit delays were set to 750 ns for ACD, 200 ns for TKR, and none for CAL, and never changed}. 

Once a trigger is issued, the optimal time delays for readout, corresponding to maximum amplitudes in the detectors, are set by the {\em Latch Delays}. \modified{On orbit delays were set to 200 ns for ACD, 2450 ns for TKR, and 2500 ns for CAL, and never changed}.
The dependence of the overall efficiency on the latch delays varies little around the optimal point, e.g., a shift of $\pm 10$ ticks, or 500 ns, around the optimal latch delay for the CAL causes a decrease in the MIP peak amplitude of only 1\%.

As illustrated in Figure~\ref{fig:trigger-rate}, the long-term average trigger rate is steady overall, although subject to environmental changes (see \secref{subsec:orbital}). At the other end of the data chain, the steady flux of bright pulsars such as Vela reported in the 4FGL catalog further demonstrates that the trigger efficiency has not deteriorated since launch \citep{4FGL}.

%% file: timing.tex
\subsection{Timing}\label{subsec:timing}
\fermi\ has a second scientific instrument in addition to the LAT: the Gamma-ray Burst Monitor (GBM), designed to have an all-sky field of view, except as occulted by Earth, to optimally detect gamma-ray burst (GRB) transient events. The GBM  detected the gamma-ray burst GRB 170817A $1.74 \pm 0.05$ s after the LIGO gravitational wave experiment detected gravity waves from GW 170817 \citep{GW_GRB_170817A}.
The time lag between the GW event and the GRB is most likely due to the time required for particle acceleration and gamma-ray emission to occur, but were the lag due to a difference between the speed of gravity and the speed of light, $c$, that difference would be between $-3 \times 10^{-15}c$ and $+0.7 \times 10^{-15}c$.
The combined LIGO/VIRGO GW detectors system should soon start detecting dozens of GW events per year, some of which will coincide with GRBs. Should one of these cataclysms yield near-simultaneous GW and GRB arrivals at Earth, confidence in \fermi's clocks could become an important element in how an observed lag is interpreted.

\S~9 of \citet{REF:2009.OnOrbitCalib} describes pre-launch measurements of the accuracy of the timestamps assigned to LAT events. Atmospheric muons traversing both the LAT and a standalone muon detector allowed us to compare times recorded using a standalone GPS receiver with those recorded by the LAT. The timestamps were shown to be accurate to $\sim 300$ ns, root-mean-square (rms). The least significant digit of timestamp data in the data files used for astronomical analyses is $1\,\mu$s, contributing an rms of $ 300 $ ns to the pulsar results described below.

To compare an arrival time at the satellite with an event time at a different observatory, \fermi's position must also be known accurately. The satellite's Guidance \& Navigation Control system maintains an orbit model, updated periodically with positions provided by one of its two (redundant) Viceroy GPS receivers\footnote{General Dynamics’ Viceroy-4 Global Positioning System (GPS) Spaceborne Receiver}.
GPS positions close to the predicted location receive higher weights, applied with a Kalman filter.
The solution filters anomalous GPS positions, and propagates the position during GPS outages, yielding results more
accurate and robust than individual GPS measurements. Telemetry monitoring shows position rms residuals $<20$ m, 
meaning that the additional uncertainty introduced when propagating a \fermi\ time to another location is negligible. 
Here, we review telemetry data and pulsar data for the 10 years on orbit and confirm the timestamp performance.
GBM timing derives from the same spacecraft signals as the LAT timing. 
Pre-launch muon measurements also showed that LAT and GBM dates agree to better than $2\,\mu$s \citep{GBM}.
LAT's timing accuracy and stability imply that GBM's on-orbit event times are also reliable. 



\subsubsection{Telemetry monitoring of the clock stability}
The \fermi\ GPS receivers generate Pulse Per Second (PPS) signals to mark the instant of validity of their Time Tone Message (TTM), an integer number of seconds since a reference time.
The PPS signals transit through the spacecraft's UDL (Up-Down Link) module before being sent to the GBM and to the LAT GEM (Global Electronics Module).
The UDL PPS is generated by a 10 MHz oven-controlled crystal oscillator; therefore its rms accuracy is of order $100\, \mathrm{ns}/\sqrt{12} = 30 $ ns. 
Scalers in the GEM count 50 ns `ticks' from a nominally 20 MHz oscillator. 
The number of ticks $N_{PPS}$ between successive PPS signals is stored in the data streams, to provide the oscillator's true average frequency during each second. 
A LAT event trigger latches a GEM scaler counting the same 20 MHz ticks. A LAT timestamp is thus the TTM plus the fraction of a second given by the number of ticks since the most recent PPS. 
The overall timestamp accuracy is that of the PPS convolved with the variations in GEM trigger time that depend on event geometry: 
which TKR towers were crossed, whether the CAL participated in the trigger, and so on.
The quantities described here -- the measured oscillator rate, GPS lock, and others -- are part of the instrument telemetry stream
and are included in the Data Quality Monitoring described in \secref{subsec:DQM}. 

Several times per day, the Viceroys lose `lock', sometimes for several minutes, generally because of where the GPS satellites appear in the sky relative to the receiver antennae. 
The UDL PPS signal specification is to drift by no more than $\pm 0.6\, \mu$s ($\pm 12$ GEM oscillator ticks) per minute during GPS lock loss.
DQM monitoring of $N_{PPS}$ is sensitive at the level of several ticks. $N_{PPS}$ varies little, on time scales of days and weeks. 
Were a putative UDL PPS drift to exceed specifications, it would likely be noticed. 
We note that the half-day 13 $\mu$s mishap\footnote{\url{https://www.navcen.uscg.gov/pdf/gps/AirForceOfficialPressRelease.pdf}} 
in the GPS system of 2016 January 26 was reported in the LAT DQM system, and the data were flagged accordingly.


\subsubsection{Timestamp end-to-end tests using pulsars}
With pulsars we can validate the complex timing chain by comparing measured gamma-ray arrival times with predictions derived from radio telescope observations of the same pulsars. This amounts to comparing LAT clocks with ground-based observatory clocks.
We perform two tests, for clock stability and for absolute timing. The high rotational frequency $F$ of millisecond pulsars (MSPs) tests clock stability.
Gamma-ray photon statistics for MSPs limit us to 1 $\mu$s accuracy for 1-year data integrations, but confirm nevertheless that the timestamps cannot often be much worse than our published $300$ ns rms value. The high braking rate $\dot F = \frac{dF}{dt}$ of the young Crab pulsar constrains the LAT absolute time.

\begin{figure}[htbp]
  \centering
\includegraphics[width=1\textwidth]{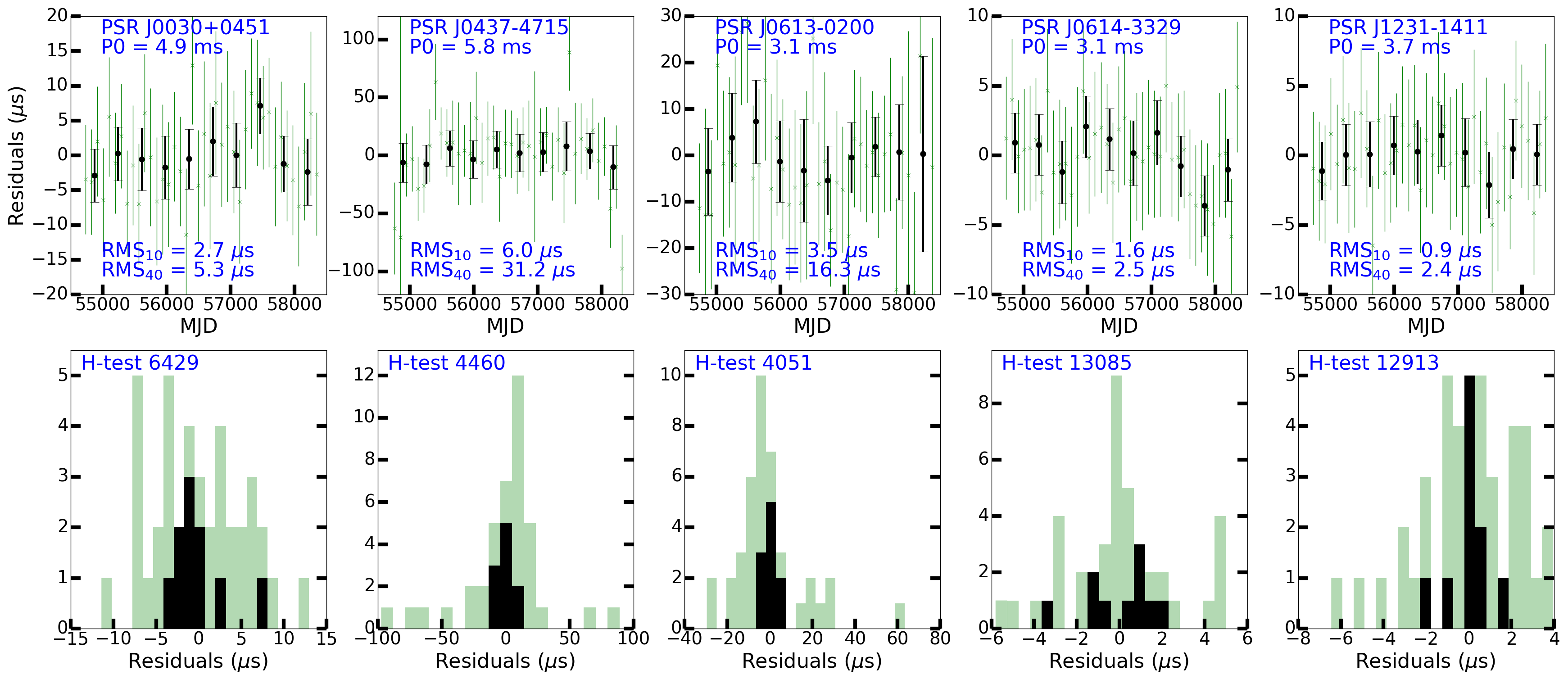}   
\caption{Stability of gamma-ray millisecond pulse arrival times compared to radio references. 
Top: Black dots show ten one-year average values, and green crosses show 40 three-month average values. The root-mean-square variations are dominated by the statistical uncertainties in both the gamma-ray and radio data.
Bottom: The same points, displayed as histograms. The H-test indicates the statistical significance of the pulsed signal (see text).}
\label{fig:RadGamResids}
\end{figure}

Figure \ref{fig:RadGamResids} summarizes our MSP results. 
Of the pulsars described by \citet{2PC}, hereafter 2PC, we chose those with short spin periods $P < 6$ ms and large H-test pulsed test statistic values \citep{DeJager2010}, 
meaning that they are bright in gamma rays and have narrow gamma-ray peaks.
The Nan\c cay radio telescope \citep[see e.g.,][]{POND} provided accurate rotation ephemerides, 
except for PSR J0437$-$4715, observed by the Parkes radio telescope \citep{ParkesMSPs}. 
We used the ephemerides to calculate a rotational phase for each gamma-ray event with the {\tt fermi} plug-in \citep{Ray2011} to {\textsc Tempo2} \citep{Hobbs2006}.
We then determined gamma-ray times-of-arrival (ToAs) using the methods described by \citet{LATtiming}.
{\textsc Tempo2} calculated the residuals shown in Figure \ref{fig:RadGamResids}, the differences between the ToAs predicted by the ephemerides and those measured with the LAT.

We repeated this twice for each MSP: once for gamma-ray pulse profiles integrated over a full year of LAT data, giving 10 ToAs, and then again for 40 ToAs.
We expect root-mean-squares ${\rm RMS_{40}} \approx 2\rm{RMS_{10}}$, roughly confirmed in Figure \ref{fig:RadGamResids}. The pulsars with the largest H-test values yield the smallest RMSs.
The observed jitter being dominated by statistics, we conclude that the brightest two pulsars best constrain LAT timing.
These are PSRs J0614$-$3329 and J1231$-$1411. 
For both, there are about 43,000 photons with energy $> 100$ MeV within $2^\circ$ of the pulsar position for 10 years of data,
and the unweighted H-test value is about 14,000. 
The annual timing variance is of the order of a microsecond, meaning that any contribution from LAT clock jitter is likely to be less than $1/\sqrt 2$ of that.

A less constraining test is provided by PSR J1959+2048, interesting nevertheless because the analysis is simpler.
One of the fastest known MSPs ($P = 1.61$ ms), it also has the narrowest known gamma-ray peak. 
With only 3 years of LAT data, \citet{GuillemotBlackWidow_2012} found a full width at half maximum of $23 \pm 11\,\mu$s. 
We repeated the study using 10 years of data and a Nan\c cay rotation ephemeris and obtained $26^{+8}_{-6}\,\mu$s. 
The uncertainty has decreased in proportion to the larger data sample, and the width agrees with the earlier measurement. The narrow width excludes timing ``accidents'' of tens of microseconds for any substantial fraction of the LAT's observation time. Unfortunately, the background level around PSR J1959+2048 is too high to allow more than a few accurate LAT ToAs.

The above observations constrain putative variations in the LAT timestamps but not their absolute accuracy.
Postulating that the radio and gamma-ray signals for PSRs J1939+2134 and J1959+2048 are emitted from the same regions near the neutron stars, at the same time,
allows a test. $P = 1.6$ ms for both MSPs, so the 2PC gamma-ray peak phases of $-0.003 \pm 0.005$ for both pulsars correspond to time offsets of 
$\delta = -5 \pm 8\,\mu$s, which could then be interpreted as a worst-case estimate of the LAT's on-orbit absolute accuracy.
However, it requires an assumption about the magnetospheric emission.

To understand how the Crab's high braking rate constrains LAT clock accuracy without such an assumption, consider an idealized pulsar whose spindown is described only by $F$ and $\dot F$.
In a time $T$ it will turn $N = FT + \frac{1}{2}\dot F T^2$ times. 
If that time is wrong by $\varepsilon$, giving recorded timestamp values $T' = T+\varepsilon$, we would count $N' = FT' + \frac{1}{2}\dot F T'^2$ turns. 
The observed phase of a gamma-ray peak relative to a radio reference would then drift as 
$(N'-N) = \dot F T \varepsilon$, and the drift in time would be $\delta(T) = P \dot F T \varepsilon$. 
Of pulsars known to be bright in gamma rays, the Crab (PSR J0534+2200) has $P=33.6$ ms and $\dot F = -3.711 \times 10^{-10}$ s$^{-2}$ and thus the largest value of $P \dot F$.
A 10 ms clock error would cause a $\approx 40\,\mu$s peak drift in 10 years.
Modeling the Crab's spindown evolution requires more parameters than just $F$ and $\dot F$, but {\textsc Tempo2} with an accurate rotation ephemeris confirmed the
adequacy of the simple expression for $\delta(T)$ for the present purposes.

\begin{figure}[htbp]
  \centering
\includegraphics[width=1\textwidth]{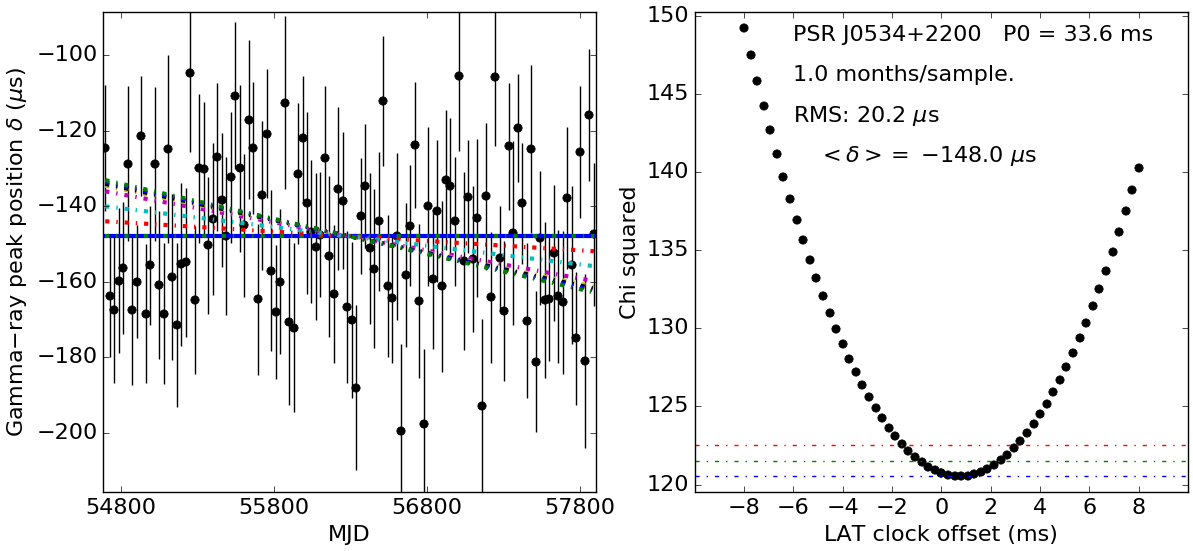} 
\caption{Accuracy of Crab gamma-ray pulse arrival times compared to a radio reference. 
Left: Black dots show monthly average arrival times. 
The solid horizontal line shows the mean offset $<\delta > = -148\, \mu$s between the gamma-ray peak and the main radio pulse.
Dashed diagonals show how the mean would vary in the presence of a putative LAT clock offset relative to UTC.
The steepest slope shown corresponds to $\delta$ drifting by $30\,\mu$s over 10 years.
For clarity, only a sample of negative slopes is shown.    
Right: Each black dot is the chi-squared of the monthly pulse arrival times compared to one of the diagonals. 
Slopes have been converted to putative LAT clock offsets (see text). 
The horizontal dashed lines show the minimum chi-squared value plus 0, 1, and 2, 
showing $\varepsilon < 3$ ms ($1\sigma$ limit).  }
  \label{fig:NoCrabDrift}
\end{figure}

Figure \ref{fig:NoCrabDrift} shows that $\varepsilon < 3$ ms, which we now explain. 
At left, we find the time lag $\delta$ between the main gamma-ray peak and the radio reference peak, for 110 one-month data integrations (the rotation ephemeris covers a bit less than the full ten years).
The mean, irrelevant for this test, is set to $\delta = -148 \, \mu$s as per the erratum\footnote{\url{https://fermi.gsfc.nasa.gov/ssc/data/access/lat/ephems/0534+2200/NewEphem/README}} to \citet{FermiCrab}. 
At right, we show the chi-squared value of these points relative to a series of values $-8 < \varepsilon < 8$ ms. 
The offset of the minimum value of chi-squared from zero varies by $\pm 2$ ms depending on details of the radio rotation ephemeris used to calculate the gamma-ray phases. 
%

The spacecraft GPS cannot be so far off: for $\varepsilon = 1$ ms,  $\varepsilon c = 300$ km, 
incompatible with the $<20$ m position accuracy cited at the beginning of this section.
The integer part of the LAT time (the TTM seconds) is thus correct, since a 1 second error would induce a huge drift in the Crab peak compared to the radio peak. 
The fractional part derives from the well-monitored scaler counts of the 20 MHz oscillator between successive PPS signals, and is thus also reliable. 
The pre-launch ground tests showed mean offsets of $\pm 300$ ns between the LAT and simple GPS times, varying from run to run.
We attribute this to inaccuracies in the simple GPS but have not demonstrated it unequivocally. 
\modified{In summary, the on-board timestamp measurements at the few microsecond level indicate that the clocks are performing as designed, confirming the published 300 ns accuracy.}

%% file: saa.tex
\subsection{South Atlantic Anomaly}\label{subsec:saa}

For mission planning and operations purposes, the boundary of the South Atlantic Anomaly (SAA) is defined by 12-vertex polygons defined separately for the LAT and GBM, specified in terms of geocentric latitude--longitude pairs. These polygons denote the geospatial location of the SAA, a region of space where geomagnetically-trapped charged particle flux densities increase dramatically. The LAT is not able to collect useful science event data inside the SAA region because of excessive dead-time, and also because the anti-coincidence rejection of charged particles is effectively lost, because the high voltages are reduced on the PMTs that read out the ACD, to avoid excessive current draw off the PMT photocathodes and so preserve the long-term performance of the PMTs. Therefore the SAA boundary location is critical to the mission planning process for defining the LAT science data acquisition start and stop commanding. The FSSC and MOC use the LAT SAA boundary polygon to supply predicted SAA entry and exit times, with the corresponding LAT stop and start commands planned to be outside the predicted SAA by 30 seconds, to allow for uncertainty between the predicted times and the actual times. The polygon definition is also uploaded to \Fermi\ and used there to supply the ``in SAA'' flag from the spacecraft flight software to the LAT flight software. 

\begin{figure}[htbp]
  \centering
  \includegraphics[width=15cm]{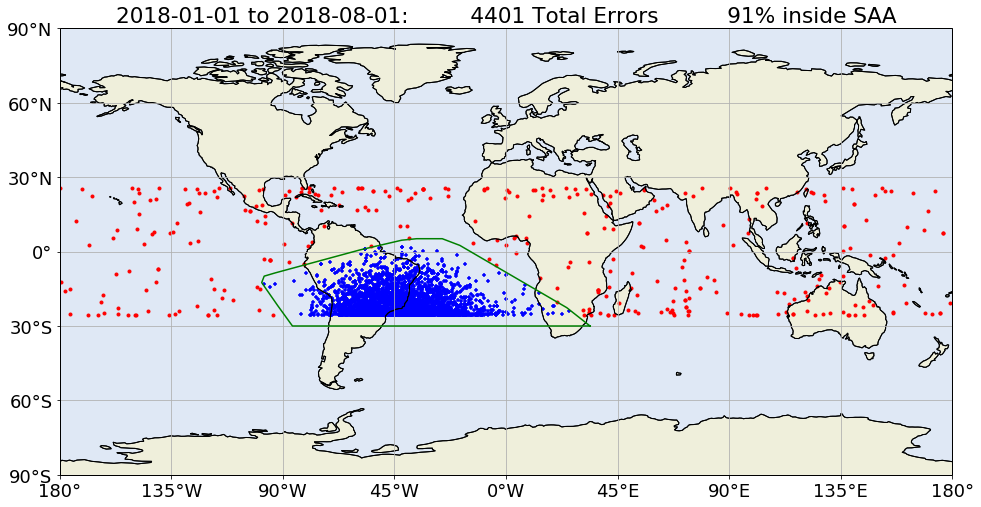}
  \caption{LAT computer memory errors inside and outside the SAA region defined for the LAT (shown by the green polygon).}
  \label{fig:memerr-map}
\end{figure}

Another notable effect of the SAA on the LAT is an associated increase in memory errors in the LAT on-board computers.
\figref{memerr-map} shows the positions of \Fermi\ when memory cell upsets in the memory electronics of the on-board LAT computers occurred, during the first 7 months of 2018. It shows that over 90\% of the LAT memory errors occurred while \Fermi\ was passing through the SAA region, which occupies about 15\% of the total \Fermi\ orbit time. This clearly demonstrates the high density of trapped charged particles within the SAA region.

\begin{figure}[htbp]
  \centering
  \includegraphics[width=15cm]{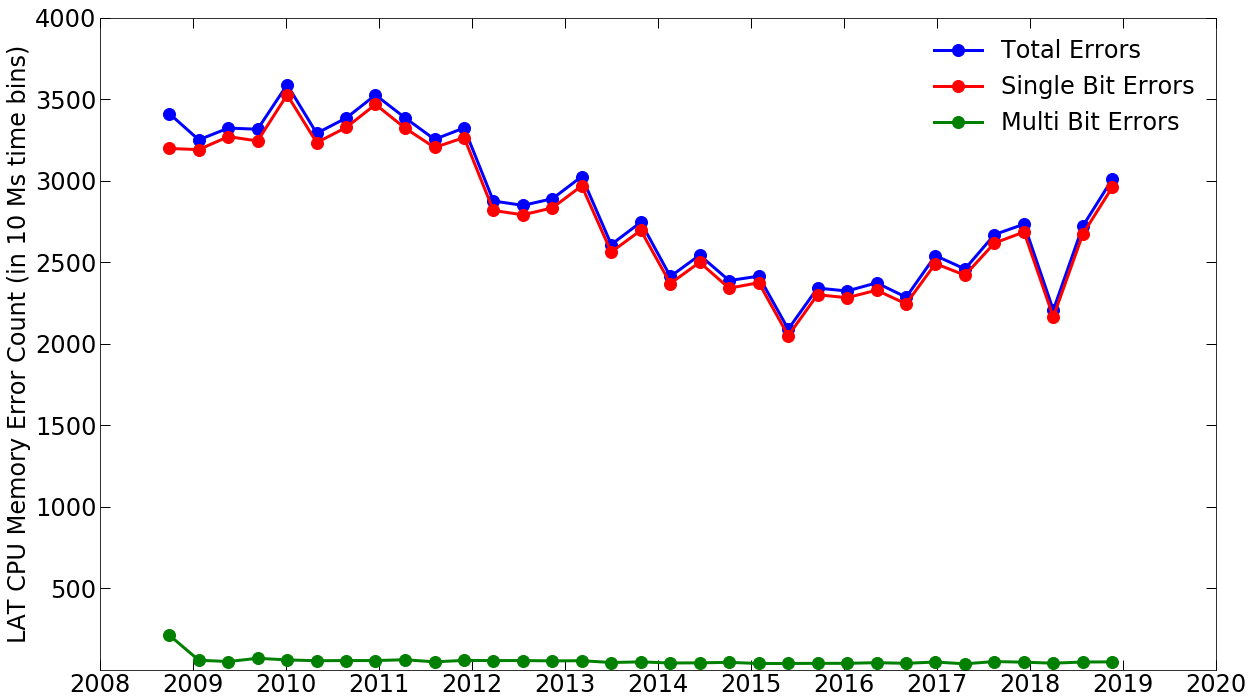}
  \caption{Counts of LAT computer memory errors in successive 10 million second time intervals during the \Fermi\ mission. The first data points in 2018 are lower than expected because that time bin includes the 17-day power-off period for the LAT in March and April 2018.}
  \label{fig:memerr-trend}
\end{figure}

\figref{memerr-trend} shows the number of LAT on-board computer memory cell upsets in successive 10 million second time intervals over the 10 year mission. The slow variation seen is another likely indication of the influence of the 11-year solar cycle on the density of trapped charged particles in the SAA, with the density being reduced during the period of maximum solar activity, which peaked in 2013-2014.

%% file: conclusion.tex
\section{Conclusion}

\modified{The LAT remains in excellent operating condition after more than ten years in space. The \Fermi\ spacecraft is also in good operating condition with no performance concerns. There are no consumables that will limit the lifetime of the LAT or \Fermi. In \Fermi's 13$^{th}$ year in orbit, continued monitoring confirms the LAT's ongoing smooth operation.}

The LAT and the \Fermi\ mission both continue to produce excellent science. \modified{The most recent NASA Senior Review} in 2019 highlighted their crucial roles in the extremely important new frontiers of multi-messenger astrophysics, that were opened by \Fermi\ and the LAT. A gamma-ray burst that was detected by GBM and followed up by the LAT was coincident with the first neutron star-neutron star merger observed by its gravitational radiation with the LIGO and Virgo detectors \citep{GW_GRB_170817A}. The LAT detected a flaring of gamma-ray emission from an active galaxy in association with the detection of a high-energy neutrino from the same direction by the IceCube neutrino detector, marking the first association between neutrino and photon emission from an active galaxy \citep{icecubeNeutrino}. These ground-breaking multi-messenger science results herald the important science still to come from \Fermi\ and the LAT, as multi-messenger capabilities and sensitivities continue to improve.

In addition to multi-messenger science, the LAT and \Fermi\ are expected to centrally participate in coming major advances in multi-wavelength studies. New observatories such as the Rubin Observatory\footnote{\url{https://project.lsst.org/}}
and the Cherenkov Telescope Array\footnote{\url{https://www.cta-observatory.org/}}
are either nearing completion or in active development and will or may overlap with the \Fermi\ mission. Data from these new observatories will leverage further use of both LAT survey data and results, and synergize with detection of transient sources by both the LAT and by the other telescopes.

\modified{The continued good performance of the LAT, and the new era of multi-messenger science that includes important contributions from \Fermi\ and the LAT, plus further enhancements by} new multi-wavelength telescopes and sky surveys coming in the near future, should produce outstanding science for the foreseeable future. 

\textbf{Acknowledgements}: The \textit{Fermi} LAT Collaboration acknowledges generous ongoing support
from a number of agencies and institutes that have supported both the
development and the operation of the LAT as well as scientific data analysis.
These include the National Aeronautics and Space Administration and the
Department of Energy in the United States, the Commissariat \`a l'Energie Atomique
and the Centre National de la Recherche Scientifique / Institut National de Physique
Nucl\'eaire et de Physique des Particules in France, the Agenzia Spaziale Italiana
and the Istituto Nazionale di Fisica Nucleare in Italy, the Ministry of Education,
Culture, Sports, Science and Technology (MEXT), High Energy Accelerator Research
Organization (KEK) and Japan Aerospace Exploration Agency (JAXA) in Japan, and
the K.~A.~Wallenberg Foundation, the Swedish Research Council and the
Swedish National Space Board in Sweden.
 
Additional support for science analysis during the operations phase is gratefully
acknowledged from the Istituto Nazionale di Astrofisica in Italy and the Centre
National d'\'Etudes Spatiales in France. This work performed in part under DOE
Contract DE-AC02-76SF00515.